\documentclass[12pt]{article}

\usepackage{textcomp}
\usepackage{chetnew}
\usepackage{amssymb,amsmath,amsfonts,mathrsfs,braket}
\usepackage[utf8]{inputenc}
\usepackage{tikz,tikz-cd,url}
\usepackage{xcolor,lmodern}
\usetikzlibrary{arrows,decorations.markings,shapes.geometric,decorations.pathmorphing}
\tikzset{snake it/.style={decorate, decoration=snake}}

\usepackage{bm}
\usepackage[colorinlistoftodos,textsize=tiny]{todonotes}

\renewcommand{\d}[1]{\ensuremath{\operatorname{d}\!{#1}}}

\def\one{{\,\hbox{1\kern-.8mm l}}}

\newcommand{\Tr}{\mathrm{Tr}}

\newcommand{\CC}{\mathcal{C}}

\newcommand{\CN}{\mathcal{N}}

\newcommand{\QQ}{\mathcal Q}
\newcommand{\QS}{\mathcal S}
\newcommand{\RR}{\mathcal R}

\allowdisplaybreaks

\def\makeatletter{\catcode`\@=11}
\makeatletter
\def\mathbox#1{\hbox{$\m@th#1$}}%
\def\math@ccstyles#1#2#3#4#5#6#7{{\leavevmode
      \setbox0\mathbox{#6#7}%
      \setbox2\mathbox{#4#5}%
      \dimen@ #3%
      \baselineskip\z@\lineskiplimit#1\lineskip\z@
      \vbox{\ialign{##\crcr
             \hfil \kern #2\box2 \hfil\crcr
             \noalign{\kern\dimen@}%
             \hfil\box0\hfil\crcr}}}}
\def\mathaccstyles{\math@ccstyles\maxdimen}
\def\maththroughstyles{\math@ccstyles{-\maxdimen}}
\def\unity%
 {\maththroughstyles{.45\ht0}\z@\displaystyle {\mathchar"006C}\displaystyle 1}

\def\AA{{\cal A}}

\def\CC{{\cal C}}

\def\EE{{\cal E}}
\def\FF{{\cal F}}

\def\II{{\cal I}}
\def\JJ{{\cal J}}
\def\KK{{\cal K}}
\def\LL{{\cal L}}
\def\MM{{\cal M}}
\def\NN{{\cal N}}
\def\OO{{\cal O}}
\def\PP{{\cal P}}
\def\QQ{{\cal Q}}
\def\RR{{\cal R}}

\def\TT{{\cal T}}

\def\YY{{\cal Y}}
\def\ZZ{{\cal Z}}

\def\IR{{\mathbb R}}
\def\IT{{\mathbb T}}

\def\IH{{\mathbb H}}

\def\L{{\boldsymbol \lambda}}

\def\half{{\frac{1}{2}}}

\def\Tr{{\rm {Tr}}}

\def\d{{\partial}}
\def\p{{\partial}}

\def\q{{\mathfrak q}}

\def\beq{\begin{equation}}
\def\eeq{\end{equation}}
\newcommand{\bea}{\begin{eqnarray}}
\newcommand{\eea}{\end{eqnarray}}
\def\bal{\begin{align}}
\def\eal{\end{align}}

\preprint{CCTP-2019-9\hfill QMUL-PH-19-30\\DCPT-19/29 \hfill
  DESY-19-189\\ ITCP-IPP-2019/8}

\title{\vspace{-1.cm}Type-B Anomaly Matching and the 6D (2,0) Theory
}

\author{Vasilis Niarchos\;$^{a,\clubsuit,}$\footnote{On leave of absence from the Department of Mathematical Sciences and Centre for Particle Theory, Durham University, Durham DH1 3LE, UK}, Constantinos Papageorgakis\;$^{b,\diamondsuit}$ and Elli Pomoni\;$^{c,\spadesuit}$}

\affiliation{$^a$ ITCP \& CCTP, Department of Physics,\\
University of Crete, 71003 Heraklion, Greece
\vspace{0.3cm} $ $\\

$^b$CRST and School of Physics and Astronomy\\ Queen Mary University of London, London E1 4NS, UK \vspace{0.3cm} $ $ \\

$^c$DESY, Theory Group, Notkestrasse 85, Building 2a, 22607 Hamburg, Germany \\

\vspace{0.3cm}
{\tt \small$^\clubsuit$niarchos@physics.uoc.gr, $^\diamondsuit$c.papageorgakis@qmul.ac.uk, 
$^\spadesuit$elli.pomoni@desy.de}}

\abstract{We study type-B conformal anomalies associated with $\frac{1}{2}$-BPS Coulomb-branch operators in 4D $\mathcal N=2$ superconformal field theories. When the vacuum preserves the conformal symmetry these anomalies coincide with the two-point function coefficients in the Coulomb-branch chiral ring. They are non-trivial functions of exactly-marginal couplings that can be determined from the $S^4$ partition function. In this paper, we examine the fate of these anomalies in vacua of the Higgs-branch moduli space, where conformal symmetry is spontaneously broken. We argue non-perturbatively that these anomalies are covariantly constant on conformal manifolds. In some cases, this can be used to show that they match in the broken and unbroken phases. Thus, we uncover a new class of data on the Higgs branch of 4D $\mathcal N=2$ conformal field theories that are exactly computable. An interesting application of this matching occurs in $\mathcal N=2$ circular quivers that deconstruct the 6D (2,0) theory on a torus. In that context, we argue that 4D supersymmetric localisation can be used to calculate non-trivial data involving $\frac{1}{2}$-BPS operators of the 6D theory as exact functions of the complex structure of the torus.}

\date{}

\begin{document}

\maketitle

\hypersetup{pageanchor=true}

\setcounter{tocdepth}{2}

\toc

\section{Introduction and Summary of Results}
\label{summary}

In even spacetime dimensions a conformal field theory (CFT) can exhibit two types of conformal anomalies, type-A and type-B \cite{Deser:1993yx}. The type-A conformal anomalies are a close analogue of the chiral anomaly---since they do not introduce a mass scale, they can be expressed in terms of a topological invariant that integrates to zero in topologically-trivial spaces. An example is the coefficient $a$ multiplying the Euler density in the Weyl anomaly of 4D CFTs. In contrast, the type-B anomalies are associated with divergences that require the introduction a scale. They manifest themselves in the Weyl transformation of the effective action in terms of a scalar density that does not integrate to zero. An example of a type-B anomaly is the coefficient $c$ multiplying the Weyl tensor squared part of the Weyl anomaly in 4D CFTs. Another example, is the coefficient of two-point functions of operators with integer scaling dimension $\Delta=\frac{D}{2}+n$, where $D$ is the spacetime dimension and $n\in \mathbb Z_{\geq 0}$. As we review in Sec.~\ref{typeB}, one way to discover this anomaly is by noting the presence of a logarithmic divergence in the two-point function in momentum space.

In a certain sense, type-A anomalies are more `rigid' quantities compared to type-B anomalies. For instance, it has been argued \cite{Schwimmer:2010za} that type-A anomalies should match in phases of a CFT with unbroken and spontaneously-broken conformal invariance. Instead, a similar argument for type-B anomalies is much harder to make.\footnote{The robustness of type-A anomalies compared to that of type-B anomalies is also manifest on their dependence on exactly-marginal couplings. For example, the Wess--Zumino consistency conditions can be used to show that the $a$ anomaly is independent of exactly-marginal couplings, but in general the $c$ anomaly may have a nontrivial dependence. We refer the reader to \cite{Nakayama:2017oye} (below Eq.~(6)) for a discussion on this point. The vast majority of the type-B anomalies that we will consider in this paper also depend non-trivially on exactly-marginal couplings.} 

The main goal of this paper is to examine non-perturbative properties of type-B anomalies for certain integer-dimension operators (the so-called Coulomb-branch operators) on the Higgs branch of 4D $\NN=2$ CFTs. In contrast to the general expectation, we argue that there are cases in this context, where these anomalies potentially match on the Higgs branch and sketch a formal, non-perturbative argument in favour of this matching. To our knowledge, there is no precedence of such anomaly matching in the literature.

More specifically, we focus on 4D $\NN=2$ superconformal field theories (SCFTs) with non-trivial superconformal manifolds, namely continuous families of $\NN=2$ SCFTs related by exactly-marginal deformations. In the conformal phase, such theories typically possess a rich spectrum of $\frac{1}{2}$-BPS Coulomb-branch operators (CBOs), whose scaling dimension is integral and protected by supersymmetry. Hence, there are natural type-B anomalies associated with the two-point functions of these operators. These two-point functions involve a chiral and an anti-chiral CBO, $\OO_I$ and $\bar \OO_J$ respectively. We call the corresponding type-B anomaly $G_{I\bar J}^{\rm CFT}$.

Additionally, each SCFT on the $\NN=2$ superconformal manifold possesses a moduli space of Higgs-branch vacua where the conformal symmetry is spontaneously broken.\footnote{4D $\NN=2$ SCFTs also possess moduli spaces of Coulomb-branch vacua parametrised by vacuum expectation values of Coulomb-branch operators. We do not consider such vacua in this paper.} In the non-conformal theory of the Higgs branch we isolate a specific contact term in the three-point function $\langle T^\mu_{~\mu}(x_1) \OO_I(x_2) \bar \OO_J(x_3) \rangle$ that leads to a Weyl anomaly. We define its dimensionless coefficient, that we call $G_{I\bar J}^{\IH}$, as the value of the type-B anomaly of interest. This coefficient captures a corresponding dilaton coupling in the effective action on the Higgs branch. More details about the definition of $G_{I\bar J}^{\IH}$ appear in Sec.~\ref{typeB} and Sec.~\ref{app:Ward}. 

Both $G_{I\bar J}^{\rm CFT}$ in the unbroken phase and $G_{I\bar J}^{\IH}$ in the broken phase are, in general, non-trivial functions of the exactly-marginal couplings that parametrise the superconformal manifold. We will argue that whenever there is at least one point on the superconformal manifold where these quantities match, the matching will necessarily extend to a finite region of couplings around this point. In other words, matching at one point, e.g.\ at weak coupling, guarantees that $G_{I\bar J}^{\rm CFT}$ and $G_{I\bar J}^{\IH}$ are the same functions of the exactly-marginal couplings in a finite region of the superconformal manifold. All the relevant concepts underlying the quantities $G_{I\bar J}^{\rm CFT}$ and $G_{I\bar J}^{\IH}$ are reviewed and defined in Secs.~\ref{setup} and \ref{app:Ward}.

The argument that we present in this paper is not a general argument for type-B anomaly matching. It relies heavily on the presence of $\NN=2$ supersymmetry and the fact that we examine type-B anomalies associated with CBOs on Higgs-branch moduli spaces. The latter is crucial for two reasons. First, the Higgs-branch moduli space is parametrised by the vacuum expectation values (vevs) of a set of $\frac{1}{2}$-BPS operators, called Higgs-branch operators (HBOs).\footnote{In standard gauge-theory examples of 4D $\NN=2$ SCFTs, like the 4D $\NN=2$ superconformal QCD, the CBOs are Casimirs of adjoint scalar fields and the HBOs are gauge-invariant mesonic operators.} Ref.\ \cite{Niarchos:2018mvl} argued that the states corresponding to HBOs have vanishing Berry phases under parallel transport on the superconformal manifold,\footnote{Equivalently, \cite{Niarchos:2018mvl} argued that the holomorphic vector bundle of Higgs-branch superconformal primaries is equipped with a flat connection.} which is a fact that we will use. Second, the argument relies crucially on a superconformal Ward identity whose integrated form receives exactly the same contributions in the unbroken and broken phases. We present a potential route towards the proof of this statement and explain why we expect it to be generally correct; an alternative and more general argument that does not rely on supersymmetry is presented in \cite{Andriolo:2022lcb}.

We now list the three ingredients that constitute our formal argument in  more detail:
\begin{itemize}

\item[(1)] We argue that derivatives of the correlation functions on the Higgs branch with respect to the exactly-marginal couplings can be defined using the same connection on the superconformal manifold that appears in the unbroken CFT phase in the context of conformal perturbation theory. This argument appears in Sec.~\ref{data}.

\item[(2)] We use a Ward identity for the superconformal currents to  motivate the statement that the anomalous contact term in the three-point function of the trace of the energy-momentum tensor with two CBOs, $\langle T^\mu_{\mu}(x_1) \OO_I(x_2) \bar \OO_J(x_3) \rangle$,  is covariantly constant on the superconformal manifold. The argument works in the same manner in the unbroken and broken phases, under the assumption that the dilatino that couples to the superconformal current on the Higgs branch cannot exhibit a massless pole on the external momentum, and contribute contact terms to the correlation functions that we consider. The Ward identities of interest are presented in Sec.~\ref{Ward}.

\item[(3)] Assuming that both $G_{I\bar J}^{\rm CFT}$ and $G_{I\bar J}^{\IH}$ are covariantly-constant quantities,\footnote{See also \cite{Andriolo:2022lcb}.} it suffices to show that they match at one point of the superconformal manifold. Then, the vanishing covariant derivatives guarantee that the matching holds in a finite range of exactly-marginal couplings. We summarise this argument in Sec.~\ref{match}. 

In Sec.~\ref{examples} we present several concrete examples (in 4D $\NN=2$ superconformal QCD, 
a 4D $\NN=2$ circular quiver and 4D $\NN=4$ SYM theory), where the tree-level matching can be established by direct computation and the non-perturbative matching follows from our general arguments. It is worth noting that a special case of type-B anomaly matching in the 4D $\NN=4$ SYM theory can also be argued independently using supersymmetry and the matching of chiral anomalies, forming a partial, independent check of our claim.
\end{itemize}

When all the assumptions are met, the proposed anomaly matching has immediate implications for the non-perturbative structure of $\NN=2$ SCFTs. The two-point function coefficients of CBOs in $\NN=2$ SCFTs (in the conformal phase) are, as we noted, non-trivial functions of the exactly-marginal coupling constants. Nevertheless, it has been recently shown  \cite{Gerchkovitz:2014gta,Gomis:2015yaa,Baggio:2014sna,Baggio:2014ioa,Gerchkovitz:2016gxx} that they can be obtained directly from the $S^4$ partition function of the theory and that they obey an integrable set of differential equations with respect to the exactly-marginal couplings, called $tt^*$ equations. In Lagrangian theories the $S^4$ partition function can be determined exactly using supersymmetric localisation techniques \cite{Pestun:2007rz}. As a result, the proposed anomaly matching implies that there is a corresponding set of data for the non-conformal theory on the Higgs branch, which is also determined by the same functions of the exactly-marginal couplings. In Sec.~\ref{localisation} we outline the supersymmetric-localisation data that determine the anomalies of interest in the conformal phase of a 4D $\NN=2$ circular-quiver theory.

The anomaly matching of CBO data on the Higgs branch of 4D $\NN=2$ SCFTs has an additional interesting application in the context of dimensional deconstruction. Some time ago it was proposed that a certain limit on the Higgs branch of a 4D $\NN=2$ circular quiver CFT---the same quiver that we analyse in Secs.~\ref{quiver} and \ref{localisation}---defines the 6D (2,0) theory of type $A$ on a 2-torus  \cite{ArkaniHamed:2001ie}. In this construction, the exactly-marginal coupling of the 4D theory controls the complex structure of the torus and the 4D CBOs map to the Kaluza-Klein (KK) modes of local $\frac{1}{2}$-BPS operators of the 6D theory. The 4D type-B anomaly matching that we present in this paper then has a striking implication: there exist data of these local 6D $\frac{1}{2}$-BPS operators that depend non-trivially on the complex structure of the torus, and yet can be computed non-perturbatively from the 4D perspective by doing a supersymmetric-localisation computation {\it in the CFT phase} of the 4D quiver. So far, most of the structure of the non-Lagrangian 6D (2,0) theory has remained a mystery, lying beyond the reach of existing techniques. Important exceptions are recent results on partition functions \cite{Kallen:2012cs,Kallen:2012va,Kallen:2012zn,Kim:2012ava,Kim:2012qf,Kim:2013nva}, chiral-algebra structures \cite{Beem:2014kka} or the superconformal-bootstrap programme \cite{Beem:2015aoa}. The results implied by the anomaly matching in this paper provide the first example of exact, non-supersymmetric data of local operators in the 6D (2,0) theory. We present the details of this relationship in Sec.~\ref{apps}.  A collection of useful facts as well as our conventions are relegated to a set of appendices at the end of the manuscript.

\vskip .3cm

\noindent{\bf Note added (December 2024):} In previous versions of this paper, we reported a mismatch between the CFT- and Higgs-phase type-B anomaly of the $N$-noded, 4D circular-quiver theory at finite $N$ and at tree level. We thank A.~Schwimmer and S.~Theisen for pointing out a computational error that led to that claim. The correct computation yields tree-level matching at all values of $N$. We have updated the current version of the paper accordingly.

\section{Review of Useful Concepts}
\label{setup}

\subsection{4D $\NN=2$ Superconformal Manifolds}
\label{conformalmanifolds}

The $\NN=2$ superconformal field theories in four spacetime dimensions are often members of continuous families of theories connected by exactly-marginal deformations that preserve the $\NN=2$ supersymmetry. The continuous space of such theories is called a superconformal manifold and is parametrised by the corresponding exactly-marginal couplings $\L = \{ \lambda^i \}$. The index $i$ labels different directions on the tangent space of the superconformal manifold.

The R-symmetry group of a 4D $\NN=2$ SCFT is $SU(2)_R \times U(1)_r$. The theory possesses two types of $\half$-BPS superconformal-primary operators: Coulomb-branch operators (CBOs), which are charged under the $U(1)_r$ symmetry but are neutral under the $SU(2)_R$, and Higgs-branch operators (HBOs), which are charged under the $SU(2)_R$ symmetry but are neutral under the $U(1)_r$. A generic $\NN=2$ theory also exhibits moduli spaces of vacua, where (a subset of) the above operators obtain a non-vanishing vev. We will return to the moduli spaces in the next subsection. For the moment we focus exclusively on the vacuum that preserves the full $\NN=2$ superconformal algebra of the theory.

The CBOs can preserve either the left- or right-chiral part of the $\NN=2$ Poincar\'e supersymmetry. We will denote the first set as $\OO_I$ and the second set as $\bar \OO_J$ (one can think of one set as the complex conjugate of the other). By default, the scaling dimension $\Delta$ of a (anti-)chiral primary CBO is related to its $U(1)_r$ charge $r$ through the relation $\Delta=|r|$ and is naturally an integer greater or equal to 2 that does not depend (generically) on the value of exactly-marginal couplings. In fact, CBOs of scaling dimension 2 are special, because a certain supersymmetric descendant of such operators has scaling dimension 4 and defines exactly-marginal deformations of the theory that preserve the $\NN=2$ supersymmetry. Therefore, for generic 4D $\NN=2$ SCFTs with a non-empty set of $\Delta=2$ CBOs there is naturally a non-vanishing superconformal manifold.\footnote{Clearly, there are examples where this is not the case, e.g. in Argyres-Douglas theories.}

Each of these superconformal primaries forms a chiral (or anti-chiral) ring under OPE multiplication. The data of the chiral ring are encoded in the two- and three-point functions
\bea
\label{setup1aa}
\langle \OO_I(x) \bar \OO_J(y) \rangle &=& \frac{G_{I\bar J}}{|x-y|^{2\Delta}}~, ~~
\\
\label{setup1ab}
\langle \OO_I(x) \OO_J(y) \bar \OO_K(z) \rangle &=& \frac{C_{I J\bar K}}{|x-y|^{\Delta_I+\Delta_J-\Delta_K} |x-z|^{\Delta_I+\Delta_K-\Delta_J} |y-z|^{\Delta_J+\Delta_K-\Delta_I}}
~.
\eea
In \eqref{setup1aa} $\Delta$ is the common scaling dimension of the operators, while in \eqref{setup1ab} $\Delta_I$ etc.\ denotes the scaling dimension of each operator. The three-point function coefficients $C_{IJ\bar K}$ are related to the OPE coefficients $C_{IJ}^L$ in the CBO chiral ring via the relation $C_{IJ \bar K} = C_{IJ}^L G_{L\bar K}$.

Clearly, as one varies the exactly-marginal couplings on a superconformal manifold one has the freedom to choose an arbitrary (possibly coupling-constant-dependent) normalisation of the operators. One of the possible normalisations, which was discussed in \cite{Baggio:2014ioa}, is to arrange for the OPE coefficients $C_{IJ}^K$ to be either one or zero. This is a special case of the so-called `holomorphic gauge'. In this case, all the non-trivial information about the structure of the CBO chiral ring is contained in the two-point function coefficients $G_{I\bar J}$, which will soon play a protagonistic role in their interpretation as type-B anomalies. These coefficients are, in general, non-trivial functions of the exactly-marginal coupling constants $\L$, the computation of which requires powerful non-perturbative tools. In recent years this has been possible using supersymmetric localisation on $S^4$. For further details on normalisation conventions and related issues we refer the reader to \cite{Baggio:2014ioa}.

The coupling-constant dependent choice of the operators on the superconformal manifold raises various practical questions. How does one sensibly relate data for different values of $\L$, and how does one write equations that behave covariantly under coupling-constant dependent redefinitions of the operators? A related question is the following: When varying the exactly-marginal couplings to relate correlation functions at nearby points on the superconformal manifold, one is instructed to compute correlation functions with integrated insertions of the exactly-marginal operators. For example, the naive derivative of an $n$-point function $\langle O_1(x_1) O_2(x_2) \cdots O_n(x_n) \rangle$ at separated points $x_1,\ldots, x_n$ with respect to the exactly-marginal coupling $\lambda^i$ is 
\beq
\label{setup2aa}
\left\langle \int d^4 x\, \Phi_i (x) O_1(x_1) O_2(x_2) \cdots O_n(x_n) \right\rangle\;,
\eeq
where $\Phi_i$ is the corresponding exactly-marginal operator. This quantity is not well defined. It exhibits ultraviolet (UV) divergences when the integrated insertion $\Phi_i$ collides with any of the other fixed insertions $O_\ell(x_\ell)$ and needs to be regularised. 

One approach to address this problem is to define the regulated version of \eqref{setup2aa} as a modified (covariant) derivative with respect to the coupling $\lambda^i$
\beq
\label{setup2ab}
\nabla_i \langle O_1(x_1) O_2(x_2) \cdots O_n(x_n) \rangle \equiv
\left\langle \int d^4 x\, \Phi_i (x) O_1(x_1) O_2(x_2) \cdots O_n(x_n) \right\rangle_{\text{regulated}}\;.
\eeq
For example, one can regulate by cutting balls of finite size around the operator insertions and removing divergent terms in the limit of vanishing cutoff, as was done in \cite{Sonoda:1991mv,Ranganathan:1992nb,Ranganathan:1993vj,Sonoda:1993dh}. Different subtraction schemes define different types of connections on the conformal manifold, some of which may not have desirable features. In this paper we follow \cite{Ranganathan:1993vj,Papadodimas:2009eu} and adopt a natural scheme where one subtracts the terms that are divergent in the spatial cutoff and keeps the finite remainder. In \cite{Baggio:2017aww} it was shown that this prescription reproduces the Berry connection in radial quantisation.

Alternatively, one can keep the definition
\beq
\label{setup2ad}
\partial_i \langle O_1(x_1) O_2(x_2) \cdots O_n(x_n) \rangle =
\left\langle \int d^4 x\, \Phi_i (x) O_1(x_1) O_2(x_2) \cdots O_n(x_n) \right\rangle
\eeq
intact and proceed to cancel divergences by modifying the OPEs with suitable contact terms. Some of these contact terms express a coupling-constant dependent operator mixing. A natural notion of connection on the conformal manifold arises in this manner from contact terms in OPEs that involve the exactly-marginal operators. This approach was advocated in \cite{Kutasov:1988xb}. 

These observations endow the conformal manifold of SCFTs with an intrinsic geometric structure. For CBOs in a 4D $\NN=2$ superconformal manifold $\MM$ this structure can be summarised in the following manner \cite{Papadodimas:2009eu,Baggio:2014ioa}. The CBOs are sections of a holomorphic vector bundle on $\MM$ and the two-point function coefficients $G_{I\bar J}$ (which are functions of the exactly-marginal couplings $\L$) can be viewed as components of a Hermitian metric on these bundles. Crucially, these bundles possess a connection $\nabla$ that defines parallel transport. Physically, parallel transport on these bundles gives the relation between  CBOs in SCFTs at different values of the exactly-marginal couplings. $\nabla$ is compatible with the metric $G$, namely the covariant derivative with respect to $\lambda^k$ vanishes
\beq
\label{setup2ac}
\nabla_k G_{I\bar J} = 0 ~ \Leftrightarrow ~ \partial_k G_{I\bar J} - (A_k)^L_I G_{L\bar J} - (A_k)^{\bar L}_{\bar J} G_{I \bar L} = 0
~.
\eeq
In these expressions $(A_k)_M^L$, $(A_k)_{\bar M}^{\bar L}$ are the components of the connection on the bundle of CBOs. The indices $L$, $\bar L$ run over the set of chiral and anti-chiral CBOs. 

Eq.\ \eqref{setup2ac} can be viewed as an automatic consequence of the definition of the connection.\footnote{For general CBOs in 4D $\NN=2$ SCFTs \eqref{setup2ac} can be proven using the regularisation prescription mentioned below Eq.~\eqref{setup2ab} and by employing the superconformal Ward identities of Sec.~\ref{Ward}. This argument is reviewed in Sec.~\ref{unbrokenphase}.} It expresses the coupling-constant dependence of the two-point functions in the scheme that defines the connection, and in that respect it is a trivial relation. However, it may be useful to appreciate that if one were to be given the connection through some independent means, then the coupling-constant dependence would follow from this equation. The role and usefulness of the $tt^*$ equations is related to this statement \cite{Papadodimas:2009eu,Baggio:2014ioa}. The $tt^*$ equations provide the curvature of the connection on the chiral ring of CBOs in terms of the two- and three-point function coefficients in the chiral ring. As we have already mentioned, in a normalisation scheme where the OPE coefficients in the chiral ring are trivialised, all the information is encoded in the two-point function coefficients. Then, the $tt^*$ equations (combined with \eqref{setup2ac}) become an integrable set of differential equations for the coupling-constant dependence of the two-point functions. A prescription based on supersymmetric localisation on $S^4$ provides a non-trivial solution to these equations \cite{Gerchkovitz:2014gta,Gomis:2015yaa,Baggio:2014sna,Baggio:2014ioa,Gerchkovitz:2016gxx}.

So far we have focussed on the geometric structure of the conformal manifold in the unbroken, conformal phase of the theory.  We will explore the role of the connection in the broken phase of the Higgs branch in Sec.~\ref{data}.

\subsection{Two-point Functions of CBOs as Type-B Anomalies}
\label{typeB}

When an even-$D$-dimensional CFT possesses operators with scaling dimension $\Delta=\frac{D}{2}+n$, $n\in \mathbb Z_{\geq 0}$, there are corresponding type-B Weyl anomalies associated with the two-point function coefficients of these operators. Since the CBOs in 4D $\NN=2$ SCFTs are naturally operators of this type we will focus the remaining discussion on them. 

As we noted in Eq.~\eqref{setup1aa}, the two-point function of two CBOs in position space is a perfectly well-defined correlation function at separated points. However, the Fourier transform of this function is not well defined and a type-B Weyl anomaly arises in momentum space as a logarithmic contribution to the two-point function
\beq
\label{setupab}
\langle \OO_I(p) \bar \OO_J(-p) \rangle \simeq (-1)^{n+1} \frac{\pi^2 G_{I\bar J}}{2^{2n} \Gamma(n+1)\Gamma(n+\frac{D}{2})} p^{2n} \log \left( \frac{p^2}{\mu^2}\right)
~,
\eeq
where $\simeq$ means that this is correct up to non-logarithmic scheme-dependent terms that we drop.
This is clearly proportional to the two-point function coefficients
$G_{I\bar J}$. The logarithm has introduced a scale $\mu$ and the conformal anomaly
\beq
\label{setupac}
\mu \frac{d}{d\mu} \langle \OO_I(p) \bar \OO_J(-p) \rangle = (-1)^{n} \frac{\pi^2 G_{I\bar J}}{2^{2n-1} \Gamma(n+1)\Gamma(n+\frac{D}{2})} p^{2n}
\eeq
translates in position space to a specific contact term
\beq
\label{setupaca}
\mu \frac{d}{d\mu} \langle \OO_I(x) \bar \OO_J(0) \rangle \propto G_{I\bar J} \Box^{n} \delta(x)
~.
\eeq
This contact term reflects an inconsistency between the Ward identities for diffeomorphism invariance and the Ward identities for Weyl invariance. The clash between these Ward identities in the conformal phase is reviewed in Sec.~\ref{unbrokenWard}.

Alternatively, one may consider the generating functional $W$ of correlation functions for CBOs. Adding spacetime-dependent sources to the action for the operators $\OO_I$, $\bar\OO_J$ 
\begin{align}
  \label{eq:7}
\delta S = \int d^4 x \, \left( t^I(x) \OO_I(x) + \bar t^J(x) \bar \OO_J(x)  \right)\;,  
\end{align}
for a theory on a curved manifold with metric $g_{\mu\nu}$, $W$ becomes a functional of $g_{\mu\nu}$ and $t^I$, $\bar t^J$. A (generalised)\footnote{The term `generalised' reflects an interesting modification of the standard geometric part of the Weyl transformation, $\delta_\sigma = 2 \sigma(x) g_{\mu\nu}(x) \frac{\delta}{\delta g_{\mu\nu}(x)}$, when the operators $\OO_I$, $\bar \OO_J$ are irrelevant. As explained in \cite{Schwimmer:2019efk}, the modification requires the inclusion of a metric $\beta$-function in the definition of $\delta_\sigma$. Otherwise, the Osborn equation \eqref{anomai} does not make sense. We refer the reader to  \cite{Schwimmer:2019efk} for further details on this point.} Weyl transformation expresses the type-B anomaly $\AA$ via Osborn's equation \cite{Osborn:1991gm}
\beq
\label{anomai}
\delta_\sigma W(\{t,\bar t\} , g_{\mu\nu}) = \int d^4 x \, \sqrt{g}\, \sigma(x) \AA\left( \{ t, \bar t \},g_{\mu\nu}\right)
~,
\eeq
where $\sigma$ is the infinitesimal parameter of the Weyl transformation and
\beq
\label{anomaj}
\AA = G_{I\bar J} t^I \Delta_c \bar t^J
\eeq
is (up to an overall numerical constant) a differential operator of the form
\beq
\label{anomak}
\Delta_c = \Box^n + {\rm curvature~terms}\;,
\eeq
with $G_{I\bar J}$ the two-point function coefficients listed above. In this form, it is clear that the type-B anomalies $G_{I \bar J}$ reflect a specific piece in the three-point function $\langle T_{\mu\nu}(x_1) \OO_I(x_2) \bar\OO_J(x_3) \rangle$. 

As an example, let us examine the case of two CBOs with scaling dimension $\Delta=4$. In momentum space the three-point function $\langle T_{\mu\nu}(p_1) \OO_I(p_2) \bar \OO_J(p_3) \rangle$ contains the term (see e.g.\ \cite{Bzowski:2018fql}, Eq.~(3.58))
\beq
\label{anomal}
\langle T_{\mu\nu}(p_1) \OO_I (p_2) \bar \OO_J (p_3) \rangle = \ldots + \frac{1}{3} \frac{\pi^2 G_{I \bar J}}{16 \Gamma(3) \Gamma(4)} p_2^2 p_3^2 \left( \eta_{\mu\nu} - \frac{(p_1)_\mu (p_1)_\nu}{p_1^2} \right)
~.
\eeq
For the corresponding three-point function of the trace of the energy-momentum tensor $T\equiv T^\mu_{~\mu}$
\beq
\label{anomala}
\langle T (p_1) \OO_I(p_2) \bar \OO_J(p_3) \rangle = \ldots + \frac{\pi^2 G_{I \bar J}}{16 \Gamma(3) \Gamma(4)} p_2^2 p_3^2 
~,
\eeq
which isolates the $G_{I\bar J} t^I \Box^2 \bar t^J$ contact term in \eqref{anomai}-\eqref{anomak}.
 
More generally, the anomalous term for operators $\OO_I$, $\bar \OO_J$ with common scaling dimension $\Delta = 2+n$ is 
\beq
\label{anomam}
\langle T (p_1) \OO_I(p_2) \bar\OO_J(p_3) \rangle = \ldots +(-1)^n \frac{\pi^2 G_{I\bar J}}{2^{2n} \Gamma(n+1)\Gamma(n+2)} p_2^n p_3^n
~.
\eeq
For a CBO that has $\Delta=2$ we get, in particular,
\beq
\label{anomao}
\langle T (p_1) \OO_I(p_2) \bar \OO_J(p_3) \rangle = \ldots + \pi^2 G_{I\bar J}
~.
\eeq
We will examine this case in more detail via several examples in Sec.~\ref{examples}.

Let us summarise the key points of the preceding discussion. In a 4D $\NN=2$ SCFT there is a chiral ring of CBOs with integer scaling dimensions that are independent of the exactly-marginal couplings on the superconformal manifold. In the vacuum of unbroken conformal invariance the two-point function coefficients $G_{I\bar J}$ of these operators play two related roles: 
\begin{itemize}
\item[$(a)$] They define a metric on the holomorphic bundles of CBOs.   
\item[$(b)$] They express a type-B Weyl anomaly that captures a particular part in three-point functions of the CBOs with the energy-momentum tensor. For three-point functions involving the  trace of the energy-momentum tensor the Weyl anomaly appears in position space as a contact term.
\end{itemize}

\subsection{Type-B anomalies on the Higgs Branch}
\label{anoHiggs}

The above statements concerned properties of the theory exclusively in a vacuum with unbroken conformal invariance. Typically, at each point $\L$ of a superconformal manifold an SCFT possesses a non-trivial moduli space of vacua labelled by the non-vanishing vevs of a set of operators. 4D $\NN=2$ SCFTs have different types of moduli spaces of vacua: Higgs-branch, Coulomb-branch and mixed Coulomb-Higgs moduli spaces. In each of these spaces the non-vanishing vev  spontaneously breaks the conformal invariance as well as part of the R-symmetry of the theory. By definition, the Higgs-branch moduli spaces are characterised by non-vanishing vevs of the Higgs-branch operators. Hence, in these vacua the $SU(2)_R$ is spontaneously broken but the $U(1)_r$ is preserved. In contrast, the Coulomb-branch moduli spaces are characterised by non-vanishing vevs of Coulomb-branch operators. Accordingly, in these vacua the $U(1)_r$ is spontaneously broken but the $SU(2)_R$ is preserved. 

In what follows we will focus on the properties of Higgs-branch vacua. We assume that over each point $\L$ of a 4D $\NN=2$ superconformal manifold $\MM$ there is a corresponding Higgs-branch moduli space, which will be denoted as $\IH_\L$ (the letter $\IH$ stands for Higgs). We will isolate a certain class of data on $\IH_\L$ and ask how they vary as we change the exactly-marginal couplings $\L$ and move across the superconformal manifold.

The data we are interested in are the type-B Weyl anomalies of CBOs. Clearly, in vacua with spontaneous breaking of conformal invariance, several effects take over and modify the behaviour of the theory across different scales, from the ultraviolet (UV) to the infrared (IR). For example, the full dependence of the two- and three-point functions of primary operators on the spacetime coordinates is now more complicated and involves non-trivially the characteristic scale of the vacuum. Moreover, typical excitations become massive and in the extreme IR composite operators can be lifted from the spectrum. For instance, from a Lagrangian perspective several elementary fields become massive with a mass set by the characteristic scale of the vacuum and the composite operators of these fields are lifted at ultra-low energies. This can also happen for certain exactly-marginal operators. In that case, the extreme-IR effective theory will be independent of those exactly-marginal couplings. On a related note, in weakly-coupled corners of the Higgs branch it is expected that the extreme-IR effective theory can be expressed in terms of abelian vector multiplets and neutral hypermultiplets. This theory obeys non-renormalisation theorems that make it independent of the exactly-marginal couplings \cite{Argyres:1996eh}.

The data of interest in this paper are associated with the full renormalisation-group (RG) flow, and will typically involve heavy operators. Most of these data cannot be defined independently in the extreme-IR theory and exhibit a non-trivial dependence on the exactly-marginal couplings. It should be noted, however, that we will also encounter examples where some of the data of interest involve massless fields. One of these cases are type-B anomalies associated with untwisted operators in the circular-quiver theory of Sec.~\ref{examples}. In that instance, the quantities of interest have a direct interpretation in the extreme-IR effective theory.

One of the crucial features of the RG running induced by the vacuum, is the fact that it leaves the form of the Ward identities intact. In the unbroken vacuum we noted that type-B anomalies express a clash between the Ward identities of diffeomorphism and Weyl transformations. In Sec.~\ref{app:Ward} we review the argument that shows the relation of these anomalies to two-point functions in the unbroken phase and extend it to vacua with spontaneous breaking of conformal invariance. In the process, we demonstrate that the anomaly survives the breaking but is no-longer directly connected to the two-point functions. This effect is a result of the modified analytic structure that correlation functions exhibit in the broken vacuum. In this context, our primary goal in the rest of the paper will be to investigate how CBO type-B anomalies behave on the Higgs branch of 4D $\NN=2$ SCFTs as a function of the exactly-marginal couplings.

Since there is no direct connection between these Weyl anomalies and the two-point functions on the Higgs branch, it is most appropriate to define the former by analysing directly the anomalous part of the three-point function between the CBOs and the trace of the energy-momentum tensor, $\langle T(x_1) \OO_I(x_2) \bar \OO_J(x_3)\rangle$. In momentum space, the anomaly can be read off in a special kinematic regime, which is summarised by the formula
\beq
\label{dataaa}
G^{\IH}_{I\bar J} = (-1)^n\frac{2^{2n}\Gamma{(n+1)\Gamma{(n+2)}}}{\pi^2 (n!)^2}  \lim_{p_1 \to 0}  \lim_{p_2,p_3 \to 0}
\left[ \frac{d}{dp_2^n} \frac{d}{dp_3^n} 
\left( \langle T(p_1) \OO_I (p_2) \bar \OO_J(p_3) \rangle \right) 
\right]\;.
\eeq
In this expression the three-point function is evaluated on a Higgs-branch vacuum of $\IH_\L$, $T=T^\mu_{~\mu}$ is the trace of the energy-momentum tensor and $\OO_I$, $\bar \OO_J$ are chiral and anti-chiral CBOs. The quantities $G^\IH_{I \bar J}$ are dimensionless functions of the exactly-marginal couplings $\L$. A priori, these functions can be different from the two-point function coefficients $G_{I\bar J}^{\rm CFT}$ that appeared in the unbroken phase, Eq.~\eqref{setup1aa}. In the following sections we will argue, however, that under certain assumptions these quantities can in fact be identical, namely $G_{I\bar J}^{\rm CFT} = G^\IH_{I \bar J}$.

This anomaly-matching statement is far from automatic. In \cite{Schwimmer:2010za} it was argued that type-A anomalies match on moduli spaces, but the expectation is that this does not generically happen with type-B anomalies. In special cases where supersymmetry relates a type-A anomaly to a type-B anomaly---and both are related to chiral anomalies---one anticipates the matching to work also for the corresponding type-B anomalies. One such example arises in $\NN=4$ SYM theory where the type-B anomaly for $\Delta= 2$, $\half$-BPS operators is believed to match on the Coulomb branch on these grounds. In Sec.~\ref{examples} we will independently confirm this expectation when the anomaly in question is viewed as a type-B anomaly on the Higgs branch of an $\NN=2$ SCFT with an adjoint hypermultiplet.

\section{Ward Identities for Diffeomorphism and Weyl Transformations}\label{app:Ward}
 
We begin the detailed discussion of type-B anomalies by exhibiting how these anomalies arise from a clash between the Ward identities of diffeomorphism and Weyl transformations. We focus on the case of scalar operators with integer scaling dimension in four-dimensional CFTs. In this section there are no assumptions of supersymmetry. For pedagogical reasons, we first review  the standard argument that exhibits the anomaly in the unbroken phase, closely following the notation of \cite{Schwimmer:2010za}. We then discuss how the argument is modified in phases with spontaneously-broken conformal symmetry.
 
Consider, for concreteness, two scalar operators $O(x)$, $\bar O(x)$
with common scaling dimension $\Delta=2+n$, $n\in \mathbb
Z_{\ge 0}$. These operators could be a chiral and an anti-chiral CBO in a
4D $\NN=2$ SCFT, but for the purposes of the present argument we do
not need to make this restriction. Let $J$ and $\bar J$ be sources of
these operators in the action. Then, the generating functional for the
correlation functions of $O$, $\bar O$ and the energy-momentum tensor $T^{\mu\nu}$ is
\beq
\label{wardaa1}
W(g,J,\bar J) = \int d^4 x d^4 y \, \Gamma^{(2)}(x,y) J(x) \bar J(y) + \int d^4 x d^4 y d^4z \, \Gamma^{(3)}_{\mu\nu} h^{\mu\nu}(x) J(y) \bar J(z) +\ldots\;,
\eeq
where $g_{\mu\nu} = \eta_{\mu\nu} + h_{\mu\nu}$ is a background-metric
perturbation and the ellipsis denotes other contributions to the generating functional.

In momentum space, the Ward identities for diffeomorphism and Weyl
transformations are respectively \cite{Schwimmer:2010za}
\beq
\label{wardab1}
q^\mu \Gamma_{\mu\nu}^{(3)}(q,k_1,k_2) = \frac{1}{2}(k_1)_\nu \Gamma^{(2)}(k_1^2) + \frac{1}{2} (k_2)_\nu \Gamma^{(2)}(k_2^2)
~,
\eeq
\beq
\label{wardac1}
\eta^{\mu\nu} \Gamma_{\mu\nu}^{(3)}(q,k_1,k_2) = \frac{\Delta}{2} \left( \Gamma^{(2)}(k_1^2) + \Gamma^{(2)}(k_2^2) \right)
~.
\eeq
These identities are valid both in the broken and unbroken phases of the CFT. The conservation of momentum yields $q=k_1+k_2$. In addition, we set
\beq
\label{wardad1}
r := k_1-k_2
~.
\eeq
Then, on general grounds the kinematical expansion of $\Gamma^{(3)}_{\mu\nu}$ takes the form
\beq
\label{wardae1}
\Gamma^{(3)}_{\mu\nu} =  \bar A \eta_{\mu\nu} + B q_\mu q_\nu + C(q_\mu r_\nu + q_\nu r_\mu) + D r_\mu r_\nu
~.
\eeq
The factors $\bar A, B,C,D$ depend on the Lorentz invariants $q^2, k_1^2,k_2^2$. In the broken phase they can also depend on the symmetry-breaking scale $v$, but we will keep this dependence implicit. Let us also define the combination
\beq
\label{wardaf1}
A := \bar A - \frac{1}{4} \left( \Gamma^{(2)}(k_1^2) + \Gamma^{(2)}(k_2^2) \right)
~.
\eeq

Inserting \eqref{wardae1} into \eqref{wardab1} we obtain
\beq
\label{wardag1}
q_\nu \left( A+ B q^2 + C q\cdot r \right) + r_\nu \left( Cq^2 + D q\cdot r - \frac{1}{4} \Gamma^{(2)}(k_1^2) + \frac{1}{4} \Gamma^{(2)}(k_2^2) \right) =0 
~.
\eeq
Setting the independent coefficients of $q_\nu$ and $r_\nu$ separately to zero yields
\beq
\label{wardaia1}
A+  q^2 B+  q\cdot r C =0
~,
\eeq
\beq
\label{wardaib1}
q^2 C + q\cdot r D - \frac{1}{4} \Gamma^{(2)}(k_1^2) + \frac{1}{4} \Gamma^{(2)}(k_2^2) =0
~.
\eeq
Finally, inserting \eqref{wardae1} into \eqref{wardac1} (and using the fact that $\Delta=2+n$) we obtain
\beq
\label{wardaic1}
4 A + q^2 B + 2 q\cdot r C + r^2 D = \frac{n}{2} \left( \Gamma^{(2)}(k_1^2) + \Gamma^{(2)}(k_2^2) \right)
~.
\eeq

\subsection{Unbroken Phase}
\label{unbrokenWard}

At this point, let us focus more specifically on the case of unbroken conformal invariance. In this phase there are no singularities at $q^2=0$, so following \cite{Schwimmer:2010za} one can set $q^2=0$ and restrict to the special kinematic regime where $k_1^2=k_2^2=k^2$. As a consequence of the last restriction 
\beq
\label{wardaica1}
q \cdot r = k_1^2 - k_2^2 = 0
~.
\eeq
In this regime, Eqs.~\eqref{wardaia1} and \eqref{wardaib1} (associated with diffeomorphism transformations) give respectively
\bea
\label{wardaicb1}
A(k^2)&=&0~,
\\
k^2 D(k^2) &=& \frac{1}{4} k^2 \frac{\partial \Gamma^{(2)}}{\partial k^2}\label{wardaicb21}
\eea
whereas Eq.~\eqref{wardaic1} (associated with Weyl transformations) yields
\beq
\label{wardaicc1}
4 A(k^2) + 4 k^2 D(k^2) = n \Gamma^{(2)}(k^2)
~.
\eeq
Inserting the precise form of the two-point function (see Eq.~\eqref{setupab})
\beq
\label{wardaicd1}
\Gamma^{(2)}(k^2) = G k^{2n} \log\left( \frac{k^2}{\mu^2} \right)
~,
\eeq
with a constant $G$ proportional to the two-point function coefficient, we find that \eqref{wardaicc1} becomes
\beq
\label{wardaice1}
4 A = - G k^{2n}
~.
\eeq
This result is in direct contradiction with Eq.~\eqref{wardaicb1}. This contradiction is the explicit manifestation of the type-B anomaly in the conformal phase.

\subsection{Broken Phase}

In the broken phase the analytic structure of the correlation functions is different \cite{Schwimmer:2010za}. Since there is a pole in $q^2$ coming from the dilaton propagator, one cannot  directly set $q^2=0$, but one can instead take $q^2\neq 0$ and consider the limit $q^2\to 0$. We will continue to discuss the kinematic regime $k_1^2=k_2^2=k^2$ with an additional limit $k^2\to 0$ taken at the end of the computation. It is also useful to notice that
\beq
\label{wardaka1}
r^2 = k_1^2 + k_2^2 - 2k_1 \cdot k_2 = 2 (k_1^2 + k_2^2) - q^2 \xrightarrow[q^2 \to 0]{} 2 (k_1^2 + k_2^2) = 4 k^2 ~.  \eeq

We observe that the $B$ term on the RHS of \eqref{wardae1} is the term
that carries the linear coupling of the energy-momentum tensor to the
dilaton. The dilaton is the massless Goldstone boson associated with
the spontaneous breaking of the conformal symmetry. There are several ways to view the linear coupling of the dilaton to the energy-momentum tensor: $(a)$ One way is by shifting classically the value of the elementary fields around the condensed vacuum (with vev $v$), which produces an expression for the energy-momentum tensor with a linear term in the dilaton, or $(b)$ Equivalently, one can argue that the effective action for the dilaton in the broken vacuum contains terms of the form
\begin{align}
\label{dilatoncouplingsBoson}
S_{eff}(\sigma) = \int d^4x \Big( - \frac{1}{2} \partial_\mu \sigma  \partial^\mu \sigma   + \frac{1}{v} \sigma(x)  T^\mu\,_\mu(x)   
+ O(v^{-3})\Big)\;,
\end{align}
where $T^{\mu\nu}$ is the original expression of the energy-momentum tensor in the unbroken phase. This action leads to the equation of motion\footnote{In this form the energy-momentum tensor is no-longer traceless. However, in terms of the full vev-dependent energy-momentum tensor of the first viewpoint $(a)$, the tracelessness condition is unaltered. For a recent discussion and review of these standard points, see for example \cite{DiVecchia:2017uqn}.}
\begin{equation}
\label{dilatoneomBosonic}
T^\mu\,_\mu(x)       =  - v \,  \Box \sigma(x)
\, .
\end{equation}

As a result, as we send $q^2 \to 0$ the $B$ term on the RHS of \eqref{wardae1} will exhibit a $1/q^2$ pole that cancels the $q^2$ in front of $B$. The resulting expression contains a term proportional to $k^{2n}$ in the low-momentum expansion
\beq
\label{wardam1}
\lim_{q^2\to 0} q^2 B \sim \widetilde G k^{2n}
\eeq
for a non-zero value $\widetilde G$. As we will show in several examples later on, this constant captures the coupling of the dilaton to the operators $O$, $\bar O$. Consequently, for the contributions \eqref{wardam1}, Eqs.~\eqref{wardaia1} and \eqref{wardaic1} become respectively
\beq
\label{wardan1}
A + \widetilde G k^{2n} = 0
~,
\eeq
\beq
\label{wardao1}
4 A + \widetilde G k^{2n} = - \bigg[  4 k^2 D - n \Gamma^{(2)}(k^2) \bigg]_{k^{2n}}
~.
\eeq
The expression $\big[ 4 k^2 D - n \Gamma^{(2)}(k^2) \big]_{k^{2n}}$ is shorthand notation for the $k^{2n}$ contribution in the combination $4 k^2 D - n \Gamma^{(2)}(k^2)$ when considering the small-$k^2$ expansion. Using \eqref{wardan1} to express $A$ in terms of $ \widetilde G k^{2n}$  \eqref{wardao1} can be recast into the form
\beq
\label{wardaoa1}
\widetilde G k^{2n} = \frac{1}{3}  \bigg[ 4 k^2 D - n \Gamma^{(2)}(k^2) \bigg]_{k^{2n}}
~.
\eeq
Employing the expression \eqref{wardaicb21}, which is still valid, we obtain
\beq
\label{wardaoa1}
\widetilde G k^{2n} = \frac{1}{3}  \bigg[ k^2 \frac{\partial \Gamma^{(2)}}{\partial k^2} - n \Gamma^{(2)}(k^2) \bigg]_{k^{2n}}
~.
\eeq
The RHS of this equation, however, vanishes because the generic $k^{2n}$ contribution to $\Gamma^{(2)}(k^2)$, 
\beq
\label{wardaob1}
\Gamma^{(2)}(k^2) \sim a k^{2n}
\eeq
for some constant $a$, cancels out. Therefore Eq.~\eqref{wardaoa1} implies that $\tilde G$ vanishes, which leads to an inconsistency. 

From this general argument,  the following  important conclusions can be reached. The type-B anomaly persists in the broken phase but its coefficient (expressed by the constant $\widetilde G$ in the above equations) is not related directly to the structure of the two-point function. However, it can be read off from the structure of the three-point function $\langle T^\mu_{~\mu}(q) \OO(k_1) \bar \OO(k_2)\rangle$ in the appropriate kinematic regime.

\section{Coupling-constant Dependence on Moduli Spaces}
\label{data}

Having established the specifics of the anomaly of interest, we would next like to understand better how it may depend on exactly-marginal couplings. In Sec.~\ref{conformalmanifolds} we observed that a proper comparison of CFT correlation functions along a conformal manifold requires the introduction of a covariant derivative. The associated connection on the bundle of operators captures details of the regularisation and operator mixing that occur in the UV when two operators collide. In what follows, we want to discuss the proper treatment of similar effects in correlation functions defined on a vacuum with spontaneously-broken conformal symmetry. Once again, we concentrate on Higgs-branch vacua in 4D $\NN=2$ SCFTs. Since we want to distil information about type-B anomalies of CBOs, we will tailor the  discussion to three-point functions of the energy-momentum tensor with two CBOs
\beq
\label{dataab}
C_{TI\bar J}(x_1,x_2,x_3) \equiv \langle T_{\mu\nu}(x_1) \OO_I (x_2) \bar \OO_J(x_3) \rangle
~,
\eeq
expressed here in position space for insertions at separated spacetime points. We will also focus on correlation functions that are evaluated in the spontaneously-broken phase.

Besides its spacetime dependence, the quantity \eqref{dataab} depends non-trivially on the exactly-marginal couplings and the vev characterising the vacuum state. To analyse the dependence on the exactly-marginal couplings one may consider derivatives of $C_{T I\bar J}$ with respect to these couplings. Repeating the logic of Sec.~\ref{conformalmanifolds} we observe that the precise, covariant definition of these derivatives requires a connection $\nabla$ on the conformal manifold
\beq
\label{dataac}
\nabla_{\lambda^i} C_{T I\bar J} = \left\langle \int d^4 x\, \Phi_i(x)\, T(x_1) \OO_I (x_2) \bar \OO_J(x_3) \right\rangle_{\text{regulated}} ~.
\eeq
We have denoted by $\Phi_i$ the exactly-marginal operator that corresponds to the coupling $\lambda^i$ and by the subscript `regulated' that potential UV divergences in the integrated four-point function have been regularised. Since these divergences are a UV effect they can be regularised with the same prescription  used in Sec.~\ref{conformalmanifolds} for the unbroken phase. In this manner, we recover a connection $\nabla$ on the moduli space, which is independent of the details of the vacuum state.

In Eq.~\eqref{dataac} we used the fact that we can vary the exactly-marginal couplings while keeping the vacuum state constant. As the choice of a vacuum amounts to the choice of a superselection sector for QFTs in more than two spacetime dimensions, this assumption may seem trivial. However, since the precise details of a theory on a non-trivial vacuum are characterised by non-vanishing vevs, one also needs to examine if these vevs can be chosen in a coupling-constant independent fashion.\footnote{In a Lagrangian formulation, the correlation functions on the moduli space can be evaluated by defining the vev of elementary fields, shifting the fields in the action around their new vacuum, and subsequently performing the path integral. In the new path integral the asymptotic behaviour of the fields at infinity is that of the trivial vacuum but the action contains extra interactions that depend on the vevs. It is therefore important, when we take derivatives with respect to the exactly-marginal couplings, to know if the extra interactions involve additional coupling-constant dependence.}

On the Higgs branch one has non-vanishing vevs for Higgs-branch operators $O_H$
\beq
\label{dataaaca}
\langle v | O_H | v \rangle = v_O ~.
\eeq
Therefore, it is important to analyse to what extent coupling-constant-independent vevs $v_O$ can be consistently chosen. We propose that such a choice can be made  on the Higgs branch because the connection on the bundle of Higgs-branch superconformal-primary operators is flat \cite{Niarchos:2018mvl} and therefore one can trivially choose the vacuum so that $\partial_{\lambda^i} v_O=0$. Under these conditions Eq.~\eqref{dataac} is correct. It should be appreciated that we had to use a special feature of the Higgs-branch moduli space to reach this conclusion.

In summary, the same connection $\nabla$ can be used to parallel transport data both in the CFT phase (where the Higgs vev vanishes) and on Higgs-branch moduli spaces (where the Higgs vev does not vanish).

\section{Superconformal Ward Identities}
\label{Ward}

The main ingredient of this section is a specific set of superconformal Ward identities. We consider these identities first in the unbroken phase of the CFT, where they are well known (see e.g.\ \cite{Baggio:2012rr,Papadodimas:2009eu}), and subsequently in the broken phase of the Higgs branch, where to our knowledge they are less explored. In this context, they can also be viewed as a particular class of superconformal soft theorems along the lines of \cite{DiVecchia:2017uqn}.

\subsection{Unbroken Phase}\label{unbrokenphase}

Recall that a 4D $\NN=2$ SCFT possesses a left-chiral supercurrent\footnote{It possesses also a right-chiral supercurrent. Since the right-chiral version of the Ward identities in this section is completely analogous to the left-chiral version, we will not repeat it explicitly.} $G^{\II\mu}_\alpha$, where $\II=1,2$ is an $SU(2)_R$ R-symmetry index, $\mu=0,1,2,3$ a spacetime index and $\alpha=\pm$ a Weyl spinor index.\footnote{Our notation for the 4D $\NN = 2$ superconformal algebra can be found in App.~\ref{app:4d}. We mostly follow the conventions of \cite{Dolan:2002zh,Gadde:2010zi}.} $G^{\II\mu}_\alpha$ is a conformal-primary operator with scaling dimension $\Delta=\frac{7}{2}$. The currents 
\beq
\label{wardaa}
j^{\II\mu}= \xi^\alpha(x) G^{\II\mu}_\alpha(x)
\eeq
are classically conserved for the conformal Killing spinors (on $\IR^4$)
\beq
\label{wardab}
\xi^\alpha(x) = \nu^\alpha + x^{\dot \alpha \alpha} \mu_{\dot \alpha}
~,
\eeq
where $\nu^\alpha$ and $\mu_{\dot \alpha}$ are arbitrary constant spinors. Setting $\nu^\alpha\neq 0$, $\mu_{\dot  \alpha}=0$ gives the left-chiral Poincar\'e supercharges $Q^\II_\alpha$ while $\nu^\alpha = 0$, $\mu_{\dot  \alpha} \neq 0$ gives the right-chiral superconformal charges $\bar S^{\II\dot \alpha}$.

Now let us consider the correlation function of $n$ operators $O_\ell$, $\ell=1,2,\ldots,n$ 
\beq
\label{wardac}
\langle O_1(x_1) \cdots O_n(x_n) \rangle
~.
\eeq
A standard derivation yields the Ward identity (see \cite{Papadodimas:2009eu} for further details)
\bea
\label{wardad}
\d_\mu\langle j^{\II\mu}(x) O_1(x_1) \cdots O_n(x_n) \rangle
&=& \sum_{\ell=1}^n \delta(x-x_\ell) \bigg[ \xi^\alpha(x_\ell) \langle O_1(x_1) \cdots [Q^\II_\alpha,O_\ell](x_\ell) \cdots  O_n(x_n) \rangle
\nonumber\\
&-& \sigma^\mu_{\alpha \dot \beta} \left(\d_\mu \xi^\alpha \right)(x_\ell) \langle O_1(x_1) \cdots [\bar S^{\II \dot \beta},O_\ell](x_\ell) \cdots  O_n(x_n) \rangle \bigg]
\, .
\eea

In the unbroken phase integrating over $x$ gives zero on the LHS of this equation. This is because we are integrating a total derivative and there is no boundary contribution from infinity for the $(n+1)$-point function $\langle j^{\II\mu}(x) O_1(x_1) \cdots O_n(x_n) \rangle$. As a result, \eqref{wardad} becomes
\bea
\label{wardae}
&&\sum_{\ell=1}^n \bigg[ \xi^\alpha(x_\ell) \langle O_1(x_1) \cdots [Q^\II_\alpha,O_\ell](x_\ell) \cdots  O_n(x_n) \rangle
\nonumber\\
&&- \sigma^\mu_{\alpha \dot \beta} \left(\d_\mu \xi^\alpha \right)(x_\ell) \langle O_1(x_1) \cdots [\bar S^{\II \dot \beta},O_\ell](x_\ell) \cdots  O_n(x_n) \rangle \bigg]
=0
~.
\eea
Taking 
\beq
\label{wardaf}
\xi^\alpha (x) = (x-x_0)^{\dot \alpha \alpha} \mu_{\dot \alpha}
\eeq
for arbitrary $x_0$, we obtain the superconformal Ward identity
\beq
\label{wardag}
\sum_{\ell=1}^n \langle O_1(x_1) \cdots [(x_\ell-x_0)^{\dot\alpha \alpha} Q^\II_\alpha - \bar S^{\II\dot \alpha} , O_\ell](x_\ell) \cdots  O_n(x_n) \rangle =0
~.
\eeq
The right-chiral version of this Ward identity is
\beq
\label{wardai}
\sum_{\ell=1}^n \langle O_1(x_1) \cdots [(x_\ell-x_0)^{\dot\alpha \alpha} \bar Q_{\II \dot \alpha} + S^{\alpha}_\II , O_\ell](x_\ell) \cdots  O_n(x_n) \rangle =0
~.
\eeq

In applications of these identities one needs to keep in mind the following key point. When the operators $O_\ell$ are superconformal primaries (in particular, when they are annihilated by the $\bar S$ supercharges) \eqref{wardag} becomes
\beq
\label{wardaj}
\sum_{\ell=1}^n \langle O_1(x_1) \cdots [(x_\ell-x_0)^{\dot\alpha \alpha} Q^\II_\alpha , O_\ell](x_\ell) \cdots  O_n(x_n) \rangle =0
~.
\eeq
Choosing $x_0$ to be one of the $x_\ell$, the corresponding operator insertion can be `hidden' from the Ward identity. Analogous statements apply to the right-chiral version \eqref{wardai}.

As a first example of a concrete application, consider the two-point function $\langle \OO_I(x_1) \bar \OO_J (x_2)\rangle$, where as above $\OO_I$ is a chiral-primary CBO and $\bar \OO_J$ an anti-chiral-primary CBO. By default, both operators are superconformal primaries and obey the shortening conditions
\beq
\label{wardaka}
[ \bar Q_{\II\dot \alpha}, \OO_I ] =0~, ~~
[ \bar S^{\II\dot \alpha}, \OO_I ] =0~,~~
[ S^\alpha_\II, \OO_I ]=0
~,
\eeq
\beq
\label{wardakb}
[ Q^\II_\alpha, \bar \OO_I ]=0~, ~~
[ \bar S^{\II\dot \alpha}, \bar \OO_I ] =0~,~~
[ S^\alpha_\II, \bar  \OO_I ]=0
~.
\eeq

We recall that the $\NN=2$-preserving, exactly-marginal deformations of a 4D $\NN=2$ SCFT can be written in the form
\beq
\label{wardal}
\delta S = \lambda^k \int d^4 x \, Q^4 \cdot \phi_k (x) + \bar \lambda^k \int d^4 x \, \bar Q^4 \cdot \bar \phi_k (x)
~,
\eeq
where $\phi_k$, $\bar \phi_k$ are chiral and anti-chiral CBOs with scaling dimension two and the notation $Q^4 \cdot \phi$ refers to the nested (anti)commutator 
\beq
\label{wardalaa}
\varepsilon_{\II \JJ} \varepsilon_{\KK \LL} \varepsilon^{\alpha \beta} \varepsilon^{\gamma \delta} 
\{ Q^\II_\alpha , [Q^\KK_\beta, \{ Q^\JJ_\gamma , [Q^\LL_\delta , \phi ]\} ]\}
~.
\eeq 
Analogous expressions apply to the complex-conjugate (anti)commutator. Suppressing the spinor indices, but keeping the $SU(2)_R$ indices explicit, it will also be convenient to use a notation of the form
\beq
\label{wardalab}
Q^4 \cdot \phi \sim Q^1 \cdot Q^1 \cdot Q^2 \cdot Q^2 \cdot \phi
\eeq

For our first observation, let us single out one of supercharges and write the exactly-marginal deformations in the following way
\beq
\label{wardam}
\delta S = \lambda^k \int d^4 x \, \{ Q^1,\YY_k \}(x) + \bar \lambda^k \int d^4 x \, \{\bar Q_1, \bar \YY_k\}(x)
~,
\eeq
where the index $k$ runs over the exactly-marginal couplings and $\YY_k$ (resp.\ $\bar \YY_k$) are obtained by stripping off one of the supercharges in the exactly-marginal operator (here we chose $Q^1$ and $\bar Q_1$). Then, by definition
\beq
\label{wardana}
\nabla_k \langle \OO_I(x_1) \bar \OO_J (x_2)\rangle = \left\langle \int d^4 x\, \{ Q^1,\YY_k \}(x)  \OO_I(x_1) \bar \OO_J (x_2) \right\rangle
~,
\eeq
\beq
\label{wardanb}
\nabla_{\bar k} \langle \OO_I(x_1) \bar \OO_J (x_2)\rangle = \left\langle \int d^4 x\, \{\bar Q_1, \bar\YY_k \}(x)  \OO_I(x_1) \bar \OO_J (x_2) \right\rangle
~.
\eeq
Applying \eqref{wardag} to the 3-point function $\langle \YY_k(x) \OO_I(x_1) \bar \OO_J (x_2) \rangle$ with $x_0=x_1$ and the right-chiral version of \eqref{wardag} to $\langle \bar \YY_k(x) \OO_I(x_1) \bar \OO_J (x_2) \rangle$ with $x_0=x_2$ we obtain that the RHS of both \eqref{wardana}, \eqref{wardanb} vanish. To summarise, in the conformal phase of an $\NN=2$ SCFT we obtain 
\beq
\label{wardao}
\nabla_k \langle \OO_I(x_1) \bar \OO_J (x_2)\rangle = 0~, ~~
\nabla_{\bar k} \langle \OO_I(x_1) \bar \OO_J (x_2)\rangle = 0
~,
\eeq
a known statement that appeared, e.g., in \cite{Baggio:2012rr,Papadodimas:2009eu}.

As a second example, which is more closely related to our discussion, consider the three-point function $\langle T_{\mu \nu}(y) \OO_I(x_1) \bar \OO_J(x_2)\rangle$ of the energy-momentum tensor $T_{\mu\nu}$ with two CBOs. As we review in Appendix \ref{app:4d}, the energy-momentum tensor is a conformal primary in the short superconformal multiplet $\hat \CC_{(0,0)}$. The bottom component in this multiplet is a scalar superconformal-primary field $\TT$ with  dimension $\Delta = 2$ and vanishing $SU(2)_R$ and $U(1)_r$ R-charges, that obeys the shortening conditions
\beq
\label{wardnewa}
(Q^\II)^2 \cdot \TT =0~, ~~ (\bar Q_\II)^2 \cdot \TT =0~, ~~ \II=1,2
~.
\eeq
The energy-momentum tensor is obtained by the successive application of one $Q^1$, one $Q^2$, one $\bar Q_1$ and one $\bar Q_2$ supercharge on the superconformal primary $\TT$. Different orders of the application of these supercharges gives operators $T_{\mu\nu}$ that differ by total derivative terms. A particular combination of these different possibilities provides the conformal primary $T_{\mu\nu}$. Equivalently,  $T_{\mu\nu}$ can be written with any ordering of supercharges  by adding the appropriate descendants. For example, one can schematically write $T_{\mu\nu}$  (suppressing spacetime indices, spinor indices and sigma-matrices on the RHS) as
\beq
\label{wardnewb}
T_{\mu\nu} = Q^1 Q^2 \bar Q_1 \bar Q_2 \TT + c_1 Q^1 \bar Q_1 \partial \TT + c_2 Q^2 \bar Q_2 \partial \TT +c_3 \partial^2 \TT\;,
\eeq
with suitable numerical constants $c_1,c_2,c_3$.

Armed with these properties of the energy-momentum tensor, we proceed to study the covariant derivatives
\beq
\label{wardbea}
\nabla_k \langle T_{\mu\nu}(y) \OO_I(x_1) \bar \OO_J (x_2)\rangle = \left\langle \int d^4 x\, Q^4 \cdot \phi_k (x)  T_{\mu\nu}(y) \OO_I(x_1) \bar \OO_J (x_2) \right\rangle
~,
\eeq
\beq
\label{wardbeb}
\nabla_{\bar k} \langle T_{\mu\nu} (y) \OO_I(x_1) \bar \OO_J (x_2)\rangle = \left\langle \int d^4 x\, \bar Q^4 \cdot \bar\phi_k (x) T_{\mu\nu}(y) \OO_I(x_1) \bar \OO_J (x_2) \right\rangle
~.
\eeq
For concreteness, let us focus on the first of these correlation functions (similar arguments apply to the second, complex-conjugate version). Repeating the steps from the preceding computation of the two-point function, one of the $Q^1$ supercharges from the exactly-marginal deformation can be singled out and moved around inside the correlation function using the superconformal Ward identities \eqref{wardag} as applied to the four-point function
\beq
\label{wardbf}
\langle Q^1 (Q^2)^2 \cdot \phi_k (x) T_{\mu\nu} (y) \OO_I(x_1) \bar \OO_J (x_2) \rangle
\eeq
with $x_0=x_1$.\footnote{In \eqref{wardbf} and the expressions that follow we will mostly suppress the spinor indices to avoid cluttering the notation.} There is obviously no contribution to this Ward identity from the action of the superconformal charges on the CBOs $\OO_I$ and $\bar \OO_J$, but there are two potential contributions of the form
\beq
\label{wardbfa}
\II_1 = \langle Q^1 (Q^2)^2 \cdot \phi_k (x) \, Q^1 \cdot T_{\mu\nu}(y) \, \OO_I(x_1) \bar \OO_J(x_2) \rangle
~,
\eeq
\beq
\label{wardbfb}
\II_2 = \langle Q^1 (Q^2)^2 \cdot \phi_k (x) \, \bar S^1 \cdot T_{\mu\nu}(y) \, \OO_I(x_1) \bar \OO_J(x_2) \rangle
\eeq
from the action of the supercurrent on the energy-momentum tensor. Explicit calculations, which we will present in a moment, show that these potential contributions are actually both zero. For the reader who is not interested in the details, this fact establishes that
\beq
\label{wardbg}
\nabla_{k} \langle T_{\mu\nu} (y) \OO_I(x_1) \bar \OO_J (x_2)\rangle = 0
~.
\eeq
Similarly, employing the complex-conjugate version of the superconformal Ward
identities  \eqref{wardaj} one obtains 
\beq
\label{wardbj}
\nabla_{\bar k} \langle T_{\mu\nu}(y) \OO_I(x_1) \bar \OO_J (x_2)\rangle = 0
~.
\eeq

Both results are consistent with the fact that all tensor structures in the three-point function $\langle T_{\mu\nu}(y) \OO_I(x_1) \bar \OO_J (x_2)\rangle$ (in the unbroken phase) are proportional to the coefficients of the two-point functions $\langle\OO_I(x_1) \bar \OO_J (x_2)\rangle$; see e.g.\ \cite{Bzowski:2018fql}. On the one hand, we established in \eqref{wardao} that the latter are covariantly constant. On the other, we have just shown that the full three-point function  $\langle T_{\mu\nu}(y) \OO_I(x_1) \bar \OO_J (x_2)\rangle$ is covariantly constant (including the part of the type-B anomaly) without using any knowledge about its structure and its relation to the two-point function of the CBOs.

\subsection*{Explicit Calculation of $\II_1 = \II_2 = 0$}

Here we present the full evaluation of \eqref{wardbfa}, \eqref{wardbfb}, starting with $\II_1$. This involves the commutator $Q^1 \cdot T_{\mu\nu}$:
\begin{itemize}
\item 

  We notice that the first two terms on the RHS of \eqref{wardnewb} cannot contribute in this commutator. Indeed, if we have an operator $\Lambda$ with the property $(Q^1)^2 \cdot \Lambda=0$, then 
\bea
\label{wardbfba}
Q^1 \cdot Q^1 \cdot \bar Q_1 \cdot \Lambda &=& Q^1 \cdot \{Q^1,\bar Q_1\} \cdot \Lambda - Q^1 \cdot \bar Q_1 \cdot Q^1 \cdot \Lambda 
\nonumber\\
&=& Q^1 \cdot \partial \Lambda - \{ Q^1, \bar Q_1 \} \cdot Q^1 \cdot \Lambda + \bar Q^1 \cdot Q^1 \cdot Q^1 \cdot \Lambda\cr
&=& Q^1 \cdot \partial \Lambda - Q^1 \cdot \partial \Lambda =0
~.
\eea
For the first term in $Q^1\cdot T_{\mu\nu}$ we apply this identity to $\Lambda = Q^2\cdot \bar Q_2 \cdot \TT$ while for the second term we apply it to $\Lambda = \partial \TT$. In both cases, the identity $(Q^1)^2 \cdot \Lambda=0$ follows from the shortening condition \eqref{wardnewa}.

\item

  For the last two terms we can use the superconformal Ward identity \eqref{wardag} once again to move around the second $Q^1$ coming from the exactly-marginal deformation inside $\II_1$. As before, there is no contribution from the action of the superconformal charges on the CBOs. There is also no contribution from $\bar S^1$ on the last two terms of $Q^1 \cdot T_{\mu\nu}$ because the former commutes with all supercharges to act directly on $\TT$, which is a primary. Additionally, there is no contribution from the action of the second $Q^1$ on the last two terms of $Q^1 \cdot T_{\mu\nu}$, because it involves the action $(Q^1)^2 \cdot \TT$ that vanishes through \eqref{wardnewa}.

\end{itemize}

The second potential contribution $\II_2$ involves the commutator $\bar S^1 \cdot T_{\mu\nu}$. This is slightly more involved, but one can again argue using the same superconformal Ward identity \eqref{wardag} that it vanishes. Because of \eqref{wardnewb} $\bar S^1 \cdot T_{\mu\nu}$ involves four terms, which we will examine in detail:

\begin{itemize}

\item Since $\bar S^1 \cdot \TT =0$, the last term $c_3 \partial^2 \TT$ in \eqref{wardnewb}  does not contribute.

\item The third term in \eqref{wardnewb} gives
\beq
\label{wardbfc}
\bar S^1 \cdot Q^2 \cdot \bar Q_2 \cdot \partial \TT = - Q^2 \cdot \bar S^1 \cdot \bar Q_2 \cdot \partial \TT = Q^2 \cdot \{ \bar S^1 , \bar Q_2 \} \cdot \partial \TT 
~,
\eeq
where in the last equality we used that $\bar S^1 \cdot \TT =0$. Using the superconformal algebra relation (see Appendix \ref{app:4d} for the explicit form with all spinor indices reinstated)
\beq
\label{wardbfd}
\{ \bar S^1 , \bar Q_2 \} = - R^1_{~2}\;,
\eeq
and the fact that the primary $\TT$ is also the highest weight of the $SU(2)_R$ representation $R^1_{~2} \cdot \TT =0$, we find $\bar S^1 \cdot Q^2 \cdot \bar Q_2 \cdot \partial \TT=0$.  Here $R{^1}_{2} = R_+$ is the raising operator of $SU(2)_R$ \cite{Dolan:2002zh}.

\item The second term in \eqref{wardnewb} yields the contribution
\beq
\label{wardbfe}
\bar S^1 \cdot Q^1 \cdot \bar Q_1 \cdot \partial \TT = - Q^1 \cdot \bar S^1 \cdot \bar Q_1 \cdot \partial \TT = Q^1 \cdot \{ \bar S^1, \bar Q_1 \} \cdot \partial \TT
~.
\eeq
From the superconformal algebra we have that
\beq
\label{wardbff}
\{ \bar S^1, \bar Q_1 \} =  \bar M + \frac{1}{2} D - R^1_{~1}
~,
\eeq
where $\bar M$ are rotation generators and $D$ the dilatation  generator. Both $\bar M$ and $R^1_{~1}$ annihilate the superconformal primary $\TT$, since it is a scalar with zero $SU(2)_R$ and $U(1)_r$ quantum numbers and here $R{^1}_{1} = R + \half r$. Moreover, $[D,\TT] = 2 \TT$. Taking into account the presence of the derivative in $\p \TT$ we end up with a potential contribution that is proportional to the correlator
\beq
\label{wardbfg}
\left\langle Q^1 (Q^2)^2 \cdot \phi_k (x) \, Q^1 \partial \TT (y) \OO_I(x_1) \bar \OO_J(x_2) \right\rangle
~.
\eeq 
Using the superconformal Ward identity \eqref{wardag} to move $Q^1$ around we find that this correlation function vanishes.

\item The final potential contribution comes from the first term in \eqref{wardnewb}
\beq
\label{wardbfi}
\bar S^1 \cdot Q^1 \cdot Q^2 \cdot \bar Q_2 \cdot \bar Q_1 \cdot \TT 
= Q^1 \cdot Q^2 \cdot \bar S^1 \cdot \bar Q_2 \cdot \bar Q_1 \cdot \TT 
~.
\eeq
We notice that 
\bea
\label{wardbfj}
\bar S^1 \cdot \bar Q_2 \cdot \bar Q_1 \cdot \TT 
&=& \{ \bar S^1, \bar Q_2 \} \cdot \bar Q_1 \cdot \TT - \bar Q_2 \cdot \{ \bar S^1, \bar Q_1 \} \cdot \TT
\nonumber\\
&=& -  R^1_{~2} \cdot \bar Q_1 \cdot \TT -  \bar Q_2 \cdot \left( \bar M + \frac{1}{2} D - R^1_{~1} \right) \cdot \TT
\nonumber\\
&=& - [R^1_{~2}, \bar Q_1] \cdot \TT +  \bar Q_2 \cdot \TT
\nonumber\\
&=&  \bar Q_2 \cdot \TT +  \bar Q_2 \cdot \TT = 2  \bar Q_2 \cdot \TT
~.
\eea
Hence, the contribution to $\bar S^1 \cdot T_{\mu\nu}$ is proportional to $Q^1 \cdot Q^2 \cdot \bar Q_2 \cdot \TT$, which leads to the correlation function
\beq
\label{wardbfk}
\left \langle Q^1 (Q^2)^2 \cdot \phi_k(x) \, Q^1 \cdot Q^2 \cdot \bar Q_2 \cdot \TT (y) \, \OO_I(x_1) \, \bar\OO_J(x_2) \right \rangle
~.
\eeq
Using the superconformal Ward identity \eqref{wardag} once again, we move the second $Q^1$ around. There is no contribution from the resulting action of $Q^1$ or $\bar S^1$  on the CBOs. From the action  of  $Q^1$ on the $Q^1 \cdot Q^2 \cdot \bar Q_2 \cdot \TT$ factor there is no contribution, because both $Q^1$s move past $Q^2$ and $\bar Q_2$ and $(Q^1)^2\cdot \TT =0$. From the action of $\bar S^1$ on $Q^1 \cdot Q^2 \cdot \bar Q_2 \cdot \TT (y)$ we also obtain zero because 
\bea
\label{wardbfl}
\bar S^1 \cdot Q^1 \cdot Q^2 \cdot \bar Q_2 \cdot \TT 
&=& Q^1 \cdot Q^2 \cdot   \bar S^1 \cdot  \bar Q_2 \cdot \TT 
= Q^1 \cdot Q^2 \cdot  \{  \bar S^1,  \bar Q_2  \} \cdot \TT 
\nonumber\\
&=& - Q^1 \cdot Q^2 \cdot R^1_{~2} \cdot \TT =0
~.
\eea

\end{itemize}

\subsection{Broken Phase}

We are now in position to explore similar properties in the broken phase on the Higgs branch. In general, Ward identities are operator equations and retain the same form independently of the vacuum. Nevertheless, when a Ward identity is applied for a current whose corresponding symmetry is broken on the vacuum, some care needs to be taken with the asymptotic behaviour of the fields at infinity. The latter phenomenon is typically responsible for extra contributions that are absent in the unbroken phase.

More specifically, consider the superconformal Ward identity \eqref{wardad}, which is of central importance to our discussion. In this form the Ward identity continues to hold on the Higgs branch, despite the fact that the Higgs vevs break the superconformal symmetry. As has been already noted, in order to derive \eqref{wardae} one needs to integrate $x$ over the whole of $\IR^4$. The integral of the RHS of \eqref{wardad}, which involves contact terms, is the same in the broken and unbroken phases, but the LHS can be different if there is a boundary term for the $(n+1)$-point function $\langle j^{I\mu}(x) O_1(x_1) \cdots O_n(x_n)\rangle$ at $|x|\to \infty$. In the unbroken phase all correlation functions run sufficiently fast to zero at infinity and a boundary term is absent. In the broken phase the supermultiplet that contains the dilaton couples linearly to the supermultiplet that contains the energy-momentum tensor and supercurrents--c.f. Eq.~\eqref{dilatoncouplings}---to create a massless pole in the analytic structure of the $(n+1)$-point function in momentum space. It is precisely this pole that can generate potential contributions otherwise absent in the unbroken phase. For the case at hand, the following observations can be made.

In momentum space the integrated LHS of \eqref{wardad} corresponds to taking a low-momentum limit. In this limit the above massless pole dominates the correlator and yields the contribution
\beq
\label{wardca}
\lim_{p\to 0} p_\mu \langle v | j^{I\mu}(p) O_1 (q_1)\cdots O_n(q_n) | v \rangle \simeq \lim_{p\to 0} p_\mu \langle v | j^{I\mu}(p) | \chi^\alpha(p) \rangle\, \frac{p_{\alpha\dot \alpha}}{p^2} \, \langle \bar \chi^{\dot \alpha}(p) | O_1 (q_1)\cdots O_n(q_n) | v \rangle
~,
\eeq
where $\chi^\alpha$ is the Weyl dilatino that couples to the supercurrent and propagates as an intermediate massless state. We are using bra-ket notation, with the appropriate time-ordering prescriptions left implicit, while $|v\rangle$ is a Higgs-branch vacuum state. In momentum space Eq.\ \eqref{wardaa} reads\footnote{In this argument we focus again on the left-chiral supercurrents. Analogous statements can be made in an obvious manner for the right-chiral parts of the supercurrents.}
\beq
\label{wardcb}
j^{\II \mu}(p) = \left( \nu^{\alpha} - i \mu_{\dot\alpha} \frac{\d}{\d p_{\alpha \dot\alpha}}  \right) G^{\II \mu}_\alpha(p)
~.
\eeq
Also, let us denote the linear coupling 
\beq
\label{wardcc}
\langle v | G_\beta^{\II \mu}(p) | \chi^\alpha \rangle = f^\II p^\mu \delta^\alpha_\beta\;,
\eeq
with a non-vanishing coefficient $f^\II$ that is left unspecified, and the last factor in \eqref{wardca}
\beq
\label{wardcd}
\FF^{\dot \alpha} \equiv \langle \bar \chi^{\dot \alpha} | O_1 (q_1)\cdots O_n(q_n) | v \rangle
~.
\eeq
Then \eqref{wardca} becomes
\bea
\label{wardce}
\lim_{p\to 0} p_\mu \langle v | j^{\II \mu}(p) O_1 (q_1)\cdots O_n(q_n) | v \rangle \simeq \lim_{p\to 0} \left(
p_{\alpha\dot \alpha} \nu^\alpha \FF^{\dot\alpha}f^\II - i \mu_{\dot \alpha} \FF^{\dot \alpha} f^\II \right)
= - i \mu_{\dot \alpha} \FF^{\dot \alpha} f^\II
~.
\eea
The first term on the RHS, which is proportional to the constant part $\nu^\alpha$ of the Killing spinor $\xi^\alpha$, represents the contribution of the Poincar\'e supercharges. This is proportional to the momentum $p$ and vanishes in the IR limit. As expected, there is no boundary contribution from this term on the LHS of \eqref{wardad} in accordance with the fact that the Poincar\'e supersymmetries are not broken on the moduli space. The second term on the RHS of \eqref{wardce} represents the non-vanishing contributions of the superconformal charges, whose corresponding symmetry is broken by the vacuum. This term expresses potentially non-vanishing effects of the symmetry breaking in the superconformal Ward identities of interest.

So far the arguments about the structure of the superconformal Ward identities \eqref{wardad} in the broken phase have been quite general. The key ingredient has been the generic linear coupling of the dilatino to the supercurrent in a superconformal theory without relying on any special properties of the moduli space or the $n$ operator insertions $O_\ell$ ($\ell=1,2,\ldots,n$). However, now we would like to specialise to the case of the four-point functions $\langle Q^4\cdot \phi_k(z) T (y) \OO_i(x_1) \bar \OO_j(x_2)\rangle$ and $\langle \bar Q^4 \cdot \bar\phi_k(z) T(y) \OO_i(x_1) \bar \OO_j(x_2)\rangle$ that appeared in the second example of Sec.~\ref{unbrokenphase}. 

Let us repeat the steps that we performed in the unbroken phase. When one applies the Ward identities \eqref{wardad} (and their complex conjugate) to the superconformal current $j^{1\mu}$ ($\bar j_1^\mu$) in order to move around a $Q^1$ (respectively, a $\bar Q_1$) from the exactly-marginal interaction, a new boundary term can arise involving five-point functions of the form
\beq
\label{wardcf}
\d_\mu\langle j^{1\mu}(x) \YY_k(z) T(y) \OO_I(x_1) \bar \OO_J(x_2)\rangle~, ~~
\d_\mu \langle \bar j_1^{\mu}(x) \bar \YY_k(z) T(y) \OO_I(x_1) \bar \OO_J(x_2)\rangle
~.
\eeq
As we argued in \eqref{wardce} (and implied for its complex conjugate version) in momentum space this  boundary term (and its conjugate version) will be proportional to the following amplitudes
\beq
\label{wardcg}
\langle \bar \chi^{\dot \alpha}(p) | \YY_k(q_1) T(q_2) \OO_i(q_3) \bar \OO_j(q_4) | v \rangle~, ~~
\langle \chi_{\alpha}(p) | \bar \YY_k(q_1) T(q_2) \OO_i(q_3) \bar \OO_j(q_4) | v \rangle
~.
\eeq
At leading order in perturbation theory such terms do not appear to have the structure needed to contribute to the contact term that captures $\nabla_a G_{I \bar J }^{\mathbb H}$ in the corresponding integrated four-point function. We expect that the absence of such contributions is generally true as an exact statement beyond perturbation theory, but we have not been able to demonstrate it rigorously in this approach.\footnote{This corrects a statement in previous versions of this article. We thank E.~Andriolo for discussions on this point. An alternative, and more powerful, argument for $\nabla_a G_{I \bar J }^{\mathbb H}=0$ is based on satisfying the Wess--Zumino consistency condition and appears in \cite{Andriolo:2022lcb}.} 

Assuming the absence of the above-mentioned contributions from the boundary term to the superconformal Ward identity \eqref{wardad}, we can repeat the previous arguments to the correlation functions of interest on the Higgs branch. Retracing the steps of the second example of Sec.~\ref{unbrokenphase} we deduce that the contact terms $\nabla_k G_{I \bar J }^{\mathbb H}$, $\nabla_{\bar k} G_{I \bar J }^{\mathbb H}$ in the correlation functions
\beq
\label{wardci}
\nabla_{k} \langle T(y) \OO_I(x_1) \bar \OO_J (x_2)\rangle ~, ~~
\nabla_{\bar k} \langle T(y) \OO_I (x_1) \bar \OO_J (x_2)\rangle 
\eeq
vanish both in the unbroken phase, as well as in the broken phase along the Higgs branch. 
This is a powerful statement with important implications that will be highlighted in the following sections.

\section{Type-B Anomaly Matching on the Higgs Branch}
\label{match}

In Sec.~\ref{anoHiggs} we denoted the value of the type-B anomaly at an arbitrary point $\L$ of the conformal manifold $\MM$ as $G_{I\bar J}^\IH(\L)$ in the broken phase of the Higgs branch $\IH_\L$ and as $G_{I\bar J}^{\rm CFT}(\L)$ in the unbroken conformal phase.

The postulated absence of contact terms in the correlation functions
\eqref{wardci} now implies that the type-B anomalies $G_{I\bar J}^\IH$ on the Higgs branch are covariantly constant across the superconformal manifold $\MM$, namely
\beq
\label{matchaa}
\nabla_a G_{I\bar J}^\IH= 0
~.
\eeq
In this equation the index $a$ can be any holomorphic or anti-holomorphic index on the tangent space of the superconformal manifold $\MM$ and expresses a covariant derivative with respect to any exactly-marginal coupling of the theory. Note that in Sec.~\ref{unbrokenphase} we also showed that 
\beq
\label{matchaa2}
\nabla_a G_{I\bar J}^{\rm CFT}= 0
\eeq
are obeyed in the conformal phase (see Eqs.\ \eqref{wardao}).\footnote{This conclusion is corroborated by the argument of \cite{Andriolo:2022lcb} based on Wess--Zumino consistency of the corresponding anomaly functional.} A non-perturbative formal proof of type-B anomaly matching between the broken and unbroken phases, in a range of values of the exactly-marginal couplings, can now be obtained in the following manner.

Let us assume that the anomalies match at a special point $\L^*$ of the conformal manifold, i.e.\ that 
\beq
\label{matchab}
G_{I\bar J}^\IH(\L^*) = G_{I\bar J}^{\rm CFT}(\L^*)
~.
\eeq
In a moment we will present examples of Lagrangian theories where this equality can be established at weak coupling with a tree-level computation. In those cases $\L^*$ is a weak-coupling point of the conformal manifold. We stress that \eqref{matchab} is an assumption in the present argument. It does not hold in all 4D $\NN=2$ SCFTs.

Then, the combination of the general identities \eqref{matchaa}, \eqref{matchaa2} with the assumption \eqref{matchab} yields
\beq
\label{matchac}
\nabla_a \left( G_{I\bar J}^\IH(\L) - G_{I\bar J}^{\rm CFT}(\L) \right) =0~, ~~
G_{I\bar J}^\IH(\L^*) - G_{I\bar J}^{\rm CFT}(\L^*)=0
~.
\eeq
The first equality implies that an arbitrary number of covariant derivatives on $G_{I\bar J}^\IH(\L) - G_{I\bar J}^{\rm CFT}(\L)$ vanish at $\L=\L^*$. Therefore, combined with the second equality we infer that an arbitrary number of ordinary derivatives of $G_{I\bar J}^\IH(\L) - G_{I\bar J}^{\rm CFT}(\L)$ also vanishes at $\L=\L^*$, namely 
\beq
\label{matchacAA}
\p_{a_1}\cdots\p_{a_n} \left( G_{I\bar J}^\IH - G_{I\bar J}^{\rm CFT} \right) |_{\L=\L^*}=0
\eeq 
for arbitrary integer $n$.\footnote{Alternatively, one can show that the system of first-order differential equations in \eqref{matchac} admits only the trivial solution.} As a result, we conclude that the quantities $G_{I\bar J}^\IH(\L) - G_{I\bar J}^{\rm CFT}(\L)$ vanish at all points $\L$ (at least in a finite neighbourhood of $\L^*$) establishing the type-B anomaly matching on the Higgs branch non-perturbatively, for finite values of the exactly-marginal couplings. This argument works independently in each sector of CBOs with the same scaling dimension.

The above statement can be practically very powerful: It allows for the evaluation of the quantities $G_{I\bar J}$ at generic points of the Higgs branch by means of a corresponding CFT computation in the unbroken phase at the origin of the moduli space. The latter can be performed by taking suitable derivatives of the $S^4$ partition function with respect to the exactly-marginal couplings 
\cite{Gerchkovitz:2014gta,Gomis:2015yaa,Baggio:2014sna,Baggio:2014ioa,Gerchkovitz:2016gxx}. In theories with a known Lagrangian formulation the $S^4$ partition function can be further reduced, using supersymmetric localisation methods \cite{Pestun:2007rz}, to a finite-dimensional integral.

In Sec.~\ref{apps} we will employ these considerations to a specific 4D $\NN=2$ circular quiver that deconstructs the 6D $\NN=(2,0)$ theory on a torus, to  obtain certain predictions for a new class of exactly-computable data in the 6D $(2,0)$ theory.

\section{Examples at Tree-Level}
\label{examples}

We now present three examples of Lagrangian theories where type-B anomalies on the Higgs branch are evaluated with a simple tree-level computation at weak coupling.\footnote{Here we will perform the computation in components. It could also be performed conveniently in $\NN=1$ superspace language using superspace Feynman diagrams, which could be useful in explicit higher-loop computations. We hope to return to an explicit higher-loop demonstration of the proposed type-B anomaly matching in future work.} The first example concerns $\NN=2$ superconformal QCD (SCQCD). This is a simple case where the anomaly matching can be established at tree level and therefore holds nonperturbatively according to our general argument. The second example involves a certain 4D $\NN=2$ superconformal circular quiver with $N$ nodes,
for which we also find a match at tree level. The latter plays a crucial role in the deconstruction of 6D $\NN=(2,0)$ theory on $\mathbb T^2$ that will be discussed in Sec.~\ref{apps}. The final, third example revisits (and extends) the issue of type-B anomaly matching in 4D $\NN=4$ SYM theory. This provides a check of the statements in Secs.~\ref{Ward}, \ref{match}.

\subsection{$\NN=2$ SCQCD}
\label{scqcd}

\begin{figure}[t]
  \begin{tikzpicture}
    \tikzset{square/.style={regular polygon,regular polygon sides=4, inner sep = 0}}
\draw (0,0);
\draw[thick] (8.3,0) node[circle,inner sep=3pt,draw] {$k$};

\draw (8.6-1.3,-0.6) node {$\tilde Q_2$};
\draw (8.6-1.3,0.6) node {$Q_2$};
\draw (8.6+1.1,0.6) node {$Q_1$};
\draw (8.6+1.1,-0.6) node {$\tilde Q_1$};

\draw[thick] (8.6,-0.1) to (9.5,-0.1);
\draw[thick,<-] (9.48,-0.1) to (10.38,-0.1);

\draw[thick] (10.38,0.1) to (9.48,0.1);
\draw[thick,<-] (9.48,0.1) to (8.6,0.1);

\draw[thick] (10.8,0) node[square,inner sep=4pt,draw] {$k$};

\draw[thick] (10.8-5,0) node[square,inner sep=4pt,draw] {$k$};

\draw[thick,->] (8.6-2.4,-0.1) to (9.5-2.4,-0.1);
\draw[thick] (9.48-2.4,-0.1) to (10.38-2.4,-0.1);

\draw[thick,->] (10.38-2.4,0.1) to (9.48-2.4,0.1);
\draw[thick] (9.48-2.4,0.1) to (8.6-2.4,0.1);

\end{tikzpicture}
\caption{The quiver diagram of $\NN=2$ superconformal QCD.}
\label{SCQCD}
\end{figure}
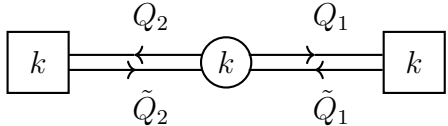

The field content of $\NN=2$ SCQCD theory is captured  by the quiver diagram in Fig.~\ref{SCQCD}. The circular node depicts the gauge group of $\NN=2$ SYM theory, which we take here to be $SU(k)$.  The $\NN=2$ SYM theory is coupled to $2k$ hypermultiplets. In the quiver of Fig.~\ref{SCQCD} we have chosen to depict explicitly only a $U(k) \times U(k)$ part of the full $U(2k)$ flavour symmetry group using two square nodes. The links reflect pairs of fundamental $\NN=1$ chiral multiplets that build up two sets of $\NN=2$ hypermultiplets. 

It is convenient to denote the hypermultiplets as $(q_\II)^a_{\mathfrak i}$, where $a=1,\ldots, k$ is a colour index, ${\mathfrak i}=1,\ldots, 2k$ is a flavour index and $\II=1,2$ is an $SU(2)_R$ index. In terms of the $\NN=1$ chiral fundamentals 
\beq
\label{1nodeai}
q_1 = \bigg\{ {Q_1 \atop Q_2} { \mathfrak i=1,\ldots,k \atop ~~~~\mathfrak  i=k+1,\ldots,2k}
\eeq
\beq
\label{1nodeaj}
q_2 = \bigg\{ { \overline{\tilde Q}_1 \atop \overline{\tilde Q}_2}  {\mathfrak i=1,\ldots,k \atop ~~~~ \mathfrak i=k+1,\ldots,2k}
~.
\eeq
The $\NN=2$ SCQCD theory has a multi-dimensional Higgs-branch moduli space \cite{Argyres:1996eh}. In this subsection, we consider the specific direction where
\beq
\label{1nodeveva}
\langle (q_1)^a_{\mathfrak i} \rangle = v \delta_{\mathfrak i}^a ~, ~~
\langle (q_2)^a_{\mathfrak i} \rangle = 0
\eeq
using the same letter $q_\II$ for the bottom scalar components of the corresponding supermultiplets.\footnote{We hope it will be clear from the context when we will be referring to a component versus a full superfield.}

We will be working in components following the conventions of
\cite{Gadde:2009dj,Gadde:2010zi}. The Lagrangian of the theory can be written as
$\LL=\LL_V+\LL_H$, where $(V)$ denotes the vector-multiplet part and
$(H)$ the hypermultiplet part. We remind the reader that besides the gauge field, whose field strength is $F_{\mu\nu}$, the $\NN=2$ vector multiplet possesses a complex scalar field $\varphi$ and Weyl fermions $\lambda^\II$ all in the adjoint representation of the gauge group. The hypermultiplets contain, beside the complex scalar fields $q_\II$, corresponding fermions $\psi, \tilde \psi$. For reference, in this notation,
\bea
\label{1nodegaga}
\LL_V &=& - \Tr \bigg[ \frac{1}{4} F_{\mu\nu}F^{\mu\nu} + i \bar\lambda_\II \bar \sigma^\mu D_\mu \lambda^\II + D^\mu \varphi \overline{D_\mu \varphi}
\nonumber\\
&&+i g \sqrt{2} \left( \epsilon_{\II \JJ} \lambda^\II \lambda^\JJ \bar
  \varphi -  \epsilon^{\II\JJ} \bar\lambda_\II \bar \lambda_\JJ \varphi \right) + \frac{g^2}{2} [\varphi,\bar\varphi]^2 \bigg]
~,
\eea
\bea
\label{1nodeaga}
\LL_H &=&- \bigg[ D^\mu \bar q^\II D_\mu q_\II + i \bar \psi \bar \sigma^\mu D_\mu \psi + i \tilde \psi \bar\sigma^\mu D_\mu \bar{\tilde \psi}
+i \sqrt 2 g \left( \varepsilon^{\II\JJ} \bar \psi \bar \lambda_\II q_\JJ - \varepsilon_{\II\JJ} \bar q^\II \lambda^\JJ \psi \right)
\nonumber\\
&&+ i \sqrt 2 g( \tilde \psi \lambda^\II q_\II - \bar q^\II \bar \lambda_\II \bar{\tilde \psi}
+ \tilde \psi \varphi \psi - \bar \psi \bar \varphi \bar{\tilde \psi})
+g^2 \bar q_\II (\bar \varphi \varphi + \varphi \bar \varphi ) q^\II + g^2 V(q) \bigg]
~.
\eea
Here $V(q)$ is a quartic expression in the $q$s, the detailed form for which will not be important; for the explicit form of $V(q)$
we refer the reader to \cite{Gadde:2010zi}. The $D_\mu$ are standard
gauge-covariant derivatives. 

To analyse the Higgs branch we implement the following shift on the hypermultiplets
\beq
\label{1nodeak}
(q_1)^a_{\mathfrak i} \to v \delta^a_{\mathfrak i} + (q_1)^a_{\mathfrak i}
~, ~~
(q_2)^a_{\mathfrak i} \to (q_2)^a_{\mathfrak i} 
~.
\eeq
After the shift, the following terms in \eqref{1nodeaga} are especially important for the purposes of our computation
\beq
\label{1nodeam}
\mathcal L_H \supset
- g^2 v \Tr \left[ (\bar\varphi \varphi + \varphi \bar\varphi ) (Q_1 + \bar Q_1)\right] - 2 g^2 v^2 \Tr \left[\varphi \bar\varphi \right]
~.
\eeq
The last term shows that the adjoint complex scalar field $\varphi$ has mass squared $m^2 = 2g^2 v^2$.

A crucial aspect of the spontaneous breaking of conformal invariance on the moduli space is the associated massless Goldstone boson, the dilaton. In the present example, the dilaton takes the form 
\beq
\label{1nodean}
\sigma :=   \Tr\left[ Q_1 + \bar Q_1 \right]
~.
\eeq
As we noted previously, the dilaton has a linear coupling to the energy-momentum tensor $T_{\mu\nu}$ (see e.g.\ \cite{Schwimmer:2010za}) of the form
\begin{equation}
\begin{split}
\label{Ts}
\begin{tikzpicture}
\draw (0,0);
\draw[snake it,double,thick] (0,0) to (1.8,0);
\draw[dashed,thick] (1.8,0) to (3.45,0);
\draw (5.7,0) node {$ \sim - \frac{1}{6} v (q_\mu q_\nu - \eta_{\mu\nu} q^2) \sigma~.$};
\draw (0.4,0.5) node {$T_{\mu\nu}$};
\draw (0.4,-0.5) node {$q$};
\draw (3.3,0.5) node {$\sigma$};
\filldraw (1.8,0) circle (2pt);
\end{tikzpicture}
\end{split}
\end{equation}
This coupling can be seen most easily in the classical expression of $T_{\mu\nu}$ (after the shift \eqref{1nodeak}) as a term
\beq
\label{1nodeao}
- \frac{1}{6} v \left( \d_\mu \d_\nu - \eta_{\mu\nu} \Box \right) \sigma
~,
\eeq
which is linear in the vev $v$. 

The dilaton also has a cubic coupling to the massive adjoint scalar fields $\varphi$, which follows immediately from the first trace on the RHS of \eqref{1nodeam}
\begin{equation}
\begin{split}
\label{sppvertex}
\begin{tikzpicture}
\draw (0,0);
\draw[dashed,thick] (0,0) to (1.5,0);
\draw[thick] (1.5,0) to (2.7,1.2);
\draw[thick] (1.5,0) to (2.7,-1.2);
\draw (0.3,0.3) node {$\sigma$};
\draw (3,1.1) node {$\varphi$};
\draw (3,-1.1) node {$\bar\varphi$};
\draw (4.5,0) node {$ \sim - \frac{2}{k} g^2 v~.$};
\end{tikzpicture}
\end{split}
\end{equation}
The $1\over k$ factor follows from isolating the trace part of $Q_1+\bar Q_1$ in \eqref{1nodeam}.

Our main goal here will be to evaluate, at leading order in the Yang-Mills coupling $g$, the three-point function $\langle T(q) \OO_2(p) \bar\OO_2(\bar p) \rangle$ in the vanishing-momentum limit, where
\beq
\label{CBO2}
\OO_2 = \Tr[ \varphi^2 ]
\eeq
is the single Coulomb-branch operator of this theory with scaling dimension $\Delta=2$.

The computation involves the Feynman diagram of Fig.\ \ref{triangle1},\footnote{This is in complete analogy with the corresponding case of the $\NN=4$ SYM theory analysed in Ref.\  \cite{Schwimmer:2010za}.} which receives the following contributions:

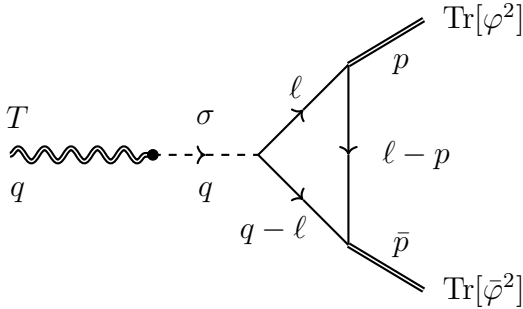
\begin{figure}[!th]
\begin{tikzpicture}
\draw (0,0);
\draw[snake it,double,thick] (5,0) to (6.9,0);

\draw[dashed,thick,->] (6.9,0) to (7.6,0);
\draw[dashed,thick] (7.6,0) to (8.3,0);

\draw[thick,->] (8.3,0) to (8.9,0.6);
\draw[thick] (8.9,0.6) to (9.5,1.2);

\draw[thick,->] (8.3,0) to (8.9,-0.6);
\draw[thick] (8.9,-0.6) to (9.5,-1.2);

\draw[thick,->] (9.5,1.2) to (9.5,0);
\draw[thick] (9.5,0) to (9.5,-1.2);

\draw[thick,double] (9.5,1.2) to (10.5,1.8);
\draw[thick,double] (9.5,-1.2) to (10.5,-1.8);

\draw (5.1,0.5) node {$T$};
\draw (5.1,-0.5) node {$q$};
\draw (7.6,0.5) node {$\sigma$};
\draw (7.6,-0.5) node {$q$};
\draw (8.8,0.9) node {$\ell$};
\draw (8.5,-1) node {$q-\ell$};
\draw (10.2,1.2) node {$p$};
\draw (10.2,-1.2) node {$\bar p$};
\draw (11.3,1.8) node {$\Tr[\varphi^2]$};
\draw (11.3,-1.8) node {$\Tr[\bar \varphi^2]$};
\draw (10.4,0) node {$\ell-p$};

\filldraw (6.9,0) circle (2pt);
\end{tikzpicture}
\caption{The Feynman diagram that determines the leading contribution to $G_{2\bar 2}^\IH$ in the 4D $\NN=2$ SCQCD theory.}
\label{triangle1}
\end{figure}

\begin{itemize}
\item The $T$-$\sigma$ coupling \eqref{Ts} gives the factor $-\frac{1}{6} q^2 v(-3)= \frac{v q^2}{2}$.
\item The dilaton propagator gives the factor $\frac{2ik}{q^2}$, which cancels the above $q^2$.
\item There is a factor $(-i)(-\frac{2}{k}g^2 v)$ from the vertex $\sigma \varphi \bar\varphi$ \eqref{sppvertex}.
\item There is a one-loop momentum integration that yields the factor
\beq
\label{1nodeap}
I = \int \frac{d^4 \ell}{(2\pi)^4} \frac{i}{\ell^2 - m^2} \frac{i}{(q-\ell)^2 - m^2} \frac{i}{(\ell-p)^2 - m^2} \xrightarrow[ p,q \to 0]{} - \frac{1}{(4\pi)^2} \frac{1}{2 m^2}
~,
\eeq
where at the RHS of this expression we evaluated the limit of the integral at vanishing momentum, $p,q\to 0$.
\item The contraction of gauge indices in the triangle give the extra factor $4(k^2-1)$. From the point of view of the index contractions it is twice as if one is evaluating the tree-level contribution to the two-point function $\langle \Tr[\varphi^2](p) \Tr[\bar\varphi^2](-p)\rangle$, which involves the Feynman diagram
\begin{equation}
\begin{split}
\label{22}
\begin{tikzpicture}
\draw (0,0);
\draw[double,thick] (0,0) to (1.5,0);
\draw[double,thick] (4,0) to (5.5,0);
\draw (-0.9,0) node {$\Tr[\varphi^2]$};
\draw (6.4,0) node {$\Tr[\bar\varphi^2]$};
\draw[thick] (1.5,0) .. controls(2.75,0.7) .. (4,0);
\draw[thick] (1.5,0) .. controls(2.75,-0.7) .. (4,0);
\end{tikzpicture}
\end{split}
\end{equation}
\end{itemize}
Collecting all these factors we obtain the following result
\beq
\label{1nodeaq}
G^{\mathbb H}_{2\bar 2} = \frac{1}{\pi^2}\lim_{q,p \to 0}
\langle T(q) \OO_2(p) \bar\OO_2(\bar p) \rangle 
= \frac{2(k^2-1)}{(2\pi)^4}\;,
\eeq
at leading order in the gauge coupling $g$.

This result can be compared against the coefficient of the two-point function $\langle \Tr[\varphi^2]\Tr[\bar\varphi^2]\rangle$ in the CFT phase. Specifically, for the CFT in position space one has (see e.g.\ \cite{Penati:2000zv})
\beq
\label{1nodear}
\langle  \Tr[\varphi^2](x) \Tr[\bar\varphi^2(0) ]\rangle 
= \frac{2(k^2-1)}{(2\pi)^4} \frac{1}{|x|^4}
~.
\eeq
In momentum space, this result takes the form 
\beq
\label{1nodeas}
\langle  \Tr[\varphi^2](p) \Tr[\bar\varphi^2(-p) ]\rangle 
= - \frac{2(k^2-1)}{(4\pi)^2} \log(p^2)
~
\eeq
and from these equations we read off $G^{\rm CFT}_{2\bar 2} = \frac{2(k^2-1)}{(2\pi)^4}$.
As a result, in this case we have verified explicitly the tree-level matching of type-B anomalies
\beq
\label{22match}
G^{\text{CFT}}_{2\bar 2} = G^{\mathbb H}_{2\bar 2}
~.
\eeq
We expect that analogous steps produce a similar matching of type-B anomalies for all the CBOs in this theory. These CBOs are freely generated as multi-trace products of the single-trace Casimirs $\Tr [\varphi^l]$. We will return to this aspect in future work.

\subsection{$\NN=2$ Circular Quivers}
\label{quiver}

\begin{figure}[t]
\begin{center}
\begin{tikzpicture}
\draw[thick,dashed] (0,1) to (0,-1);
\draw[thick] (0,-0.1) to (0.5,-0.1);
\draw[thick,<-] (0.5,-0.1) to (1,-0.1);
\draw[thick,->] (0,0.1) to (0.5,0.1);
\draw[thick] (0.5,0.1) to (1,0.1);
\draw[thick] (1.33,0) node[circle,inner sep=3pt,draw] {$k$};
\draw[thick] (1.66,-0.1) to (2.2,-0.1);
\draw[thick,<-] (2.2,-0.1) to (2.7,-0.1);
\draw[thick,->] (1.66,0.1) to (2.2,0.1);
\draw[thick] (2.2,0.1) to (2.7,0.1);
\draw[thick] (3.03,0) node[circle,inner sep=3pt,draw] {$k$};
\draw[thick] (3.36,-0.1) to (3.9,-0.1);
\draw[thick,<-] (3.9,-0.1) to (4.4,-0.1);
\draw[thick,->] (3.36,0.1) to (3.9,0.1);
\draw[thick] (3.9,0.1) to (4.4,0.1);
\draw[thick] (5.2,0) node {$\cdots \cdots$};
\draw[thick] (6,-0.1) to (6.5,-0.1);
\draw[thick,<-] (6.5,-0.1) to (7,-0.1);
\draw[thick,->] (6,0.1) to (6.5,0.1);
\draw[thick] (6.5,0.1) to (7,0.1);
\draw[thick] (7.33,0) node[circle,inner sep=3pt,draw] {$k$};
\draw[thick] (7.66,-0.1) to (8.2,-0.1);
\draw[thick,<-] (8.2,-0.1) to (8.7,-0.1);
\draw[thick,->] (7.66,0.1) to (8.2,0.1);
\draw[thick] (8.2,0.1) to (8.7,0.1);
\draw[thick] (9.6,0) node {$\cdots \cdots$};
\draw[thick] (10.4,-0.1) to (10.9,-0.1);
\draw[thick,<-] (10.9,-0.1) to (11.4,-0.1);
\draw[thick,->] (10.4,0.1) to (10.9,0.1);
\draw[thick] (10.9,0.1) to (11.4,0.1);
\draw[thick] (11.73,0) node[circle,inner sep=3pt,draw] {$k$};
\draw[thick] (12.06,-0.1) to (12.6,-0.1);
\draw[thick,<-] (12.6,-0.1) to (13.1,-0.1);
\draw[thick,->] (12.06,0.1) to (12.6,0.1);
\draw[thick] (12.6,0.1) to (13.1,0.1);
\draw[thick] (13.43,0) node[circle,inner sep=3pt,draw] {$k$};
\draw[thick] (13.76,-0.1) to (14.3,-0.1);
\draw[thick,<-] (14.3,-0.1) to (14.8,-0.1);
\draw[thick,->] (13.76,0.1) to (14.3,0.1);
\draw[thick] (14.3,0.1) to (14.8,0.1);
\draw[thick,dashed] (14.8,1) to (14.8,-1);
\draw (8.3,0.6) node {$Q^{(\alpha)}$};
\draw (8.3,-0.6) node {$\tilde Q^{(\alpha)}$};
\draw (6.8,0.6) node {$Q^{(\alpha-1)}$};
\draw (6.8,-0.6) node {$\tilde Q^{(\alpha-1)}$};
\end{tikzpicture}
\end{center}
\caption{The circular $\NN=2$ superconformal quiver. Each of the $N$ nodes denotes an $SU(k)$ $\NN=2$ vector multiplet. The links are $\NN=1$  bifundamental chiral multiplets. The dashed perpendicular lines at the left and right ends are identified.}
\label{fig:quiver}
\end{figure}
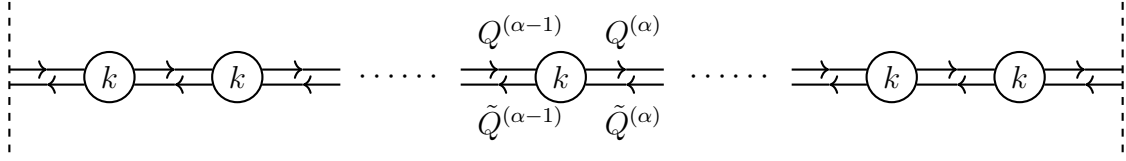

Our second example involves the circular superconformal quiver of Fig.~\ref{fig:quiver}. Every two consecutive nodes in this quiver are linked by a pair of $\NN=1$ chiral bifundamental fields $Q^{(\alpha)}$, $\tilde Q^{(\alpha)}$ that make up an $\NN=2$ bifundamental hypermultiplet, while there are also adjoint chiral superfields $\Phi^{(\alpha)}$ that are part of the $\NN = 2$ vector multiplet. By a slight abuse of notation we will continue denoting the bottom components of $Q^{(\alpha)}$, $\widetilde Q^{(\alpha)}$ with the same symbol, while  $\Phi^{(\alpha)}$ will have bottom component $\varphi^{(\alpha)}$. The node label $\alpha$ takes the $N$ discrete values $-\lfloor N/2\rfloor+1,-\lfloor N/2\rfloor+2,\ldots,\lfloor N/2\rfloor$. The gauge group at each node is $SU(k)$ and we focus on the case where all the gauge couplings $g^{(\alpha)}$ are equal, $g^{(\alpha)} \equiv g$. This particular point  on the space of couplings is usually referred to as the `orbifold point', since this version of the theory can be obtained as an $\NN=2$ preserving $\mathbb Z_N$ orbifold of $\NN=4$ SYM theory with gauge group $SU(kN)$.

For our purposes it is convenient to introduce the following notation. We set $\q = e^{2\pi i/N}$ and perform the discrete Fourier transformations:
\bea
\label{nnodeaa}
Q^{(\alpha)} = \frac{1}{\sqrt N} \sum_\beta \q^{\alpha \beta} \widehat Q^{(\beta)}
~,\\
\widetilde Q^{(\alpha)}=  \frac{1}{\sqrt N} \sum_\beta \q^{\alpha \beta} \widehat {\widetilde Q}^{(\beta)}
~,\\
\varphi^{(\alpha)} =  \frac{1}{\sqrt N} \sum_\beta \q^{\alpha \beta} \hat \varphi^{(\beta)}
\label{nnodeaa3}
~,
\eea 
where $\beta$ is summed over all the quiver nodes. For the Hermitian conjugates we set 
\beq
\label{nnodeab}
\bar Q^{(\alpha)} = \frac{1}{\sqrt N} \sum_\beta \q^{-\alpha \beta} \widehat {\bar Q}^{(\beta)}
\eeq
and similarly for the remaining scalars. For the real and imaginary parts of the Fourier coefficients $\widehat Q$ we introduce the notation
\bea
\label{nnodeac}
\widehat Q^{(\alpha)} = \frac{1}{\sqrt 2} \frac{1}{g} \left( X^{(\alpha)} + i Y^{(\alpha)} \right)
~,\\
\widehat {\bar Q}^{(\alpha)} = \frac{1}{\sqrt 2} \frac{1}{g} \left( X^{(-\alpha)} - i Y^{(-\alpha)} \right)
~.
\eea

This theory has a rich moduli space of vacua. Here, we will only be interested in a very specific direction along the Higgs branch, where in terms of the elementary fields we consider vevs
\beq
\label{nnodeaca}
\langle Q^{(\alpha)} \rangle=\frac{v}{\sqrt 2} \one_{k\times k} ~, ~~
\langle \widetilde Q^{(\alpha)} \rangle =0
~,
\eeq
with $\one_{k\times k}$ denoting the $k\times k$ identity matrix, in analogy with the vev \eqref{1nodeveva} of Sec.~\ref{scqcd}. In this case, the dilaton $\sigma$ is proportional to the trace part of $X^{(0)}$. Specifically
\beq
\label{nnodeas}
\sigma = \frac{\sqrt{2N}}{g} \Tr \left[ X^{(0)} \right]
~.
\eeq

Let us highlight some of the pertinent features of the classical action of this quiver; we use the conventions of \cite{Lambert:2012qy}. $\NN=2$ supersymmetry requires at each node $\alpha$ the $\NN=1$ superpotential
\beq
\label{nnodeacb}
W^{(\alpha)} = - i \sqrt 2 g \Tr \left[ \widetilde Q^{(\alpha)} \Phi^{(\alpha)} Q^{(\alpha)} - Q^{(\alpha)} \Phi^{(\alpha+1)} \widetilde Q^{(\alpha)} \right]
~.
\eeq
The corresponding scalar-potential $V$ receives two contributions $V=V_F+V_D$. The $F$-term contributions are 
\beq
\label{nnodead}
V_F = \sum_\alpha \Tr\left[ \bar F_{Q^{(\alpha)}} F_{Q^{(\alpha)}} 
+ \bar F_{\widetilde Q^{(\alpha)}} F_{\widetilde Q^{(\alpha)}}
+ \bar F_{\varphi^{(\alpha)}} F_{\varphi^{(\alpha)}} \right]
~,
\eeq
with
\bea
\label{nnodeag}
F_{Q^{(\alpha)}} = - i \sqrt 2 g \left( \widetilde Q^{(\alpha)} \varphi^{(\alpha)} - \varphi^{(\alpha+1)} \widetilde Q^{(\alpha)} \right)
~,
\eea
\bea
\label{nnodeai}
F_{\widetilde Q^{(\alpha)}} = - i \sqrt 2 g \left( \varphi^{(\alpha)} Q^{(\alpha)}  -  Q^{(\alpha)} \varphi^{(\alpha+1)} \right)
~,
\eea
\bea
\label{nnodeaj}
F_{\varphi^{(\alpha)}} = - i \sqrt 2 g \left( Q^{(\alpha)} \widetilde Q^{(\alpha)} - \widetilde Q^{(\alpha-1)} Q^{(\alpha-1)} \right)\;,
\eea
while the D-term contributions are
\beq
\label{nnodeae}
V_D = \frac{g^2}{2} \sum_\alpha D^{(\alpha)A}  D^{(\alpha)A}
~,
\eeq
where
\bea
\label{nnodeaf}
D^{(\alpha)A} &=& \Tr \bigg[ T^A \bigg( \left[ \varphi^{(\alpha)}, \bar \varphi^{(\alpha)} \right] + Q^{(\alpha)} \bar Q^{(\alpha)} - \widetilde {\bar Q}^{(\alpha)} \widetilde Q^{(\alpha)}
\nonumber\\
&&~~~~~\qquad\qquad - \bar Q^{(\alpha-1)} Q^{(\alpha-1)} + \widetilde Q^{(\alpha-1)} \widetilde {\bar Q}^{(\alpha-1)} \bigg) \bigg]
~.
\eea
In this expression $A$ is an $SU(k)$ Lie-algebra index and $T^A$ the generators of the Lie algebra with normalisation $\Tr [T^A T^B] =\delta^{AB}$. In \eqref{nnodeaf} we assume canonical kinetic terms for all the vector and chiral multiplets.

When expanding around the vevs \eqref{nnodeaca}, one obtains expressions of the following form from the D-terms
\beq
\label{nnodeal}
D^{(\alpha)A} = D_0^{(\alpha)A} + \frac{v}{g\sqrt N} \sum_\beta \q^{\alpha\beta} \left(1-\q^{-\beta}\right) \Tr \left[ T^A X^{(\beta)} \right]
~,
\eeq
where $D_0^{(\alpha)A}$ is 0$^{\text{th}}$ order in the $v$ expansion. Feeding this expansion into $V_D$ gives 
\bea
\label{nnodeam}
V_D &=& \frac{g^2}{2} \sum_\alpha D_0^{(\alpha)A}  D_0^{(\alpha)A}
+ \frac{v g}{\sqrt N} \sum_{\alpha,\beta} \q^{\alpha \beta}\left( 1- \q^\beta \right) D_0^{(\alpha)A} \Tr\left[ T^A X^{(\beta)} \right] 
\nonumber\\
&&+ \frac{v^2}{2} \sum_\alpha \left( 1- \q^\alpha\right) \left( 1- \q^{-\alpha} \right) \Tr \left[ T^A X^{(\alpha)} \right] \Tr \left [ T^A X^{(-\alpha)} \right]
~.
\eea
We can further recast this expression by using the identity
\beq
\label{nnodean}
\Tr \left[ T^A X \right] \left[ T^A Y \right ] = \Tr \left[ XY \right] - \frac{1}{k} \Tr \left[ X \right] \Tr \left[ Y \right]
~,
\eeq
but we will not write explicitly the resulting expressions. Instead, we will refer the interested reader to \cite{Lambert:2012qy} for related computations. The observation that is needed here is that the field $X^{(0)}$ does not appear in a cubic scalar coupling and does not receive a mass, in agreement with the fact that it is directly related to the dilaton, \eqref{nnodeas}.

The F-term part of the potential, $V_F$, provides more interesting contributions. We note the following useful expansions
\beq
\label{nnodeao}
F_{\widetilde Q^{(\alpha)}} = F_{0\widetilde Q^{(\alpha)}} 
- \frac{i v g}{\sqrt N} \sum_\beta \q^{\alpha \beta} \left( 1- \q^\beta\right) \hat \varphi^{(\beta)}
~,
\eeq
\beq
\label{nnodeap}
\bar F_{\widetilde Q^{(\alpha)}} = \bar F_{0\widetilde Q^{(\alpha)}}
+ \frac{i v g}{\sqrt N} \sum_\beta \q^{\alpha \beta} \left( 1- \q^\beta\right) \hat {\bar \varphi}^{(-\beta)}
~,
\eeq
where $F_{0\widetilde Q^{(\alpha)}}$, $\bar F_{0\widetilde Q^{(\alpha)}}$ are 0$^{\text{th}}$ order expressions in the $v$ expansion. Then,
\bea
\label{nnodeaq}
&&V_F \supset \sum_\alpha \Tr \left[ \bar F_{0 \widetilde Q^{(\alpha)}} F_{0\widetilde Q^{(\alpha)}} \right]
\nonumber\\
&&~~~~~~ +\frac{vg}{\sqrt N} \sum_{\alpha,\alpha'} \left( 1- \q^\alpha \right) \Tr \left[ \left( X^{(-\alpha -\alpha')}-i Y^{(-\alpha -\alpha')} \right) \left( \hat {\bar \varphi}^{(-\alpha')} \hat \varphi^{(\alpha)} - \q^{\alpha'} \hat \varphi^{(\alpha)} \hat {\bar \varphi}^{(-\alpha')} \right) \right]
\nonumber\\
&& ~~~~~~ +\frac{vg}{\sqrt N} \sum_{\alpha,\alpha'} \left( 1- \q^\alpha \right) \Tr \left[ \left( X^{(-\alpha -\alpha')}+i Y^{(-\alpha -\alpha')} \right) \left( \hat {\bar \varphi}^{(-\alpha)} \hat \varphi^{(\alpha')} - \q^{\alpha'} \hat \varphi^{(\alpha')} \hat {\bar \varphi}^{(-\alpha)} \right) \right]
\nonumber\\
&& ~~~~~~ + v^2 g^2 \sum_\alpha \left( 1- \q^\alpha \right) \left( 1- \q^{-\alpha} \right) \Tr \left[ \hat {\bar \varphi}^{(\alpha)} \hat \varphi^{(\alpha)} \right]
~.
\eea
From this formula we read off the mass 
\beq
\label{nnodear}
m_\alpha^2 = 2 v^2 g^2  \left( 1- \q^\alpha \right) \left( 1- \q^{-\alpha} \right) 
\eeq
for the modes $\hat \varphi^{(\alpha)}$. In addition, we read off a cubic coupling between the dilaton and the modes $\hat \varphi^{(\alpha)}$
\beq
\label{nnodeat}
V_F \supset - \frac{\sqrt 2 v g^2}{k N} \sum_\alpha \left( 1- \q^\alpha \right) \left( 1- \q^{-\alpha} \right) \sigma\, \Tr \left[ \hat \varphi^{(\alpha)} \hat {\bar \varphi}^{(\alpha)} \right]
~.
\eeq

This information is all we need to perform the computation of the three-point function $\langle T \OO \bar \OO\rangle$ at tree-level for general CBOs $\OO$. The vector multiplet at each quiver node contains an adjoint complex scalar field $\varphi^{(\alpha)}$ from which we can build the Casimir 
\beq
\label{nnodeata}
\OO_\ell^{(\alpha)} = \Tr[(\varphi^{(\alpha)})^\ell]
~.
\eeq
This is a CBO with scaling dimension $\Delta = \ell$.\footnote{We can also consider multi-trace products of the Casimir, which are also CBOs, but we will not do this explicitly in this paper. The single-trace operators are generators of the Coulomb-branch chiral ring.} For this theory it is customary to define the operators $\widehat \OO_\ell^{(\alpha)}$ via the discrete Fourier transform
\beq
\label{nnodeatb}
\OO_\ell^{(\alpha)} = \frac{1}{\sqrt N} \sum_\beta \q^{\alpha\beta} \widehat \OO_\ell^{(\beta)}
~.
\eeq
The operator $\widehat \OO_\ell^{(0)}$, which has vanishing discrete-Fourier momentum, is called an `untwisted' CBO operator of the quiver theory. All the other operators are `twisted' operators.

For calculations in the broken phase it is useful to express $\widehat \OO_\ell^{(\alpha)}$ as a sum over traces of the discrete-Fourier-transformed elementary fields $\hat \varphi^{(\alpha)}$ (see Eq.~\eqref{nnodeaa3})
\beq
\label{nnodebk}
\widehat \OO^{(\alpha)}_\ell =
\frac{1}{\sqrt N} \sum_\beta \q^{-\alpha \beta} \Tr\left[ \left( \varphi^{(\beta)} \right)^\ell \right]
= \frac{1}{N^{\frac{\ell-1}{2}}} \sum_{\alpha_1,\ldots,\alpha_{\ell-1}} \Tr \left[ \left( \prod_{n=1}^{\ell-1} \hat \varphi^{(\alpha_{n})} \right) \hat \varphi^{(\alpha - \sum_{m=1}^{\ell-1} \alpha_m )} \right]
~.
\eeq

As an illustration of the above structure, let us focus henceforth on $\Delta=2$ CBOs, which can be cast in the following form
\beq
\label{nnodeau}
\widehat \OO^{(\alpha)}_2 = \frac{1}{\sqrt N} \sum_{\alpha'} \Tr \left[ \hat \varphi^{(\alpha')} \hat \varphi^{(\alpha - \alpha')} \right]
~.
\eeq
Our purpose is to evaluate the tree-level contribution to the type-B anomaly in $\langle T \widehat \OO_2^{(\alpha)} \overline{\widehat \OO}_2^{(\alpha)} \rangle$ in the broken phase. As in the previous subsection, this involves the computation of a Feynman diagram similar to that of Fig.\ \ref{triangle1}. We will call the tree-level result $G^{(\alpha)\IH}_{2\bar 2}$ when it refers to the operator $\widehat \OO^{(\alpha)}_2$.

First, let us consider separately the case where the external CBO is untwisted, i.e.\ the case $\alpha=0$ in \eqref{nnodeau}. We obtain
\bea
\label{nnodeba}
G^{(0)\IH}_{2\bar 2} &=& \frac{1}{\pi^2}\left(\frac{v}{2\sqrt 2} q^2\right) \left( \frac{2Nk i}{q^2} \right)  \left( -i \frac{- \sqrt 2 v g^2}{kN} \right) 
\nonumber\\
&& \times 
\sum_{\alpha' \neq 0}  \left( 1- \q^{\alpha'} \right) \left( 1- \q^{-\alpha'} \right)  \frac{1}{N} \frac{-1}{(4\pi)^2} \frac{1}{2v^2 g^2} \frac{8 (k^2-1)}{2 \left( 1- \q^{\alpha'} \right) \left( 1- \q^{-\alpha'} \right) } 
\nonumber\\
&=& \frac{2(k^2-1)}{(2\pi)^4} \left( 1- \frac{1}{N} \right)
~.
\eea
The result in the conformal phase is $G^{(0){\rm CFT}}_{2\bar 2}= \frac{2(k^2-1)}{(2\pi)^4}$ and at first sight we observe a mismatch. However, the anomaly receives an extra contribution from the IR chiral ring and the anomalies in fact match; for an updated discussion see \cite{Niarchos:2020nxk}.

For the twisted sector CBOs we can compute the anomaly in a similar fashion. In this case, the calculation requires the value of the integral
\beq
\label{nnodebb}
I = \int \frac{d^4 \ell}{(2\pi)^4} \frac{i}{\ell^2 - m_1^2}    \frac{i}{\ell^2 - m_1^2}  \frac{i}{\ell^2 - m_2^2} 
~.
\eeq
For $m_1=m_2=m$, which occurs in $\II^{(0)}$, we obtain 
\beq
\label{nnodebc}
I = - \frac{1}{(4\pi)^2} \frac{1}{2m^2}
~.
\eeq
For $m_1 = m\neq 0$ and $m_2=0$ we obtain
\beq
\label{nnodebd}
I = - \frac{1}{(4\pi)^2} \frac{1}{m^2}
~.
\eeq
More generally, when both $m_1, m_2 \neq 0$ we obtain
\beq
\label{nnodebe}
I =  - \frac{1}{(4\pi)^2} \frac{m_1^2 - m_2^2 - m_2^2 \log \left( \frac{m_1^2}{m_2^2} \right)}{(m_1^2 - m_2^2)^2}
~.
\eeq
Collecting all the contributions to the diagram for $\alpha\neq 0$ we find
\beq
\label{nnodebf}
G^{(\alpha)\IH}_{2\bar 2} =  \frac{1}{(2\pi)^4} \frac{2 (k^2-1)}{N}  \left[ 2 + \sum_{\alpha' \neq \, 0, \alpha} \mathfrak L(\alpha'; \alpha) \right]
~,
\eeq
where
\begin{align}
\begin{split}
\label{nnodebg}
&\hspace{-0.7cm}
\mathfrak L(\alpha';\alpha)
= \left [ \left( 1- \q^{\alpha'} \right) \left( 1- \q^{-\alpha'} \right) - \left( 1- \q^{-\alpha+\alpha'} \right) \left( 1- \q^{\alpha-\alpha'} \right) \right]^{-2}
\\
&\hspace{-0.7cm}
\Bigg\{ \left( 1- \q^{\alpha'} \right) \left( 1- \q^{-\alpha'} \right)
\Bigg[ \left( 1- \q^{\alpha'} \right) \left( 1- \q^{-\alpha'} \right) - \left( 1- \q^{-\alpha + \alpha'} \right) \left( 1- \q^{\alpha -\alpha'} \right) 
\\
&\hspace{4cm}
- \left( 1- \q^{-\alpha + \alpha'} \right) \left( 1- \q^{\alpha-\alpha'} \right) \log \frac{\left( 1- \q^{\alpha'} \right) \left( 1- \q^{-\alpha'} \right)}{\left( 1- \q^{-\alpha + \alpha'} \right) \left( 1- \q^{\alpha-\alpha'} \right)} \Bigg]
\\
&\hspace{-0.7cm}
+\left( 1- \q^{-\alpha + \alpha'} \right) \left( 1- \q^{\alpha -\alpha'} \right)
\Bigg[ \left( 1- \q^{-\alpha + \alpha'} \right) \left( 1- \q^{\alpha-\alpha'} \right) - \left( 1- \q^{\alpha'} \right) \left( 1- \q^{-\alpha'} \right) 
\\
&\hspace{4.5cm}
+ \left( 1- \q^{\alpha'} \right) \left( 1- \q^{-\alpha'} \right) \log \frac{\left( 1- \q^{\alpha'} \right) \left( 1- \q^{-\alpha'} \right)}{\left( 1- \q^{-\alpha + \alpha'} \right) \left( 1- \q^{\alpha-\alpha'} \right)} \Bigg] 
\Bigg\}
\\
&\hspace{1cm}
=1
~.
\end{split}
\end{align}
The term 1 inside the parenthesis of the RHS of \eqref{nnodebf} comes from the case where one of the propagators inside the triangle is massless. There is no case where only massless fields, or two massless fields run inside the loop. The remaining contributions, which are captured by the sum over the quantity $\mathfrak L$, involve only massive propagators. For all $\alpha'\neq 0, \alpha$ we obtain $\mathfrak L(\alpha';\alpha)=1$. As a result we find that
\beq
\label{nnodebi}
G^{(\alpha)\IH}_{2\bar 2} =   \frac{2 (k^2-1)}{(2\pi)^4} 
\eeq
independent of $\alpha$.\footnote{A previous version of the paper reported an erroneous extra factor of $(1-1/N)$ in \eqref{nnodebi}.
}

In comparison, the tree-level two-point function in the conformal phase
\beq
\label{nnodebj}
\langle \widehat \OO^{(\alpha)}_2 ~ \overline {\widehat \OO}^{(\alpha)}_2 \rangle
=\frac{1}{N} \sum_{\alpha',\alpha''} \left \langle \Tr\left[ \hat \varphi^{(\alpha')} \hat \varphi^{(\alpha-\alpha')} \right] ~  \Tr\left[ \hat {\bar \varphi}^{(\alpha'')} \hat {\bar \varphi}^{(\alpha-\alpha'')} \right] \right \rangle
\eeq
yields
\beq
\label{nnodebjtree}
G^{(\alpha){\rm CFT}}_{2\bar 2} =  \frac{2 (k^2-1)}{(2\pi)^4}
~,
\eeq
which is also independent of the discrete momentum $\alpha$ and
agrees exactly with \eqref{nnodebi} on the Higgs branch. This fact will be very useful in Sec.\ \ref{apps}.

We expect that there is a corresponding matching of type-B anomalies for all the single-trace CBOs $\widehat \OO^{(\alpha)}_\ell$ and their multi-trace versions. We aim to return to this point in the future.

\subsection{$\NN=4$ SYM Theory}
\label{sym}

The case of the $\NN=4$ SYM theory can be treated as a special case of an $\NN=2$ theory. The $\NN=4$ SYM theory has an enhanced $SO(6)_R\supset SU(2)_R\times U(1)_r$ R-symmetry and six real adjoint scalar fields $\varphi_1,\ldots,\varphi_6$ charged under it. The $\NN=2$ SCA can be embedded inside the $\NN=4$ algebra so that the $\NN=2$ $SU(2)_R$ symmetry rotates the $\varphi_1,\ldots,\varphi_4$ fields, while the $\NN=2$ $U(1)_r$ the $\varphi_5,\varphi_6$. In the context of this embedding, the $\NN=4$ Coulomb branch can be split into an $\NN=2$ Higgs-branch direction and an $\NN=2$ Coulomb-branch direction. Correspondingly, the Higgs branch is characterised by vevs of the scalars $\varphi_1,\ldots,\varphi_4$, while the Coulomb branch by vevs of the scalars $\varphi_5,\varphi_6$. To repeat the computations of the previous subsections we focus on $\half$-BPS chiral-primary operators that are single- (or multi-) trace Casimirs of the scalar fields $\varphi_5,\varphi_6$. These fields are $\NN=2$ CBOs under the above superconformal-algebra embedding. The type-B anomalies of these operators on the Higgs branch can then be computed at tree-level as in Secs.~\ref{scqcd}, \ref{quiver}, to establish the tree-level matching of type-B anomalies on the $\NN=4$ Coulomb branch.

This computation was performed in \cite{Schwimmer:2010za} for $\Tr[\varphi_5 \varphi_6]$ type-B anomalies in the $\NN=4$ $SU(2)$ theory, where the Coulomb-branch direction was taken to be associated with $\varphi_1$ vevs. The authors of \cite{Schwimmer:2010za} verified at tree level that the corresponding type-B anomalies of the broken and unbroken phases match. Our arguments in this paper are improving the analysis of \cite{Schwimmer:2010za} in the following directions:
\begin{itemize}
\item[(a)] The arguments of Sec.~\ref{match} can be used to argue that the above matching holds non-perturbatively at finite Yang-Mills coupling. This conclusion is consistent with expectations, based on the relationship between type-B and chiral anomalies for this high degree of supersymmetry. Therefore, the $\NN= 4$ example could be viewed as a mild check of the general claims of Secs.~\ref{Ward}, \ref{match}.
\item[(b)] One can use our approach to extend the tree-level matching to more general $\half$-BPS operators of the $\NN=4$ SYM theory with $\Delta>2$. In that case the arguments of Secs.~\ref{Ward}, \ref{match} would imply the non-perturbative matching for type-B anomalies that are not related to chiral anomalies. 
\end{itemize}

\section{The Type-B Anomaly from SUSY Localisation}\label{localisation}

We now turn to the explicit calculation of the type-B anomalies for the Lagrangian $\NN =2$ theories that we discussed in Sec.~\ref{examples} using supersymmetric localisation. 

\subsection{The Partition Function on $S^4_{\epsilon_1,\epsilon_2}$}

The partition function of Lagrangian $\NN=2$ $SU(k)$ SCFTs $\ZZ_{S^4_{\epsilon_1,\epsilon_2}}(\tau, \bar \tau, m;\epsilon_1, \epsilon_2)$ on the  ellipsoid $S^4_{\epsilon_1,\epsilon_2}$
is explicitly known from the localisation calculation of \cite{Pestun:2007rz,Hama:2012bg}: 
\begin{align} 
  \ZZ_{S^4_{\epsilon_1,\epsilon_2}}(\tau, \bar \tau, m;\epsilon_1, \epsilon_2) = \int [ da ] \left| Z_{\rm 4D}(\tau, a,m;\epsilon_1 , \epsilon_2) \right|^2\;.
\end{align}
Here we use the notation $\tau= \{\tau^{(\alpha)}\}$ for the set of
 marginal gauge couplings, with the index $\alpha$ counting the number of
 gauge groups,  $a= \{a^{(\alpha)}_b\}$ for the set of Coulomb-branch
 parameters, with the index  $b = 1,...,k$ labelling the Coulomb-branch parameters within a given colour group. $m$ denotes the set of all masses. In the case of SCQCD we have only one gauge group with one marginal coupling $\tau$, $k$ traceless Coulomb-branch parameters
  $a= \{a_b\}$
  and $N_f = 2k$ fundamental masses $m = \{m^{{\mathfrak i}} \}$ with ${\mathfrak i}=1,\ldots, 2k$ a flavour index. In the case of  the circular $\NN=2$ superconformal quiver with $N$ $SU(k)$ nodes, $m=m_{\rm bif} = \{m_{\rm bif}^{(\alpha)} \}$ is the set of $N$ bifundamental-hypermultiplet masses. 

The holomorphic half of the integrand of the partition function  $Z_{\rm 4D}(\tau, a,m;\epsilon_1 , \epsilon_2)$ is the  IR 
 Coulomb-branch partition function  on $\mathbb R^4_{\epsilon_1,\epsilon_2}$, which factorises as
\begin{align}
\label{eq:4DQuiverPF}
Z_{\rm 4D}&=Z_{\text{4D,cl}}  Z_{\text{4D,1-loop}} Z_{\text{4D,inst}}\; .
\end{align}
The one-loop part 
\begin{align}
 Z_{\text{4D,1-loop}}( a,m_{\rm bif};\epsilon_1 , \epsilon_2) =\prod_{\alpha}  Z_{\text{1-loop}}^{\text{vec}} (a ;\epsilon_1 , \epsilon_2)  
  Z_{\text{1-loop}}^{\text{bif}}(a,m_{\rm bif}\; ;\epsilon_1 , \epsilon_2)\;,
\end{align}
contains vector and hypermultiplet contributions,
\begin{align}
  Z^{\text{bif}}_{\text{1-loop}}(a,m_{\rm bif};\epsilon_1 , \epsilon_2) &=\prod_{b,c=1}^k\Gamma_2\left(a_b^{(\alpha)}-a_c^{(\alpha+1)} -m_{\rm bif}^{(\alpha)}+\frac{\epsilon_+}{2}\big|\epsilon_1,\epsilon_2\right)\;,\cr
                                                                          Z^{\text{vec}}_{\text{1-loop}}(a;\epsilon_1 , \epsilon_2)&=\prod_{b,c=1}^k \Gamma_2\left(a_b^{(\alpha)}-a_c^{(\alpha)}\big|\epsilon_1,\epsilon_2\right)^{-1}\;,
\end{align}
where $\epsilon_+ = \epsilon_1 + \epsilon_2$ and $\Gamma_2(x|\epsilon_1,\epsilon_2) $ is the Barnes double-Gamma function.
Here we only  give the contribution of a bifundamental hypermultiplet, from which  the fundamental is easily obtained. Similarly, the instanton part is given by \cite{Alday:2009aq}
\begin{align}
  \label{eq:instantonPF}
 Z_{\text{4D,inst}}(\tau, a,m_{\rm bif };\epsilon_1 , \epsilon_2) = \sum_{\nu} \prod_{\alpha}   \mathbf{q}_{(\alpha)}^{\sum_{b=1}^k| \nu^{(\alpha)}_{b} |} 
   Z_{\text{inst}}^{\text{vec}} (a; \nu )  
  Z_{\text{inst}}^{\text{bif}} (a,m_{\rm bif}; \nu )
\end{align}
where $\mathbf{q}_{(\alpha)} = e^{2\pi i \tau^{(\alpha)} }$. A  Young diagram $\nu^{(\alpha)}_{b}$ appears for each of the Coulomb moduli  $a^{(\alpha)}_{b} $, collectively denoted by $\nu =\{\nu_b^{(\alpha)}\}$, while the instanton number is given in terms of the total number of boxes of the Young diagram $| \nu_b^{(\alpha)} |$.
The vector and hypermultiplet instanton contributions respectively read
\begin{align}
  Z_{\text{inst}}^{\text{bif}} (\tau, a,m_{\rm bif}; \nu)&=\prod_{b,c=1}^k{\rm N}_{\nu_{c}^{(\alpha+1)}\nu_{b}^{(\alpha)}}\left(a_b^{(\alpha)} -a^{(\alpha+1)}_c -m_{\rm bif}^{(\alpha)}-\frac{\epsilon_+}{2}\right)\;,\cr
 Z_{\text{inst}}^{\text{vec}} (a,m_{\rm bif}; \nu )&=\prod_{b,c=1}^k{\rm N}_{\nu_{c}^{(\alpha)}\nu_{b}^{(\alpha)}}\left(a^{(\alpha)}_b -a^{(\alpha)}_c\right)^{-1}\;,
\end{align}
where the functions $\textrm{N}_{\lambda\mu}(a)$ involved above are defined as
\begin{align}\label{Nfunctions}
\textrm{N}_{\lambda\mu}(a)=\prod_{(i,j)\in \lambda}\left[a+\epsilon_1(\lambda_i-j+1)+\epsilon_2(i-\mu_j^t)\right]
 \prod_{(i,j)\in \mu}\left[a+\epsilon_1(j-\mu_i)+\epsilon_2(\lambda_j^t -i+1)\right]\;.
\end{align}
Specifically, for the circular $\NN=2$ superconformal quiver, the  Coulomb-branch partition function \eqref{eq:4DQuiverPF} takes the form \cite{Hayling:2017cva}
\begin{align}
\label{eq:4DQuiverPFpertinst}
&  Z_{\text{4D,cl}} = \exp \left[ - \frac{2\pi i }{\epsilon_1 \epsilon_2} \sum_{\alpha =-\lfloor\frac{N}{2}\rfloor+1}^{\lfloor\frac{N}{2}\rfloor} \tau^{(\alpha)} \sum_{b=1}^k \left( a_b^{(\alpha)} \right)^2 \right]\;,\cr
& Z_{\text{4D,1-loop}} 
=
\prod_{i,j=1}^{\infty}\prod_{\alpha =\lfloor -\frac{N}{2}\rfloor+1}^{\lfloor \frac{N}{2}\rfloor }\prod_{1\leq b\leq c\leq k}
\frac{\left[  a^{(\alpha)}_{b} - a_c^{(\alpha+1)} -  m_{\rm bif}^{(\alpha)}  +\epsilon_1(j-1/2)-\epsilon_2(i-1/2)\right]  
  }{    
\left[a^{(\alpha+1)}_{b} - a^{(\alpha+1)}_{c}  +\epsilon_1 i-\epsilon_2 (j-1)\right]
}
\cr 
&\qquad \qquad\times\prod_{1\leq b<c\leq k}
\frac{\left[ a_b^{(\alpha+1)}  - a^{(\alpha)}_{c} + m_{\rm bif}^{(\alpha)}
+\epsilon_1(j-1/2)-\epsilon_2(i-1/2)\right]
}{
\left[a^{(\alpha)}_{b} - a^{(\alpha)}_{c}   +\epsilon_1(i-1)-\epsilon_2 j\right]}\;,
\cr
 & Z_{\text{4D,inst}} 
 =
 \sum_{\nu}\prod_{\alpha =\lfloor -\frac{N}{2}\rfloor+1}^{\lfloor \frac{N}{2}\rfloor }  \mathbf{q}_{(\alpha)}^{\sum_{b=1}^k|\nu_{b}^{(\alpha)}|} 
 \prod_{1\leq b\leq c\leq k}
 \frac{{\rm N}_{\nu_{c}^{(\alpha+1)}\nu_{b}^{(\alpha)}}
 \left( a^{(\alpha)}_{b} - a_c^{(\alpha+1)} -  m_{\rm bif}^{(\alpha)} -\frac{\epsilon_+}{2}\right)
 }{
 {\rm N}_{\nu_{c}^{(\alpha+1)}\nu_{b}^{(\alpha+1)}}\left(a^{(\alpha+1)}_{b} - a^{(\alpha+1)}_{c}  \right)}
 \cr
  &\qquad \qquad \times  \prod_{1\leq b<c\leq k}
  \frac{{\rm N}_{\nu_{c}^{(\alpha)}\nu_{b}^{(\alpha+1)}}\left(    a_b^{(\alpha+1)}  - a^{(\alpha)}_{c} + m_{\rm bif}^{(\alpha)}  -\frac{\epsilon_+}{2}\right)}{{\rm N}_{\nu_{c}^{(\alpha)}\nu_{b}^{(\alpha)}}\left(a^{(\alpha)}_{b} - a^{(\alpha)}_{c} -\epsilon_+\right)}\;.
\end{align}
Moreover, complex conjugation is implemented by:
\begin{align}
  \label{eq:31}
\overline{ Z_{\rm 4D}}(\tau , a,m_{\rm bif};\epsilon_1 , \epsilon_2) := Z_{\rm 4D} (\bar \tau, -a,-m_{\rm bif};-\epsilon_1 , -\epsilon_2)\;.  
\end{align}
Before we turn to the correlation functions of CBOs, a final note is in order.
In the case of $\mathcal{N}=4$ SYM, for  both the perturbative and the instanton part the contributions of the vector and  hypermultiplet cancel against each other\footnote{This fact can be seen almost immediately from \eqref{eq:4DQuiverPFpertinst} after setting $N=1$ and $ m_{\rm bif}=-\frac{\epsilon_+}{2}$. For the perturbative part a slight reorganisation of the poles is required.} and the partition function simply becomes
\begin{align}
\label{eq:N=4pf}
  \mathcal Z^{\mathcal{N}=4}_{S^4_{\epsilon_1,\epsilon_2}}(\tau, \bar \tau, m;\epsilon_1, \epsilon_2)=\int  \left(\prod_{b=1}^kda_b\right)\delta(\sum_{b=1}^ka_b)\,e^{-2\pi\text{Im}\tau \sum^N_{b=k} a^2_b}   \prod_{b<c}(a_b-a_c)   \,.
\end{align}

\subsection{Correlation Functions of CBOs}\label{CBOloc}

Following the prescription of  \cite{Gerchkovitz:2014gta,Gerchkovitz:2016gxx}, the above partition functions can be used to calculate the two-point function coefficients $G_{I \bar J}$ for the flat-space theory in the unbroken phase---for similar work see \cite{Pini:2017ouj}. For concreteness, let us  focus on dimension-two CBOs, in which case the corresponding two-point function coefficients are given by
\begin{equation}
\label{CBOlocaa}
G_{\alpha\bar{\beta}} =  \frac{\partial^2  \log \ZZ_{S^4_{\epsilon_1,\epsilon_2}}(\tau^{(\alpha)}, \bar \tau^{(\beta)}, m;\epsilon_1, \epsilon_2)}{\partial \tau^{(\alpha)}\partial \bar{\tau}^{(\beta)}}\;.
\end{equation}
Note that by deforming the $S^4$ partition function with irrelevant couplings $t_{\ell}$ that source higher-dimension CBOs and taking derivatives with respect to those couplings, one can also determine more general two-point functions in the chiral ring following \cite{Gerchkovitz:2016gxx}.

For $SU(k)$ $\mathcal{N}=4$ SYM there is only one marginal coupling and hence the single component of the two-point function matrix in \eqref{CBOlocaa} can be calculated to be\footnote{Compared to Sec.~\ref{examples}, the two-point functions in this section are evaluated in a normalisation of the elementary fields $\phi^{(\alpha)}$ where $\phi^{(\alpha)}_{\rm here} = \frac{\pi}{{\rm Im}\tau^{(\alpha)}} \phi^{(\alpha)}_{{\rm Sec}.\ref{examples}}$.}
\begin{align}
\label{eq:correfunctionscalN4}
G^{\mathcal{N}=4} = \frac{\partial^2}{{\partial \tau\partial \bar{\tau}}}\log \ZZ^{\mathcal{N}=4}_{S^4_{\epsilon_1,\epsilon_2}}=\frac{k^2-1}{8}\frac{1}{(\text{Im}\tau)^2}\,.
\end{align}
This result is exact and does not receive additional corrections in the coupling.

For the circular $\NN=2$ superconformal-quiver theory with $N$ $SU(k)$ nodes,  one can compute the leading-order corrections at weak coupling and in the planar limit $k\to \infty$. Following the steps performed in Sec.~4 and App.~C of \cite{Mitev:2015oty}---see also \cite{Mitev:2014yba}---a lengthy calculation shows that, to first nontrivial order, the non-zero matrix elements of the two-point function matrix are 
\begin{align}\label{Zammetric}
&G^{\text{quiver}}_{\alpha \beta}=2\pi^2\left\{\begin{array}{ll}g_\alpha^4 +12 \left(g_{\alpha-1}^2 + g_{\alpha+1}^2- 3 g_\alpha^2\right) \zeta (3) g_\alpha^6 + \mathcal{O}(g^{12})& \text{ for }\alpha=\beta\\6g_\alpha^4g_\beta^4\zeta(3) + \mathcal{O}(g^{12})& \text{ for } \alpha = \beta \pm 1
\end{array}\right. \, .
\end{align}
The $g_\alpha$  appearing above are rescaled versions of the 't Hooft coupling $g_\alpha^2 = \frac{k}{4 \pi}\frac{1}{(\text{Im}\tau^{(\alpha)})}$. Note that \eqref{Zammetric} receives corrections only from nearest-neighbour nodes. If one were to extend this calculation to the next loop order, the correction would be proportional to $\zeta(5)$ and would include contributions from next-to-nearest-neighbour nodes and so on.

The two-point function coefficients comprise a symmetric matrix that obeys 
\begin{align}
\label{eq:ZrsymmetryoftheZamMetric}
  \Omega  G^{\text{quiver}}(g_1,g_2,\ldots, g_N)=  G^{\text{quiver}}(g_2,g_3\ldots, g_1)\Omega\,,
\end{align}
$i.e.$ commutes with the shift matrix $\Omega$ up to a cyclic permutation in the couplings 
\begin{align}
\Omega_{\alpha \beta}=\delta_{\alpha+1,\beta}\,.
\end{align} 
In agreement with inheritance theorems \cite{Bershadsky:1998mb,Bershadsky:1998cb} we conjecture that this property  persists to higher orders in perturbation theory. It implies that $G$ and $\Omega$ are simultaneously diagonalisable at the orbifold point where  $g_\alpha=g_\beta\equiv g$. The shift matrix can be diagonalised into the clock matrix
\begin{align}
Q = \text{diag}(\mathfrak q^{\lfloor \frac{N}{2}\rfloor-N +1},\ldots,\mathfrak q^{-1},\mathfrak q^0, \mathfrak q^1, \ldots, \mathfrak q^{\lfloor\frac{N}{2}\rfloor})\,,
\end{align}
where $\mathfrak q = e^{\frac{2 \pi i }{N}}$, through a similarity transformation with  $\frac{1}{\sqrt N}\mathfrak q^{\alpha \beta}$. Note that this is the same matrix that implements the discrete Fourier transform on the circular-quiver fields of Sec.~\ref{quiver}.

In this way one finds that the two-point function coefficient for  untwisted dimension-two CBOs\footnote{This is the $\alpha = 0$ case in \eqref{nnodebj}.} is given at the orbifold point by
\begin{align}
  \label{eq:8}
  G_{\text{untwisted}} = \frac{2\pi^2g^4}{N^2} = \frac{k^2}{8}\frac{1}{N^2(\text{Im}\tau)^2}\;.
\end{align}
Note that this is precisely equal to the ${\mathcal{N}=4}$ SYM result, where $g_{YM}^2 = \frac{g^2}{N}$. This result can be interpreted as the $c$ anomaly, as expected by inheritance arguments  \cite{Bershadsky:1998mb,Bershadsky:1998cb}, because in $\mathcal N=4$ SYM $\Tr [\varphi^2]$ is the highest-weight state in the superconformal multiplet that contains the energy-momentum tensor.

\section{Application to the Deconstruction of the 6D (2,0) Theory}
\label{apps}

We close with an application of the above technology in the context of dimensional deconstruction: the procedure of obtaining compact extra dimensions in a certain limit of circular-quiver gauge theories \cite{ArkaniHamed:2001ca,Hill:2000mu}.

\subsection{Deconstructing the 6D (2,0) Theory}

In \cite{ArkaniHamed:2001ie} it was conjectured that the full 6D $A_{k-1}$ (2,0) theory on $\IT^2$ can be recovered via deconstruction. One starts with the 4D $\mathcal N=2$ circular-quiver superconformal gauge theory with $N$ $SU(k)$ nodes at the `orbifold point' in the space of couplings, $g^{(\alpha)} = g$. This picture suggests that the $SL(2,\mathbb Z)$-duality group of $\NN = 4$ SYM descends to a symmetry on the circular-quiver theory. This duality group is $\pi_1(\mathcal M_{1,N})$, where $\mathcal M_{1,N}$ is the moduli space of smooth Riemann surfaces of genus one with $N$ distinct and unordered  marked points \cite{Witten:1997sc}.

Deconstruction proceeds by taking this theory onto the Higgs branch, where a vev $v$ to the bifundamental hypermultiplets results in the gauge-symmetry breaking $SU(k)^N\to SU(k)$, and considering the limit
\begin{align}
  \label{eq:1}
  g\to \infty\;,\qquad v\to \infty \;,\qquad N\to \infty\;,
\end{align}
while keeping the ratios
\begin{align}
  \label{eq:2}
  \frac{g}{v}:= R_5\to \text{fixed}\;,\qquad \frac{N}{g v}:= R_6\to\text{fixed}\;.
\end{align}
From such an operation one recovers the massive spectrum of a 6D (2,0) theory on $\IT^2 = S^1_{R_5} \times S^1_{R_6}$: the KK spectrum on $S^1_{R_5}$ is obtained directly, while that on $S^1_{R_6}$ can be inferred from the former via the 4D duality transformation that takes $g\to \frac{N}{g}$. Note that the amount of supersymmetry has doubled at the end of the deconstruction process. In \cite{ArkaniHamed:2001ie} these considerations were also complemented by a string-theory construction. In the additional limit $R_6\to 0$, there exists a Lagrangian description in terms of a weakly-coupled 5D $\mathcal N = 2$ SYM on $S^1_{R_5}$, with $g_{\text{5D} }^2\propto R_6$ \cite{Lambert:2012qy}. In this description, it was proposed in \cite{Lambert:2014jna} that units of momentum along $S^1_{R_6}$ are recovered by dressing the 5D SYM operators with instanton operators; see also \cite{Tachikawa:2015mha}. 

A quantitative check of the deconstruction proposal for the (2,0) theory was performed in \cite{Hayling:2017cva}. Through the application of a set of deconstruction-inspired replacement rules, the full $S^4\times \IT^2$ partition function was obtained from the circular-quiver partition function on $S^4$, which was calculated using supersymmetric localisation.\footnote{For a more general application of this procedure to other pairs of theories related by deconstruction see also \cite{Hayling:2018fmv, Hayling:2018fgy}.}

The above four-dimensional starting point for deconstruction is precisely the setup presented in Sec.~\ref{quiver}, with the vev given by \eqref{nnodeaca}. Using the relation with deconstruction, we will next interpret the 4D type-B anomalies of Sec.~\ref{quiver} in the context of the 6D (2,0) theory on $\IT^2$. We remind the reader that, from the 4D circular-quiver point of view, the type-B anomalies in question arise from specific coefficients in the low-momentum limit ($p_1,q\to 0$) of three-point functions
\begin{align}\label{correlator}
\langle {T^\mu}_\mu(q) \widehat{\OO}^{(\alpha)}_I (p_1) \overline{\widehat{\OO}_J^{(\alpha)}}(p_2)\rangle
\end{align}
of the trace of the 4D energy-momentum tensor with a chiral twisted CBO $\widehat{\OO}^{(\alpha)}_I$ and an anti-chiral twisted CBO $\overline{\widehat{\OO}_J^{(\alpha)}}$. 

For concreteness, in what follows we will focus on the case of the scaling dimension two operators $\OO^{(\alpha)}_2$ and their complex-conjugates. Using the anomaly matching that was argued in the previous sections, the corresponding type-B anomalies in \eqref{correlator} can be evaluated exactly via the computation of related two-point function coefficients at the conformal point, using supersymmetric localisation as in Sec.~\ref{CBOloc}. The result is a function of the fixed combination
\begin{align}
  \frac{N}{g^2} = \frac{R_5}{R_6} \;,
\end{align}
and does not depend on the vev $v$. Hence, the anomaly is a robust
quantity along the Higgs branch and survives intact in the deconstruction limit \eqref{eq:1}.

\subsection{4D CBOs in Deconstruction}

The first step in interpreting \eqref{correlator} in terms of the 6D theory involves the determination of the uplift of the 4D operators appearing in the correlator. In Sec.~\ref{quiver} we defined the $\widehat \OO^{(\alpha)}_2$ through the equation
\begin{align}\label{DFT}
  \widehat\OO_2^{(\alpha)} = \frac{1}{\sqrt N} \sum_{\alpha'} \Tr [ \hat \varphi^{(\alpha')}\hat \varphi^{(\alpha-\alpha')}]\;, 
\end{align}
where the $\hat \varphi^{(\alpha)}$ are discrete-Fourier modes of the adjoint scalars in the $\alpha^{\text{th}}$ quiver node
\begin{align}
  \varphi^{(\alpha)} = \frac{1}{\sqrt N} \sum_\beta e^{ \frac{2 \pi i}{N}\alpha \beta }\hat\varphi^{(\beta)}\;.
\end{align}
The real and imaginary parts of $\hat \varphi^{(\alpha)}$ can be interpreted as the $\alpha^{\text{th}}$ KK  modes along the  $S^1_{R^5}$ direction of two of the real scalars in the 6D free-tensor multiplet compactified  on $\IT^2$ \cite{Lambert:2012qy}.
Therefore, the composite operators \eqref{DFT}, which are dimension-2 twisted/untwisted  CBOs from the 4D viewpoint, are also expected to admit a KK-mode interpretation. We will next argue that the  $\widehat \OO^{(\alpha)}_2$ are $S^1_{R_5}$ KK modes for scalar operators that are R-symmetry descendants of the superconformal-primary operator in the (2,0) stress-tensor multiplet.

We recall that the unitary, irreducible representations of the 6D superconformal algebra can be
labelled by their $SO(2)\times SO(6)$ (conformal) and $SO(5)$ (R-symmetry) quantum numbers \cite{Minwalla:1997ka}. These have been explicitly constructed in \cite{Buican:2016hpb,Cordova:2016emh}; we will be using the notation of \cite{Beem:2014kka,Buican:2016hpb}.\footnote{Our conventions are summarised in App.~\ref{app:6d}.} In that language, a special class of unitary irreducible representations that will appear momentarily are the $\frac{1}{2}$-BPS multiplets of type $\mathcal D$.

A full R-symmetry module can be generated starting with the highest-weight state of the $SO(5)$ R-symmetry and acting with a sequence of the lowering operators, $\RR_{1,2}^-$. For a given $SO(5)$ representation, the (positive) Dynkin labels for the state at each stage, $[d_1, d_2]$, denote that one can act $d_1$ times with $\RR_1^-$ and $d_2$ times with $\RR_2^-$. The $SO(5)$ labels for the resulting state can then be obtained by subtracting those of the simple roots $\omega_1 = (2,-2)$, $\omega_2 = (-1,2)$ for each action with  $\RR_1^-$ and $\RR_2^-$ respectively. E.g.\ the 6D supercharges $\QQ_{{\bf A} a}$, which transform in the spinor representation ([0,1] or $\bf 4$) of $SO(5)$, are related in the following way \cite{Beem:2014kka}
\begin{align}\label{Qsup}
  \QQ_{{\bf 1} a}\xrightarrow{\RR_2^-} \QQ_{{\bf 2} a} \xrightarrow{\RR_1^-}\QQ_{{\bf 3} a} \xrightarrow{\RR_2^-}\QQ_{{\bf 4} a}\;.
\end{align}

Let us apply this action to the $\frac{1}{2}$-BPS multiplet $\mathcal D[0,0,0;d_1,0]$ of
the flat-space (2,0) theory.\footnote{In this language, the free-tensor multiplet is denoted as $\mathcal D[0,0,0;1,0]$, while the stress-tensor multiplet as $\mathcal D[0,0,0;2,0]$.} Here the $[0,0,0;d_1,0]$ labels are [Lorentz; R-symmetry] Dynkin labels for the highest-weight (superconformal-primary) state, which we will call $|\psi\rangle$. Using the above arguments, it is easy to determine that the primary for a $\frac{1}{2}$-BPS multiplet of type $\mathcal D$ satisfies
\begin{align}\label{Rsymm}
  (\RR_1^-)^{d_1 + 1} |\psi\rangle = 0\;.
\end{align}
By definition it is also annihilated by the supercharges
\begin{align}\label{BPS}
  \QQ_{{\bf A} a} |\psi\rangle = 0 ~, ~~ {\bf A}={\bf 1}, {\bf 2}
\end{align}
Combining Eqs.~\eqref{Qsup}, \eqref{BPS} and \eqref{Rsymm} one can obtain the following additional relations
\begin{align}\label{one}
  (\RR_1^-)^{d_1} \QQ_{{\bf 1} a}|\psi\rangle = 0                                                                          \implies \QQ_{{\bf 1} a}|\psi'\rangle  = 0                                                                             
\end{align}
as well as 
\begin{align}\label{two}
  (\RR_1^-)^{d_1+1} \QQ_{{\bf 2} a}|\psi\rangle  =0
                                                                                             \implies \QQ_{{\bf 3} a}|\psi'\rangle  = 0
\end{align}
with $|\psi'\rangle \equiv (\RR_1^-)^{d_1}|\psi\rangle$. 

Note that the 6D SCA admits a 4D $\mathcal N=2$ superconformal subalgebra; we describe this embedding in App.~\ref{app:6d}. Using the dictionary of Table~\ref{Tab:supercharges}, the shortening conditions \eqref{one} and \eqref{two} become
\begin{align}\label{4DBPS}
  Q_{\alpha}^1|\psi'\rangle  = 0 \;, \qquad Q^2_{\alpha}|\psi'\rangle  =0
\end{align}
with $SU(2)_R$ and $U(1)_r$ charges $R =0$, $ r= d_1$ respectively.

We have therefore determined that the $d_1$-th descendant of the superconformal primary in the 6D $\mathcal D[0,0,0;d_1,0]$ $\frac{1}{2}$-BPS multiplet has an interpretation as a scalar with charges $R =0$, $r = d_1$, which is also annihilated by the supercharges $Q_{\alpha}^1, Q_{\alpha}^2$  of a 4D $\mathcal N=2$ subalgebra of the 6D (2,0) SCA. These are precisely the properties characterising a CBO in a 4D superconformal theory.

Although deconstruction does not reproduce the (2,0) theory in flat space Eqs.~\eqref{4DBPS} still hold; compactification of the (2,0) theory on $\IT^2$ breaks the superconformal supersymmetry, but preserves Poincar\'e  as well as R symmetry. Following  \cite{Hayling:2017cva}, we will identify the Poincar\'e supersymmetry of the 4D $\mathcal N=2$ subalgebra with the Poincar\'e supersymmetry of the 4D quiver theory on the Higgs branch. In this sense, the CBOs of the 4D quiver with dimension $\Delta_{4\text{D}} = |r|=|d_1|$, deconstruct the $d_1$-th R-symmetry descendants of the $\mathcal D[0,0,0;d_1,0]$ primary, once the (2,0) theory is placed on $\IT^2$.

So far we have focused only on Lorentz and R-symmetry quantum numbers, which are shared by both the untwisted and twisted-sector 4D CBOs. However, we also need to account for the two integers labelling the KK spectrum of the (2,0) theory on $\IT^2$:
\begin{itemize}
\item {\it Momentum on $S^1_{R^5}$}:
The 4D CBOs carry different charges under the discrete $\mathbb Z_N$ symmetry of the $N$-noded quiver. In the deconstruction limit $N\to \infty$ and $\mathbb Z_N \to U(1)$. It is natural to identify the untwisted sector CBOs, which are uncharged under this symmetry, with the s-wave KK mode associated with $S^1_{R_5}$, and the twisted-sector CBOs with the higher modes.

  \item {\it Momentum on $S^1_{R^6}$}: 
    To obtain the KK spectrum along $S^1_{R_6}$, the deconstruction proposal of \cite{ArkaniHamed:2001ie} relied on the duality symmetry that takes $g\to g^\prime =\frac{N}{g}$. In order to recover CBOs that carry momentum along this direction one would therefore have to use the dual weakly-coupled description in the 4D quiver. The action of the S-duality transformation on chiral-primary operators (and their two- and three-point functions) in $\mathcal N=4$ SYM has been discussed in \cite{Gomis:2009xg}. An analysis along similar lines for circular-quiver gauge theories obtained by orbifolding $\NN =4$ SYM would presumably lead to the operators of interest. It would be interesting to further investigate this direction in the future.

\end{itemize}

We conclude that the operators \eqref{DFT} correspond to the 2nd R-symmetry descendant of the superconformal primary in the $\mathcal D[0,0,0;2,0]$ (stress-tensor) $\frac{1}{2}$-BPS multiplet, carrying $\alpha$ units of momentum along $S^1_{R^5}$ and no momentum along $S^1_{R_6}$, after the 6D (2,0) theory has been put on $\IT^2 = S^1_{R_5} \times S^1_{R_6}$. The extension of this relationship to descendants of the 6D $\mathcal D[0,0,0;d_1,0]$ primary and 4D CBOs with higher scaling dimensions is straightforward.

\subsection{An Anomaly for the (2,0) Theory on $\IT^2$}

We are now in position to assign a 6D interpretation to the full
correlator \eqref{correlator}. Consider the connected generating
functional of correlation functions for local operators in a 6D CFT on
$\mathcal M_6$, $W(\lambda^i, g)$, where $g_{mn}$ $(m,n=0,1,\ldots,5)$
is the metric on $\mathcal M_6$ and the couplings $\lambda^i$ are space-dependent sources for integer-dimension operators $\widehat\OO_i$.  As in Sec.~\ref{typeB}, a local 6D Weyl transformation yields \cite{Osborn:1991gm,Schwimmer:2019efk}
\begin{align}
  \label{eq:4}
  \delta_\sigma W = \int d^6x\; \sqrt g \; \sigma(x^m) \mathcal A(\{\lambda^i, g\})\;,
\end{align}
where $\AA$ is a local anomaly functional. $\AA$ includes type-B anomalies that can be traced, for example, in the correlation functions of the form $\langle {T^m}_
  m \widehat \OO \widehat \OO \rangle$.

We would like to apply this framework to the (2,0) theory on $\mathbb
R^4 \times S^1_{R_5}\times S^1_{R_6}$. In doing so, it is convenient
to KK expand the source terms in $W$ in terms of equivalent 4D
couplings of the form $\lambda^i_{(-\alpha,-\beta)}(x^\mu)
\widehat\OO^{(\alpha,\beta)}_i(x^\mu)$, where $\mu = 0,...,3$. Let us further set all, except the two coefficients $\lambda^i_{(\alpha,0)}$, $\lambda^i_{(-\alpha,0)}$ with some fixed $\alpha$, to zero and restrict the local Weyl transformation along the four flat, non-compact directions, with parameter $\sigma(x^\mu)$. This restricted Weyl transformation in the 6D theory will yield, according to \eqref{eq:4}, a type-B anomaly
\begin{align}
  \label{eq:5}
  \delta_\sigma W = \int d^4x \; \sigma(x^\mu) \mathcal A(\{\lambda^i\})\;
  .
\end{align}
This anomaly can be determined by studying the correlator $\langle
{(T_{{\rm 6D}})^\mu}_{\mu} \widehat\OO^{(\alpha,0)}_i
\widehat\OO^{(-\alpha,0)}_i\rangle$, where ${{(T_{{\rm 6D}})}^\mu}_\mu
$ is a partial trace of the 6D energy-momentum tensor $T^{mn}_{\rm
  6D}$ along the four non-compact directions of the background
spacetime. By identifying this partial trace with the trace of the energy-momentum tensor $T_{\rm 4D}^{\mu\nu}$ of the 4D quiver gauge theory, and the 6D $\widehat\OO^{(\alpha,0)}$, $\widehat\OO^{(-\alpha,0)}$ operators with the 4D $\widehat\OO^{(\alpha)}$, $\overline{\widehat\OO^{(\alpha)}}$ CBOs, we are able to identify the above 6D type-B anomalies with the 4D type-B anomalies of the circular-quiver gauge theory in dimensional deconstruction.

To summarise, the coefficient of a certain term in the correlator \eqref{correlator} in the 4D circular-quiver gauge theory captures a type-B anomaly arising from the response of the connected generating functional of correlation functions of the $A_{k-1}$ (2,0) theory on $\IT^2$ to Weyl transformations along the four non-compact directions. This anomaly concerns the $\alpha^{\text{th}}$ KK mode along the $S^1_{R_5}$ of $\frac{1}{2}$-BPS scalar operators which are R-symmetry descendants in the short multiplets $\mathcal D[0,0,0;d_1,0]$. In the special case $d_1=2$ and the operators in \eqref{DFT}, these scalar operators are the 2nd R-symmetry descendants of the primary of the stress-tensor multiplet $\mathcal D[0,0,0;2,0]$. 

As a final consistency check of this picture, we note that we can successfully track the type-B anomaly all the way from the UV to the IR fixed point of the RG flow generated by the deconstruction limit. As one approaches the IR, the KK modes of the (2,0) theory on $\IT^2$ become increasingly more massive, with only the zero modes surviving at very low energies. The extreme IR theory is 4D $\NN = 4$ SYM theory. Consequently, in this limit the only operators that survive are the ones in the untwisted sector of the UV superconformal circular-quiver. The type-B anomalies associated with these operators are expected to match those of the $\NN = 4$ SYM theory. This can be explicitly verified for the case of $\Delta =2$ untwisted operators. Indeed, we can easily deduce from the formulae of the previous sections (see Eq.~\eqref{eq:8}) that the type-B anomaly of untwisted CBOs in the quiver reproduces the $c$ anomaly of $\NN = 4$ SYM theory.


\ack{ \bigskip We would like to thank E.~Andriolo, K.~Papadodimas, A.~Schwimmer and
  S.~Theisen for very helpful discussions and especially Z.~Komargodski for correspondence and comments on the manuscript. C.P.\ is supported by the Royal Society through a University Research Fellowship, grant number UF120032. E.P.’s work is supported by the German Research Foundation (DFG) via the Emmy Noether program “Exact results in Gauge theories”.}

\begin{appendix}

\section{4D $\mathcal{N}=2$ Superconformal Algebra Conventions}\label{app:4d} 

In this appendix we collect useful facts about the conventions used in Sec.~\ref{Ward} and \ref{match} concerning the 4D $\mathcal{N}=2$ SCA. We mostly follow the presentation of \cite{Dolan:2002zh,Gadde:2009dj,Beem:2013sza}.

The Lie superalgebra governing the dynamics  of 4D $\mathcal{N}=2$ SCFTs is $\mathfrak{su}(2,2|2)$. Superconformal primaries are labelled as $\ket{\Delta; j,\bar{j} ; R, r}$, where $\Delta$ is their conformal dimension, $j$ and $\bar{j}$ their Lorentz quantum numbers for $\mathfrak{su}(2)_{\alpha}\oplus \mathfrak{su}(2)_{\dot{\alpha}} \in \mathfrak{su}(2,2)$, $R$ is the Cartan of the  $\mathfrak{su}(2)_R$ R-symmetry while $r$ is the charge under the $\mathfrak{u}(1)_r$ R-symmetry. There are eight Poincar\'e and eight superconformal supercharges, denoted by $Q^{\II}_{\alpha}$,  $\bar{Q}_{\II}^{\dot{\alpha}}$ and $S_{\II \alpha}$,  $\bar{S}^{\II \dot\alpha}$, where $\II=1,2$ is an $SU(2)_R$ R-symmetry index and $\alpha,{\dot\alpha} =\pm$ a spinor index of $\mathfrak{su}(2)_{\alpha}\oplus \mathfrak{su}(2)_{\dot{\alpha}}$.  The superconformal primaries are annihilated by all $S_{\II \alpha}$, $\bar{S}^{\II \dot\alpha}$ and the special conformal generators ${K}_\mu$. We make use of the following non-vanishing commutation relations of the 4D SCA Euclidean generators in the main text \cite{Dolan:2002zh}:
\begin{align}\label{anticomms}
 \left\{ \bar{S}^{\II \dot\alpha} \, ,  \bar{Q}_{\JJ \dot\beta} \right\}  &=  \delta^{\II}_{\JJ} \left( \frac{1}{2}  \delta^{\dot\alpha}\,_{\dot\beta} D + \bar{{M}}^{\dot\alpha}\,_{\dot\beta}\right) - \delta^{\dot\alpha}\,_{\dot\beta}{R}^{\II}\,_{\JJ} \, \cr
\left[   \bar{{M}}^{\dot\alpha}\,_{\dot\beta} \, ,  \bar{Q}_{\II \dot\gamma}  \right]
 &= -\delta^{\dot\alpha}_{\dot\gamma}  \bar{Q}_{\II \dot\beta}  + {1\over 2} \delta^{\dot\alpha}_{\dot\beta} \bar{Q}_{\II \dot\gamma} \;,\cr
\left[ D \, ,  \bar{Q}_{\II \dot\alpha}  \right] &=  {1\over 2}   \bar{Q}_{\II \dot\alpha}\;,\\
\left[  {R}^{\II}\,_{\JJ} \, ,  \bar{Q}_{\KK \dot\alpha}  \right] 
 &= \delta^{\II}_{\KK}  \bar{Q}_{\JJ \dot\alpha}  -    {1\over 4}   \delta^{\II}_{\JJ}  \bar{Q}_{\KK \dot\alpha}\;, \cr
 \left [ P_{\alpha \dot\alpha}, \bar{Q}_{\JJ \dot\beta} \right ] &= 0 ~,~~
\left [ P_{\alpha \dot\alpha},  \bar{S}^{\II \dot\beta} \right] = - \delta^{\dot \beta}_{\dot \alpha} Q_\alpha^\II
\, . \nonumber
 \end{align}
Here the $\bar{ M}$, $ R$ and $D$ are rotation, R-symmetry and Hamiltonian generators respectively.

In this work we explicitly use two maximally-short, irreducible superconformal representations.
The first one is the chiral $\EE_{r}$, whose highest-weight superconformal primary obeys the shortening condition $\Delta = r $ coming from the relation ${Q}_{\II \, \dot{\alpha}} \ket{\Delta; j,\bar{j} ; R, r}=0$ for all $\II$,$\dot{\alpha}$. In Lagrangian theories, like the 4D $\NN=2$ SCQCD theory, the highest-weight superconformal primary can be expressed as a Casimir $\Tr {\bar \varphi}^\ell$ with $r= \ell$. One can also consider multiple traces of Casimirs. $\bar \varphi$ is the adjoint complex scalar field in the vector multiplet. The highest-weight operators of  $\EE_{r}$ parametrise the Coulomb branch.
\begin{table}\centering
  \begin{tabular}{l|cccccc}
$\Delta$\\
$\ell$ & $0_{\left(0,0\right)}$\\
$\ell+\frac{1}{2}$ &  & $\frac{1}{2}_{\left(0,\frac{1}{2}\right)}$\\
$\ell+1$&  &  & $0_{\left(0,1\right)}, {1_{\left(0,0\right)}}$\\
$\ell+\frac{3}{2}$ &  &  &  & $\frac{1}{2}_{\left(0,\frac{1}{2}\right)}$\\
$\ell+2$ &  &  &  &  &  & $0_{\left(0,0\right)}$\\
\hline
    $r$ & $\ell$&$ \ell-\frac{1}{2}$ & $\ell-1$ &$ \ell-\frac{3}{2}$ &  & $\ell-2$
  \end{tabular}
  \caption{The quantum numbers for states in the $\EE_{r}$ multiplet.} \label{tab:emult}
\end{table}
The multiplets $\EE_{2}$ contain the (Lagrangian) marginal deformations  of  $\mathcal{N}=2$ theories as the lowest-weight states with $\Delta =4$ and $r=0$. Schematically, marginal deformations can be denoted $\delta\mathcal{L}_k=   \lambda^k  \; Q^4 \cdot \Tr \varphi_k^2$ in Lagrangian theories. The various states in this multiplet are summarised in Tab.~\ref{tab:emult}.  We use the notation $R_{( j,\bar{j})}$ to label the $\mathfrak{su}(2)_R$ R-symmetry and the Lorentz quantum numbers of the different elements in the multiplet, while their conformal dimension $\Delta$ and their $\mathfrak{u}(1)_r$ R-symmetry are given on the vertical and the horizontal axes of Tab.~\ref{tab:emult}, respectively.

The second short superconformal representation is the $\hat{\CC}_{(0,0)}$ multiplet with shortening condition $\Delta = 2$ summarised in Tab.~\ref{tab:cmult}. It contains the stress-energy tensor $T_{\mu \nu}$ (the state $0_{\left( 1,1 \right)}$ with $\Delta=4$), the supercurrents $G^{\II \mu}_{\alpha}$  and  $\bar{G}_{\II \mu}^{\dot\alpha}$  (the states $\frac{1}{2}_{\left( 1, \frac{1}{2} \right)}$ and $\frac{1}{2}_{\left(  \frac{1}{2},1 \right)}$  with $\Delta=\frac{7}{2}$) and the $SU(2)_{R}$ and $U(1)_{r}$ R-symmetry currents $J^{SU(2)_R}_{\mu}$, $J^{U(1)_r}_{\mu}$ ($1_{(\frac{1}{2},\frac{1}{2})}$ and $0_{\left(  \frac{1}{2}, \frac{1}{2} \right)}$ respectively) of the $\NN=2$ theory. 
\begin{table}[t]\centering
    \begin{tabular}{l|ccccc}
$ \Delta $      &                                                                              \\
 &&&&\\
 $2$          &   &  &$0_{\left( 0,0 \right)}$                                         &\\
 &&&&\\
$ \frac{5}{2}$ &    &       $ \frac{1}{2}_{\left(  \frac{1}{2}, 0 \right)}   $& & $ \frac{1}{2}_{\left( 0, \frac{1}{2} \right)}$      \\
 &&&&\\
 3   &      $ { 0_{\left( 1,0 \right)}}$ &  &     ${ {1_{(\frac{1}{2},\frac{1}{2})}}}  ,\,  0_{\left(  \frac{1}{2}, \frac{1}{2} \right)} $  &  &  $0_{\left( 0,1 \right)} $ \\
 &&&&\\
$\frac{7}{2}$  &                &      $  {\frac{1}{2}_{\left( 1, \frac{1}{2} \right)} }$  & &  $\frac{1}{2}_{\left(  \frac{1}{2},1 \right)} $     \\
&&&&\\
 4  &            &      &   $ 0_{\left( 1,1 \right)} $\\
 &&&$-0_{(0,0)},\,-1_{(0,0)}$&\\
 &&&&\\
$ \frac{9}{2}$& & $-\frac{1}{2}_{(\frac{1}{2},0)}$&&$-\frac{1}{2}_{(0,\frac{1}{2})}$\\
 &&&&\\
 5 &&&$-0_{(\frac{1}{2},\frac{1}{2})}$&\\
 &&&&\\
 \hline
$  r$               &   1  & $\frac{1}{2} $      &   0    & $-\frac{1}{2}$  &-1
    \end{tabular}
    \caption{The quantum numbers for the states in the $\hat{\CC}_{(0,0)}$ multiplet.}\label{tab:cmult}
\end{table}
The   $\hat{\CC}_{(0,0)}$ multiplet has as its superconformal primary a $\Delta=2$ scalar operator $\mathcal{T}$. In Lagrangian theories $\TT = \bar\varphi  \varphi -   \MM_{{\bf {1}}}$, where $\MM_{{\bf {1}}}$ is a mesonic operator.\footnote{To obtain this precise form of the eigenvector of the dilatation operator, a one-loop calculation is needed \cite{Gadde:2010zi}.} 
The states with minuses correspond to null vectors or equations of motion which must be removed from the multiplet. In the $\hat{\CC}_{(0,0)}$ case 
the null vectors $-0_{(0,0)}$ and $-1_{(0,0)}$ at $\Delta=4$ correspond to the conservation equations $\partial^{\mu} J^{U(1)_r}_{\mu} = 0$ and $\partial^{\mu} J^{SU(2)_R}_{\mu} = 0$, at  $\Delta = \frac{9}{2}$ to the conservation of the supercurrent, while at $\Delta = 5$ to $\partial^{\mu} T_{\mu\nu} = 0$.

When conformal symmetry is broken, the null vectors form an $\mathcal{N}=2$ supersymmetry  multiplet, coupled to the $\mathcal{N}=2$ dilaton multiplet and are no-longer conserved.
The dilaton $\sigma$ and the dilatino $\chi^{\II}_{\alpha}$ are related by supersymmetry through $[Q^{\II}_{\alpha}, \sigma]= \chi^{\II}_{\alpha}$ and  $[\bar{Q}_{\II}^{\dot\alpha}, \sigma] = \bar{\chi}_{\II}^{\dot\alpha}$ and couple to the {\it null vectors} of the $\hat{\CC}_{(0,0)}$ multiplet as
\begin{align}\label{dilatoncouplings}
S_{eff}(\sigma , \chi) = \int d^4x \Big( - \frac{1}{2} \partial_\mu \sigma  \partial^\mu \sigma  & + \frac{1}{v} \sigma(x)  T^\mu\,_\mu(x)   
  + i \bar{\chi}_{\II \dot\alpha} \partial^{{\dot\alpha} \alpha}  \chi_{\alpha}^{\II} \cr
  & + \frac{1}{v} \bar{\chi}_{\II \dot \alpha} \bar\sigma_{\mu}^{\dot\alpha\alpha}  G^{\II \mu}_{ \alpha}   + \frac{1}{v}  \chi_{\alpha}^{\II}  \sigma^{\mu\alpha\dot\alpha} \bar{G}_{\II \mu \dot \alpha}
+ O(v^{-3})\Big)
\end{align}
leading to the following equations of motion:
\begin{equation}
 T^\mu\,_\mu(x)       =  - v \,  \Box \sigma(x)
 \quad \mbox{and} \quad  
 G^{\II \mu}_{ \alpha}   =  i v \partial^\mu \chi^{\II}_{ \alpha} (x)\,.
\end{equation}

\section{6D (2,0) Superconformal Algebra Conventions}\label{app:6d} 

In this appendix we collect some of the conventions that we use in Sec.~\ref{apps} pertaining to the SCA for the 6D (2,0) theory. A more complete account can be found in App.~A of \cite{Beem:2014kka}.

In Lie superalgebra notation the 6D (2,0) SCA is denoted as $\mathfrak{osp}(8^*|4)$. The associated superconformal primaries are designated $\ket{\Delta;c_1,c_2,c_3;d_1,d_2}$. They are labelled by their conformal dimension $\Delta$, their Lorentz quantum numbers for $\mathfrak{su(4)}$ in the Dynkin basis $c_i$ and their R-symmetry quantum numbers in the Dynkin basis $d_i$. Each of these primaries is in one-to-one correspondence with a highest weight labelling irreducible representations of the maximal compact subalgebra $\mathfrak{so}(6)\oplus\mathfrak{so}(2)\oplus\mathfrak{so}(5)_R\subset\mathfrak{osp}(8^*|4)$.  There are sixteen Poincar\'e and sixteen superconformal supercharges, denoted by $\QQ_{{\bf A} a}$ and $\QS_{{\bf A} \dot a}$, where $\dot a ,a=1,\ldots,4$ are (anti)fundamental indices of $\mathfrak{su}(4)$ and ${\bf A} =1,\ldots,4$ a spinor index of $\mathfrak{so}(5)_R$. There are also six momenta $\PP_m$ and special conformal generators $\mathcal{K}_m$, where $m$ is a vector index of the Lorentz group, $m =0,\ldots,5$. The superconformal primary is annihilated by all $\QS_{{\bf A} \dot a}$ and $\mathcal{K}_m$. The unitary irreducible representations for this SCA were classified in \cite{Minwalla:1997ka} and explicitly constructed in \cite{Buican:2016hpb,Cordova:2016emh}; see also \cite{Beem:2014kka}. 

\begin{table}[t]
\begin{center}
\begin{tabular}{|c|c|c|c|c||c|}
\hline
Charge $\QQ_{{\bf A}a}$ & $h_1,h_2,h_3$ & $(j,\bar j)$ & $R$ & $\hat{r}$ &~$\mathfrak{su}(2,2|2)$~\tabularnewline
\hline
\hline
$\QQ_{{\bf1}1}$ &  $+++$ &   $(+, 0)$ & $+$  &  $+$ & $Q^1_{+}$ 		\tabularnewline
$\QQ_{{\bf2}1}$ &  $+++$ &   $(+, 0)$ & $+$  &  $-$ & 			 			\tabularnewline
$\QQ_{{\bf3}1}$ &  $+++$ &   $(+, 0)$ & $-$  &  $+$ & $Q^2_{+}$ 		\tabularnewline
$\QQ_{{\bf4}1}$ &  $+++$ &   $(+, 0)$ & $-$  &  $-$ & 			 			\tabularnewline
\hline
\hline
$\QQ_{{\bf1}2}$ &  $+--$ &   $(-, 0)$ & $+$  &  $+$ & $Q^1_{-}$ 			\tabularnewline
$\QQ_{{\bf2}2}$ &  $+--$ &   $(-, 0)$ & $+$  &  $-$ & 					\tabularnewline
$\QQ_{{\bf3}2}$ &  $+--$ &   $(-, 0)$ & $-$  &  $+$ & $Q^2_{-}$ 		\tabularnewline
$\QQ_{{\bf4}2}$ &  $+--$ &   $(-, 0)$ & $-$  &  $-$ & 					\tabularnewline
\hline
\hline
$\QQ_{{\bf1}3}$ &  $-+-$ &   $(0, +)$ & $+$  &  $+$ &  					\tabularnewline
$\QQ_{{\bf2}3}$ &  $-+-$ &   $(0, +)$ & $+$  &  $-$ &  $\bar Q_{2\dot+}$
                                                      \tabularnewline
$\QQ_{{\bf3}3}$ &  $-+-$ &   $(0, +)$ & $-$  &  $+$ &  					\tabularnewline
$\QQ_{{\bf4}3}$ &  $-+-$ &   $(0, +)$ & $-$  &  $-$ &  $\bar Q_{1\dot+}$\tabularnewline
\hline
\hline
$\QQ_{{\bf1}4}$ &  $--+$ &   $(0, -)$ & $+$  &  $+$ &  					\tabularnewline
$\QQ_{{\bf2}4}$ &  $--+$ &   $(0, -)$ & $+$  &  $-$ &  $\bar Q_{2\dot-}$	\tabularnewline
$\QQ_{{\bf3}4}$ &  $--+$ &   $(0, -)$ & $-$  &  $+$ &  					\tabularnewline
$\QQ_{{\bf4}4}$ &  $--+$ &   $(0, -)$ & $-$  &  $-$ &  $\bar Q_{1\dot-}$	\tabularnewline
\hline
\end{tabular}
\end{center}
\caption{\label{Tab:supercharges}Supercharge summary for the 6D (2,0) SCA and its 4D $\NN =2 $ subalgebra. All orthogonal-basis quantum numbers have magnitude $\half$. The four-dimensional
  subalgebra acts on the $h_2$ and $h_3$ planes.}
\end{table}

In order to connect the 6D (2,0) theory to the circular quiver via deconstruction in Sec.~\ref{apps}, we need to identify a 4D $\NN =2 $ subalgebra of $\mathfrak{su}(2,2|2)\subset \mathfrak{osp}(8^*|4)$ \cite{Beem:2014kka}. First, let us fix our conventions for the generators of various maximal and Cartan subalgebras of the bosonic symmetries in 6D.

There is a maximal subalgebra $\mathfrak{su}(2)_R \oplus
\mathfrak{u}(1)_{\hat{r}} \subset \mathfrak{usp}(4)\simeq
\mathfrak{so}(5)$ for the R symmetry. This is the subalgebra under which the $\bf 5$ of
$\mathfrak{usp}(4)\simeq \mathfrak{so}(5)$
decomposes as ${\bf 5}\rightarrow {\bf 3}_0\oplus {\bf 1}_{+1}\oplus
{\bf 1}_{-1}$. The generators $R$ and $\hat{r}$ define the orthogonal basis of weights for $\mathfrak{so}(5)$, and are related to the $\mathfrak{so}(5)$ Dynkin weights $d_1$ and $d_2$ according to
\begin{align}
d_1=R-\hat{r}~,\qquad d_2=2\hat{r}~.
\end{align}
Note, that this is in conventions where the $\bf 5$ of
$\mathfrak{so}(5)$, which has Dynkin label $[1,0]$, corresponds to
$R = 1$, $\hat{r}=0$.

The orthogonal basis for the Cartan subalgebra of $\mathfrak{so}(6)$
is given by the generators of rotations in the three orthogonal planes
in $\mathbb R^6$, $\mathcal L_i$. We denote the eigenvalues of these generators by
$h_i$, $i= 1,...,3$. These orthogonal-basis quantum
numbers (the orthogonal-basis of $\mathfrak{so}(6)$ weights) are
related to the Dynkin basis $[c_1,c_2,c_3]$ of $\mathfrak{su}(4)$
according to:
\begin{align}
  h_1=\frac12 c_1+ c_2+\tfrac12 c_3~,\quad h_2=\tfrac12 c_1+\tfrac12 c_3~,\quad h_3=\tfrac12 c_1-\tfrac12 c_3~\;.
\end{align}

The four-dimensional $\CN=2$ superconformal algebra $\mathfrak{su}(2,2|2)$ can then be embedded such that the four-dimensional rotation symmetry is $\mathfrak{su}(2)_\alpha\oplus \mathfrak{su}(2)_{\dot \alpha}$ and the four-dimensional R-symmetry  is $\mathfrak{su}(2)_R\oplus {\rm diag}[\mathfrak{u}(1)_{\hat{r}},\mathfrak{u}(1)_{\LL_1}]$. The precise map between the supercharges for this embedding is shown in Table \ref{Tab:supercharges}.

\end{appendix}



\bibliography{anomaly_matching}

\begin{thebibliography}{10}
\ifx\href\asklfhas\newcommand{\href}[2]{#2}\fi
\ifx\arxivref\asklfhas\newcommand{\arxivref}[2]{\href{http://arxiv.org/abs/#1}{#2}}\fi
\ifx\doiref\asklfhas\newcommand{\doiref}[2]{\href{http://dx.doi.org/#1}{#2}}\fi
\parskip 0pt
\normalsize

\bibitem{Deser:1993yx}
S.~Deser \& A.~Schwimmer,
\textit{``{Geometric classification of conformal anomalies in arbitrary
  dimensions}''},
\doiref{10.1016/0370-2693(93)90934-A}{Phys.~Lett. \textbf{B309}, 279
  (1993)\ignorespaces}\ignorespaces,
\normalsize{\texttt{\arxivref{hep-th/9302047}{hep-th/9302047}}}\ignorespaces
\bibitem{Schwimmer:2010za}
A.~Schwimmer \& S.~Theisen,
\textit{``{Spontaneous Breaking of Conformal Invariance and Trace Anomaly
  Matching}''},
\doiref{10.1016/j.nuclphysb.2011.02.003}{Nucl.~Phys. \textbf{B847}, 590
  (2011)\ignorespaces}\ignorespaces,
\normalsize{\texttt{\arxivref{1011.0696}{arXiv:1011.0696}}}\ignorespaces
\bibitem{Nakayama:2017oye}
Y.~Nakayama,
\textit{``{Can we change $c$ in four-dimensional CFTs by exactly marginal
  deformations?}''},
\doiref{10.1007/JHEP07(2017)004}{JHEP \textbf{1707}, 004
  (2017)\ignorespaces}\ignorespaces,
\normalsize{\texttt{\arxivref{1702.02324}{arXiv:1702.02324}}}\ignorespaces
\bibitem{Niarchos:2018mvl}
V.~Niarchos,
\textit{``{Geometry of Higgs-branch superconformal primary bundles}''},
\doiref{10.1103/PhysRevD.98.065012}{Phys.~Rev. \textbf{D98}, 065012
  (2018)\ignorespaces}\ignorespaces,
\normalsize{\texttt{\arxivref{1807.04296}{arXiv:1807.04296}}}\ignorespaces
\bibitem{Andriolo:2022lcb}
E.~Andriolo, V.~Niarchos, C.~Papageorgakis \& E.~Pomoni,
\textit{``{Covariantly Constant Anomalies on Conformal Manifolds}''},
\normalsize{\texttt{\arxivref{2210.10891}{arXiv:2210.10891}}}\ignorespaces
\bibitem{Gerchkovitz:2014gta}
E.~Gerchkovitz, J.~Gomis \& Z.~Komargodski,
\textit{``{Sphere Partition Functions and the Zamolodchikov Metric}''},
\doiref{10.1007/JHEP11(2014)001}{JHEP \textbf{1411}, 001
  (2014)\ignorespaces}\ignorespaces,
\normalsize{\texttt{\arxivref{1405.7271}{arXiv:1405.7271}}}\ignorespaces
\bibitem{Gomis:2015yaa}
J.~Gomis, P.-S. Hsin, Z.~Komargodski, A.~Schwimmer, N.~Seiberg \& S.~Theisen,
\textit{``{Anomalies, Conformal Manifolds, and Spheres}''},
\doiref{10.1007/JHEP03(2016)022}{JHEP \textbf{1603}, 022
  (2016)\ignorespaces}\ignorespaces,
\normalsize{\texttt{\arxivref{1509.08511}{arXiv:1509.08511}}}\ignorespaces
\bibitem{Baggio:2014sna}
M.~Baggio, V.~Niarchos \& K.~Papadodimas,
\textit{``{Exact correlation functions in $SU(2)$ $\mathcal N=2$ superconformal
  QCD}''},
\doiref{10.1103/PhysRevLett.113.251601}{Phys.~Rev.~Lett. \textbf{113}, 251601
  (2014)\ignorespaces}\ignorespaces,
\normalsize{\texttt{\arxivref{1409.4217}{arXiv:1409.4217}}}\ignorespaces
\bibitem{Baggio:2014ioa}
M.~Baggio, V.~Niarchos \& K.~Papadodimas,
\textit{``{tt$^{*}$ equations, localization and exact chiral rings in 4d $
  \mathcal{N} $ =2 SCFTs}''},
\doiref{10.1007/JHEP02(2015)122}{JHEP \textbf{1502}, 122
  (2015)\ignorespaces}\ignorespaces,
\normalsize{\texttt{\arxivref{1409.4212}{arXiv:1409.4212}}}\ignorespaces
\bibitem{Gerchkovitz:2016gxx}
E.~Gerchkovitz, J.~Gomis, N.~Ishtiaque, A.~Karasik, Z.~Komargodski \& S.~S.
  Pufu,
\textit{``{Correlation Functions of Coulomb Branch Operators}''},
\normalsize{\texttt{\arxivref{1602.05971}{arXiv:1602.05971}}}\ignorespaces
\bibitem{Pestun:2007rz}
V.~Pestun,
\textit{``{Localization of gauge theory on a four-sphere and supersymmetric
  Wilson loops}''},
\doiref{10.1007/s00220-012-1485-0}{Commun.~Math.~Phys. \textbf{313}, 71
  (2012)\ignorespaces}\ignorespaces,
\normalsize{\texttt{\arxivref{0712.2824}{arXiv:0712.2824}}}\ignorespaces
\bibitem{ArkaniHamed:2001ie}
N.~Arkani-Hamed, A.~G. Cohen, D.~B. Kaplan, A.~Karch \& L.~Motl,
\textit{``{Deconstructing (2,0) and little string theories}''},
\doiref{10.1088/1126-6708/2003/01/083}{JHEP \textbf{0301}, 083
  (2003)\ignorespaces}\ignorespaces,
\normalsize{\texttt{\arxivref{hep-th/0110146}{hep-th/0110146}}}\ignorespaces
\bibitem{Kallen:2012cs}
J.~Källén \& M.~Zabzine,
\textit{``{Twisted supersymmetric 5D Yang-Mills theory and contact
  geometry}''},
\doiref{10.1007/JHEP05(2012)125}{JHEP \textbf{1205}, 125
  (2012)\ignorespaces}\ignorespaces,
\normalsize{\texttt{\arxivref{1202.1956}{arXiv:1202.1956}}}\ignorespaces
\bibitem{Kallen:2012va}
J.~Källén, J.~Qiu \& M.~Zabzine,
\textit{``{The perturbative partition function of supersymmetric 5D Yang-Mills
  theory with matter on the five-sphere}''},
\doiref{10.1007/JHEP08(2012)157}{JHEP \textbf{1208}, 157
  (2012)\ignorespaces}\ignorespaces,
\normalsize{\texttt{\arxivref{1206.6008}{arXiv:1206.6008}}}\ignorespaces
\bibitem{Kallen:2012zn}
J.~Källén, J.~A. Minahan, A.~Nedelin \& M.~Zabzine,
\textit{``{$N^3$-behavior from 5D Yang-Mills theory}''},
\doiref{10.1007/JHEP10(2012)184}{JHEP \textbf{1210}, 184
  (2012)\ignorespaces}\ignorespaces,
\normalsize{\texttt{\arxivref{1207.3763}{arXiv:1207.3763}}}\ignorespaces
\bibitem{Kim:2012ava}
H.-C. Kim \& S.~Kim,
\textit{``{M5-branes from gauge theories on the 5-sphere}''},
\doiref{10.1007/JHEP05(2013)144}{JHEP \textbf{1305}, 144
  (2013)\ignorespaces}\ignorespaces,
\normalsize{\texttt{\arxivref{1206.6339}{arXiv:1206.6339}}}\ignorespaces
\bibitem{Kim:2012qf}
H.-C. Kim, J.~Kim \& S.~Kim,
\textit{``{Instantons on the 5-sphere and M5-branes}''},
\normalsize{\texttt{\arxivref{1211.0144}{arXiv:1211.0144}}}\ignorespaces
\bibitem{Kim:2013nva}
H.-C. Kim, S.~Kim, S.-S. Kim \& K.~Lee,
\textit{``{The general M5-brane superconformal index}''},
\normalsize{\texttt{\arxivref{1307.7660}{arXiv:1307.7660}}}\ignorespaces
\bibitem{Beem:2014kka}
C.~Beem, L.~Rastelli \& B.~C. van~Rees,
\textit{``{$ \mathcal{W} $ symmetry in six dimensions}''},
\doiref{10.1007/JHEP05(2015)017}{JHEP \textbf{1505}, 017
  (2015)\ignorespaces}\ignorespaces,
\normalsize{\texttt{\arxivref{1404.1079}{arXiv:1404.1079}}}\ignorespaces
\bibitem{Beem:2015aoa}
C.~Beem, M.~Lemos, L.~Rastelli \& B.~C. van~Rees,
\textit{``{The (2, 0) superconformal bootstrap}''},
\doiref{10.1103/PhysRevD.93.025016}{Phys.~Rev. \textbf{D93}, 025016
  (2016)\ignorespaces}\ignorespaces,
\normalsize{\texttt{\arxivref{1507.05637}{arXiv:1507.05637}}}\ignorespaces
\bibitem{Sonoda:1991mv}
H.~Sonoda,
\textit{``{Composite operators in QCD}''},
\doiref{10.1016/0550-3213(92)90675-2}{Nucl.~Phys. \textbf{B383}, 173
  (1992)\ignorespaces}\ignorespaces,
\normalsize{\texttt{\arxivref{hep-th/9205085}{hep-th/9205085}}}\ignorespaces
\bibitem{Ranganathan:1992nb}
K.~Ranganathan,
\textit{``{Nearby CFTs in the operator formalism: The Role of a connection}''},
\doiref{10.1016/0550-3213(93)90136-D}{Nucl.~Phys. \textbf{B408}, 180
  (1993)\ignorespaces}\ignorespaces,
\normalsize{\texttt{\arxivref{hep-th/9210090}{hep-th/9210090}}}\ignorespaces
\bibitem{Ranganathan:1993vj}
K.~Ranganathan, H.~Sonoda \& B.~Zwiebach,
\textit{``{Connections on the state space over conformal field theories}''},
\doiref{10.1016/0550-3213(94)90436-7}{Nucl.~Phys. \textbf{B414}, 405
  (1994)\ignorespaces}\ignorespaces,
\normalsize{\texttt{\arxivref{hep-th/9304053}{hep-th/9304053}}}\ignorespaces
\bibitem{Sonoda:1993dh}
H.~Sonoda,
\textit{``{Connection on the theory space}''},
International~Conference~on~Strings,~Berkeley,~California \textbf{{May 24-29,
  1993}}, 0154 (1993)\ignorespaces\ignorespaces,
\normalsize{\texttt{\arxivref{hep-th/9306119}{hep-th/9306119}}}\ignorespaces
\bibitem{Papadodimas:2009eu}
K.~Papadodimas,
\textit{``{Topological Anti-Topological Fusion in Four-Dimensional
  Superconformal Field Theories}''},
\doiref{10.1007/JHEP08(2010)118}{JHEP \textbf{1008}, 118
  (2010)\ignorespaces}\ignorespaces,
\normalsize{\texttt{\arxivref{0910.4963}{arXiv:0910.4963}}}\ignorespaces
\bibitem{Baggio:2017aww}
M.~Baggio, V.~Niarchos \& K.~Papadodimas,
\textit{``{Aspects of Berry phase in QFT}''},
\doiref{10.1007/JHEP04(2017)062}{JHEP \textbf{1704}, 062
  (2017)\ignorespaces}\ignorespaces,
\normalsize{\texttt{\arxivref{1701.05587}{arXiv:1701.05587}}}\ignorespaces
\bibitem{Kutasov:1988xb}
D.~Kutasov,
\textit{``{Geometry on the Space of Conformal Field Theories and Contact
  Terms}''},
\doiref{10.1016/0370-2693(89)90028-2}{Phys.~Lett. \textbf{B220}, 153
  (1989)\ignorespaces}\ignorespaces
\bibitem{Schwimmer:2019efk}
A.~Schwimmer \& S.~Theisen,
\textit{``{Osborn Equation and Irrelevant Operators}''},
\doiref{10.1088/1742-5468/ab3284}{J.~Stat.~Mech. \textbf{1908}, 084011
  (2019)\ignorespaces}\ignorespaces,
\normalsize{\texttt{\arxivref{1902.04473}{arXiv:1902.04473}}}\ignorespaces
\bibitem{Osborn:1991gm}
H.~Osborn,
\textit{``{Weyl consistency conditions and a local renormalization group
  equation for general renormalizable field theories}''},
\doiref{10.1016/0550-3213(91)80030-P}{Nucl.~Phys. \textbf{B363}, 486
  (1991)\ignorespaces}\ignorespaces
\bibitem{Bzowski:2018fql}
A.~Bzowski, P.~McFadden \& K.~Skenderis,
\textit{``{Renormalised CFT 3-point functions of scalars, currents and stress
  tensors}''},
\doiref{10.1007/JHEP11(2018)159}{JHEP \textbf{1811}, 159
  (2018)\ignorespaces}\ignorespaces,
\normalsize{\texttt{\arxivref{1805.12100}{arXiv:1805.12100}}}\ignorespaces
\bibitem{Argyres:1996eh}
P.~C. Argyres, M.~R. Plesser \& N.~Seiberg,
\textit{``{The Moduli space of vacua of N=2 SUSY QCD and duality in N=1 SUSY
  QCD}''},
\doiref{10.1016/0550-3213(96)00210-6}{Nucl.~Phys. \textbf{B471}, 159
  (1996)\ignorespaces}\ignorespaces,
\normalsize{\texttt{\arxivref{hep-th/9603042}{hep-th/9603042}}}\ignorespaces
\bibitem{DiVecchia:2017uqn}
P.~Di~Vecchia, R.~Marotta \& M.~Mojaza,
\textit{``{Double-soft behavior of the dilaton of spontaneously broken
  conformal invariance}''},
\doiref{10.1007/JHEP09(2017)001}{JHEP \textbf{1709}, 001
  (2017)\ignorespaces}\ignorespaces,
\normalsize{\texttt{\arxivref{1705.06175}{arXiv:1705.06175}}}\ignorespaces
\bibitem{Baggio:2012rr}
M.~Baggio, J.~de~Boer \& K.~Papadodimas,
\textit{``{A non-renormalization theorem for chiral primary 3-point
  functions}''},
\doiref{10.1007/JHEP07(2012)137}{JHEP \textbf{1207}, 137
  (2012)\ignorespaces}\ignorespaces,
\normalsize{\texttt{\arxivref{1203.1036}{arXiv:1203.1036}}}\ignorespaces
\bibitem{Dolan:2002zh}
F.~A. Dolan \& H.~Osborn,
\textit{``{On short and semi-short representations for four-dimensional
  superconformal symmetry}''},
\doiref{10.1016/S0003-4916(03)00074-5}{Annals~Phys. \textbf{307}, 41
  (2003)\ignorespaces}\ignorespaces,
\normalsize{\texttt{\arxivref{hep-th/0209056}{hep-th/0209056}}}\ignorespaces
\bibitem{Gadde:2010zi}
A.~Gadde, E.~Pomoni \& L.~Rastelli,
\textit{``{Spin Chains in N$=$2 Superconformal Theories: From the Z$_2$ Quiver
  to Superconformal QCD}''},
\doiref{10.1007/JHEP06(2012)107}{JHEP \textbf{1206}, 107
  (2012)\ignorespaces}\ignorespaces,
\normalsize{\texttt{\arxivref{1006.0015}{arXiv:1006.0015}}}\ignorespaces
\bibitem{Gadde:2009dj}
A.~Gadde, E.~Pomoni \& L.~Rastelli,
\textit{``{The Veneziano Limit of N = 2 Superconformal QCD: Towards the String
  Dual of N = 2 SU(N(c)) SYM with N(f) = 2 N(c)}''},
\normalsize{\texttt{\arxivref{0912.4918}{arXiv:0912.4918}}}\ignorespaces
\bibitem{Penati:2000zv}
S.~Penati, A.~Santambrogio \& D.~Zanon,
\textit{``{More on correlators and contact terms in N=4 SYM at order g**4}''},
\doiref{10.1016/S0550-3213(00)00633-7}{Nucl.~Phys. \textbf{B593}, 651
  (2001)\ignorespaces}\ignorespaces,
\normalsize{\texttt{\arxivref{hep-th/0005223}{hep-th/0005223}}}\ignorespaces
\bibitem{Lambert:2012qy}
N.~Lambert, C.~Papageorgakis \& M.~Schmidt-Sommerfeld,
\textit{``{Deconstructing (2,0) Proposals}''},
\doiref{10.1103/PhysRevD.88.026007}{Phys.~Rev. \textbf{D88}, 026007
  (2013)\ignorespaces}\ignorespaces,
\normalsize{\texttt{\arxivref{1212.3337}{arXiv:1212.3337}}}\ignorespaces
\bibitem{Niarchos:2020nxk}
V.~Niarchos, C.~Papageorgakis, A.~Pini \& E.~Pomoni,
\textit{``{(Mis-)Matching Type-B Anomalies on the Higgs Branch}''},
\normalsize{\texttt{\arxivref{2009.08375}{arXiv:2009.08375}}}\ignorespaces
\bibitem{Hama:2012bg}
N.~Hama \& K.~Hosomichi,
\textit{``{Seiberg-Witten Theories on Ellipsoids}''},
\doiref{10.1007/JHEP09(2012)033, 10.1007/JHEP10(2012)051}{JHEP \textbf{1209},
  033 (2012)\ignorespaces}\ignorespaces,
\normalsize{\texttt{\arxivref{1206.6359}{arXiv:1206.6359}}}\ignorespaces,
[Addendum: JHEP10,051(2012)]\ignorespaces
\bibitem{Alday:2009aq}
L.~F. Alday, D.~Gaiotto \& Y.~Tachikawa,
\textit{``{Liouville Correlation Functions from Four-dimensional Gauge
  Theories}''},
\doiref{10.1007/s11005-010-0369-5}{Lett.~Math.~Phys. \textbf{91}, 167
  (2010)\ignorespaces}\ignorespaces,
\normalsize{\texttt{\arxivref{0906.3219}{arXiv:0906.3219}}}\ignorespaces
\bibitem{Hayling:2017cva}
J.~Hayling, C.~Papageorgakis, E.~Pomoni \& D.~Rodríguez-Gómez,
\textit{``{Exact Deconstruction of the 6D (2,0) Theory}''},
\doiref{10.1007/JHEP06(2017)072}{JHEP \textbf{1706}, 072
  (2017)\ignorespaces}\ignorespaces,
\normalsize{\texttt{\arxivref{1704.02986}{arXiv:1704.02986}}}\ignorespaces
\bibitem{Pini:2017ouj}
A.~Pini, D.~Rodriguez-Gomez \& J.~G. Russo,
\textit{``{Large $N$ correlation functions in $\mathcal{N}=2$ superconformal
  quivers}''},
\normalsize{\texttt{\arxivref{1701.02315}{arXiv:1701.02315}}}\ignorespaces
\bibitem{Mitev:2015oty}
V.~Mitev \& E.~Pomoni,
\textit{``{Exact Bremsstrahlung and Effective Couplings}''},
\doiref{10.1007/JHEP06(2016)078}{JHEP \textbf{1606}, 078
  (2016)\ignorespaces}\ignorespaces,
\normalsize{\texttt{\arxivref{1511.02217}{arXiv:1511.02217}}}\ignorespaces
\bibitem{Mitev:2014yba}
V.~Mitev \& E.~Pomoni,
\textit{``{Exact effective couplings of four dimensional gauge theories with
  $\mathcal N=$ 2 supersymmetry}''},
\doiref{10.1103/PhysRevD.92.125034}{Phys.~Rev. \textbf{D92}, 125034
  (2015)\ignorespaces}\ignorespaces,
\normalsize{\texttt{\arxivref{1406.3629}{arXiv:1406.3629}}}\ignorespaces
\bibitem{Bershadsky:1998mb}
M.~Bershadsky, Z.~Kakushadze \& C.~Vafa,
\textit{``{String expansion as large N expansion of gauge theories}''},
\doiref{10.1016/S0550-3213(98)00272-7}{Nucl.~Phys. \textbf{B523}, 59
  (1998)\ignorespaces}\ignorespaces,
\normalsize{\texttt{\arxivref{hep-th/9803076}{hep-th/9803076}}}\ignorespaces
\bibitem{Bershadsky:1998cb}
M.~Bershadsky \& A.~Johansen,
\textit{``{Large N limit of orbifold field theories}''},
\doiref{10.1016/S0550-3213(98)00526-4}{Nucl.~Phys. \textbf{B536}, 141
  (1998)\ignorespaces}\ignorespaces,
\normalsize{\texttt{\arxivref{hep-th/9803249}{hep-th/9803249}}}\ignorespaces
\bibitem{ArkaniHamed:2001ca}
N.~Arkani-Hamed, A.~G. Cohen \& H.~Georgi,
\textit{``{(De)constructing dimensions}''},
\doiref{10.1103/PhysRevLett.86.4757}{Phys.~Rev.~Lett. \textbf{86}, 4757
  (2001)\ignorespaces}\ignorespaces,
\normalsize{\texttt{\arxivref{hep-th/0104005}{hep-th/0104005}}}\ignorespaces
\bibitem{Hill:2000mu}
C.~T. Hill, S.~Pokorski \& J.~Wang,
\textit{``{Gauge Invariant Effective Lagrangian for Kaluza-Klein Modes}''},
\doiref{10.1103/PhysRevD.64.105005}{Phys.~Rev. \textbf{D64}, 105005
  (2001)\ignorespaces}\ignorespaces,
\normalsize{\texttt{\arxivref{hep-th/0104035}{hep-th/0104035}}}\ignorespaces
\bibitem{Witten:1997sc}
E.~Witten,
\textit{``{Solutions of four-dimensional field theories via M theory}''},
\doiref{10.1016/S0550-3213(97)00416-1}{Nucl.~Phys. \textbf{B500}, 3
  (1997)\ignorespaces}\ignorespaces,
\normalsize{\texttt{\arxivref{hep-th/9703166}{hep-th/9703166}}}\ignorespaces
\bibitem{Lambert:2014jna}
N.~Lambert, C.~Papageorgakis \& M.~Schmidt-Sommerfeld,
\textit{``{Instanton Operators in Five-Dimensional Gauge Theories}''},
\doiref{10.1007/JHEP03(2015)019}{JHEP \textbf{1503}, 019
  (2015)\ignorespaces}\ignorespaces,
\normalsize{\texttt{\arxivref{1412.2789}{arXiv:1412.2789}}}\ignorespaces
\bibitem{Tachikawa:2015mha}
Y.~Tachikawa,
\textit{``{Instanton operators and symmetry enhancement in 5d supersymmetric
  gauge theories}''},
\doiref{10.1093/ptep/ptv040}{PTEP \textbf{2015}, 043B06
  (2015)\ignorespaces}\ignorespaces,
\normalsize{\texttt{\arxivref{1501.01031}{arXiv:1501.01031}}}\ignorespaces
\bibitem{Hayling:2018fmv}
J.~Hayling, R.~Panerai \& C.~Papageorgakis,
\textit{``{Deconstructing Little Strings with $\mathcal{N}=1$ Gauge Theories on
  Ellipsoids}''},
\doiref{10.21468/SciPostPhys.4.6.042}{SciPost~Phys. \textbf{4}, 042
  (2018)\ignorespaces}\ignorespaces,
\normalsize{\texttt{\arxivref{1803.06177}{arXiv:1803.06177}}}\ignorespaces
\bibitem{Hayling:2018fgy}
J.~Hayling, V.~Niarchos \& C.~Papageorgakis,
\textit{``{Deconstructing Defects}''},
\doiref{10.1007/JHEP02(2019)067}{JHEP \textbf{1902}, 067
  (2019)\ignorespaces}\ignorespaces,
\normalsize{\texttt{\arxivref{1809.10485}{arXiv:1809.10485}}}\ignorespaces
\bibitem{Minwalla:1997ka}
S.~Minwalla,
\textit{``{Restrictions imposed by superconformal invariance on quantum field
  theories}''},
Adv.~Theor.~Math.~Phys. \textbf{2}, 781 (1998)\ignorespaces\ignorespaces,
\normalsize{\texttt{\arxivref{hep-th/9712074}{hep-th/9712074}}}\ignorespaces
\bibitem{Buican:2016hpb}
M.~Buican, J.~Hayling \& C.~Papageorgakis,
\textit{``{Aspects of Superconformal Multiplets in ${\mathrm D}>4$}''},
\doiref{10.1007/JHEP11(2016)091}{JHEP \textbf{1611}, 091
  (2016)\ignorespaces}\ignorespaces,
\normalsize{\texttt{\arxivref{1606.00810}{arXiv:1606.00810}}}\ignorespaces
\bibitem{Cordova:2016emh}
C.~Cordova, T.~T. Dumitrescu \& K.~Intriligator,
\textit{``{Multiplets of Superconformal Symmetry in Diverse Dimensions}''},
\doiref{10.1007/JHEP03(2019)163}{JHEP \textbf{1903}, 163
  (2019)\ignorespaces}\ignorespaces,
\normalsize{\texttt{\arxivref{1612.00809}{arXiv:1612.00809}}}\ignorespaces
\bibitem{Gomis:2009xg}
J.~Gomis \& T.~Okuda,
\textit{``{S-duality, 't Hooft operators and the operator product
  expansion}''},
\doiref{10.1088/1126-6708/2009/09/072}{JHEP \textbf{0909}, 072
  (2009)\ignorespaces}\ignorespaces,
\normalsize{\texttt{\arxivref{0906.3011}{arXiv:0906.3011}}}\ignorespaces
\bibitem{Beem:2013sza}
C.~Beem, M.~Lemos, P.~Liendo, W.~Peelaers, L.~Rastelli \& B.~C. van~Rees,
\textit{``{Infinite Chiral Symmetry in Four Dimensions}''},
\doiref{10.1007/s00220-014-2272-x}{Commun.~Math.~Phys. \textbf{336}, 1359
  (2015)\ignorespaces}\ignorespaces,
\normalsize{\texttt{\arxivref{1312.5344}{arXiv:1312.5344}}}\ignorespaces
\end{thebibliography}

\end{document}